\newif\ifsubmode
\submodefalse

\newif\ifprintfig
\printfigfalse

\ifsubmode
  \documentstyle[12pt,aasms4,epsf]{article}
  \received{}
  \accepted{}
  \journalid{}{}
  \articleid{}{}
\else
  \documentstyle[11pt,aaspp4,epsf]{article}
\fi

\lefthead{Verdoes Kleijn et al.}
\righthead{HST observations of nearby radio-loud early-type galaxies}

\def\etal{{et al.~}}
\def\lae{\mathrel{<\kern-1.0em\lower0.9ex\hbox{$\sim$}}}
\def\gae{\mathrel{>\kern-1.0em\lower0.9ex\hbox{$\sim$}}}
\def\deg{^{\circ}}

\def\kms{km~s$^{-1}$}

\def\HalphaNII{H$\alpha$+[NII]}
\def\Msun{\>{\rm M_{\odot}}}

\def\pc{\>{\rm pc}}
\def\kpc{\>{\rm kpc}}

\begin{document}

\title{HST OBSERVATIONS OF NEARBY RADIO-LOUD EARLY-TYPE GALAXIES\altaffilmark{1}}

\author{%
Gijs A.~Verdoes Kleijn,\altaffilmark{2,3}
Stefi A.\ Baum,\altaffilmark{3}
P.\ Tim de Zeeuw,\altaffilmark{2}
Chris P.\ O'Dea\altaffilmark{3}
}


\altaffiltext{1}{Based on observations with the NASA/ESA Hubble Space 
       Telescope obtained at the Space Telescope Science Institute, which is 
       operated by the Association of Universities for Research in Astronomy, 
       Incorporated, under NASA contract NAS5-26555.}

\altaffiltext{2}{Sterrewacht Leiden, Postbus 9513, 2300 RA Leiden, 
The Netherlands.}

\altaffiltext{3}{Space Telescope Science Institute, 3700 San Martin Drive, 
Baltimore, MD 21218.}


\ifsubmode\else
\clearpage\fi


\ifsubmode\else
\baselineskip=14pt
\fi


\begin{abstract}
  We present and analyse {\tt HST}/{\tt WFPC2} broad- and narrow-band
  observations of the central regions of 19 nearby radio-loud
  early-type galaxies. Together with two more galaxies they form a
  complete sample of Fanaroff \& Riley Type I galaxies. We obtained
  $V$- and $I$-band images and narrow-band images centered on the
  \HalphaNII\ emission lines. We use archival data for six galaxies.
  
  We describe the data reduction, give isophotal fits and analyse the
  central emission-line gas and dust distributions.  Our main
  conclusions are the following.  Although obscuration by dust
  inhibits a direct determination of central cusp slopes, the data
  suggest most but not all galaxies have shallow cores. Dust is
  detected in all but two galaxies. There is a wide variety of central
  dust morphologies, ranging from central disks to lanes and irregular
  distributions. The analysis suggests the difference between disks
  and lanes is intrinsic and not due to different viewing angles.
  Central emission-line gas is detected in all galaxies. Extended
  low-surface-brightness emission is always associated with the dust
  features. In a subsequent paper we will present a detailed analysis
  of the relation between these central properties and the nuclear
  activity.
\end{abstract}



\clearpage

  
\section{INTRODUCTION}
Galaxies with radio jets are enigmatic objects. The sizes of the radio
jets can reach up to $\sim 1$ Mpc. Yet the central `engine' which
powers the radio jets resides at the center of the galaxy in an active
region $\sim 10^{-3}$ pc in radius: the spatially unresolved AGN. The
energy for the huge radio jets (with luminosities as large as
$10^{38}$ W between 10 MHz and 100 GHz) is generally believed to be
produced by the accretion of matter onto a central super-massive black
hole (hereafter BH) (e.g., Rees 1984; see Lynden-Bell 1996 for a
historical overview). This view is supported e.g. by short time-scale
variability of the central X-ray sources detected in active galaxies
(e.g., Wagner and Witzel 1995 for a review), and by the detection of
super-massive black holes in several nearby active and normal galaxies
(e.g., Ford \etal 1998; Ho 1998; Richstone 1998; van der Marel 1998
for summaries and reviews of recent results). Within this paradigm
there remain many fundamental unanswered questions concerning the
formation and evolution of AGN in radio-loud galaxies.  For instance,
what triggers the formation of a radio jet? Why are powerful radio
jets always found in early-type galaxies rather than in spirals? Why
do we see radio jets only in a few percent of early-type galaxies?
What is the origin of the material falling on to the BH? Is it
obtained from the host galaxy itself or is it obtained by merging or
galaxy harassment? What is the `feeding' mechanism of the massive BH:
how did the accreted material loose its angular momentum? And finally,
what causes the collimation of the jets?

Radio emission is observed from the cores of a considerable fraction
(at least 20\% at $M_B=-19.5$ and 50\% at $M_B= -22$) of all
early-type galaxies (e.g., Sadler 1997 and references therein). Only a
few of these galaxies also show radio jets. Ground-based studies of
nearby galaxies with radio jets have shown that they are predominantly
ellipticals with luminosity $\geq L_{*}$ (e.g., Owen and Laing 1989).
They tend to be less flattened than normal galaxies (Hummel \etal
1983; Disney \etal 1984), and tend to have boxy isophotes (Bender
\etal 1989). At equal luminosity, S0 galaxies are a factor ten weaker
radio emitters than elliptical galaxies (Hummel \& Kotanyi 1982;
Cordey 1986). Radio-loud early-type galaxies show dust more than twice
as often than their normal counterparts (van Dokkum \& Franx 1995).
The dust distributions appear to be broadly perpendicular to the radio
jet axis (Kotanyi \& Ekers 1979; de Koff \etal 1999).

With the advent of {\tt HST} the central regions of galaxies have been
studied in the optical with unprecedented high resolution. Studies of
samples of nearby normal dust-free ellipticals reported a dichotomy in
the types of central surface brightness profiles (e.g., Jaffe \etal
1994; Lauer \etal 1995). Some galaxies have surface brightness
profiles, $I(r) \sim r^{-\gamma}$, with $\gamma > 0.5$ (i.e., a steep
cusp) while others have profiles with $\gamma < 0.3$ (i.e., a shallow
cusp).  High-luminosity galaxies (M$_{\rm B} \leq -21.5$)
predominantly have shallow cusps while lower luminosity galaxies
predominantly have steep central cusps. However, a study of the nuclei
of ellipticals with counter rotating cores (Carollo \etal 1997)
suggests a rather smooth transition zone between the two classes
instead of a bimodal distribution. The central regions of both normal
and radio-loud galaxies turned out to be a diverse and complex
environment. Small nuclear gas and dust disks with sizes of a few
hundred parsec were discovered in several galaxies (Jaffe \etal 1993;
De Juan \etal 1996; Baum \etal 1997; van der Marel \& van den Bosch
1998; de Koff \etal 1999). The central gas disks in M 87 (Ford \etal
1994) and M 84 (Bower \etal 1997) were studied in detail. They both
lie in an irregular dust and gas distribution already known from
ground-based work (e.g., Kormendy \& Stauffer 1987). Super-massive BHs
were detected by modeling the kinematics of the nuclear emission-line
disks in M 87 (Harms \etal 1994), NGC 4261 (Ferrarese \etal 1996), M
84 (Bower \etal 1998) and NGC 7052 (van der Marel \& van den Bosch
1998). Previously unknown optical jets were detected in several nearby
radio-loud galaxies (e.g., Sparks \etal 1994; Baum \etal 1997, Martel
\etal 1998).

BH masses display a roughly linear correlation with total luminosity of
the host bulge (Kormendy \& Richstone 1995; van der Marel 1998), but
there is an observational bias against detecting small BHs in big
galaxies, as well as a large scatter at fixed luminosity. The absolute
magnitude for the four nearby radio-loud galaxies in which BHs were found
varies by only 0.2 while their BH masses differ by a factor $\sim 10$
and their radio power varies by a factor $\sim 70$. 

Given the {\tt HST} results for nearby normal galaxies and the
variety of features observed in radio galaxies, it is important to
survey the central regions of nearby radio-loud galaxies in a
systematic and unbiased way. How do the central regions of radio-loud
early-types differ from the central regions of their radio-quiet and
normal counterparts?  What is the range of inner slopes of the
luminosity profiles? Are there `reservoirs' of matter such as disks of
gas and dust in all radio-loud galaxies? How do the properties of the
dust and gas distributions, such as mass, kinematics and orientation,
correlate with the properties of the radio jets? How do the inner
radio jets and the central interstellar medium (hereafter ISM)
interact? What is the relation between optical luminosity, radio
luminosity and BH mass for radio-loud galaxies?

Answers to the above questions require a systematic and
observationally unbiased approach. Therefore we defined a complete
sample of 21 nearby radio-loud early-type galaxies limited in
radio-flux and redshift (see Sec.~\ref{s_sample}). We carried out a
program to obtain images for 19 galaxies. For six of those galaxies we
use archival {\tt HST} data. We obtained broad-band $V$ and $I$ images
to study the central light distributions, color and the distribution
of dust. We also obtained narrow-band images centered on the
\HalphaNII\ emission lines to probe the emission gas at the centers of
these galaxies.

The purpose of this paper is to present the sample and the observations, to describe in
detail the data reduction and to give a preliminary analysis of the dust
and gas emission for the 19 galaxies. In a subsequent paper we will
present a detailed analysis of the properties of the centers of these
galaxies and their relation to the nuclear activity. Studies of the
same sample at radio and X-ray frequencies are described in Xu
\etal (1999a,b).

The paper is organized as follows. In Sec.~\ref{s_observations} we
introduce the sample and describe the observations and reduction of
the images. We discuss the procedure and results for the isophotal
analysis in Sec.~\ref{s_isophotes}, for the emission gas analysis in
Sec.~\ref{s_emission} and for the dust analysis in Sec.~\ref{s_dust}.
A summary of the results and conclusions is given in
Sec~\ref{s_summary}. A description for the individual galaxies is
given in Appendix \ref{s_galaxies} and details on the derivation of
optical depths is given in Appendix \ref{s_opt_depth}.

Throughout the paper we use a Hubble constant $H_0$= 75
kms$^{-1}$Mpc$^{-1}$.

\section{OBSERVATIONS AND IMAGE DATA REDUCTION}
\label{s_observations}

\subsection{The Sample}
\label{s_sample}
Our objective was to select a well-defined complete sample of nearby
galaxies with radio jets to study in a systematic way and with high
spatial resolution their central properties. The complete sample is
drawn from a catalog of 176 radio-loud galaxies constructed by Condon
\& Broderick (1988, hereafter CB88). The catalog was constructed by
position coincidence of radio identifications in the Green Bank 1400
MHz sky maps (Condon \& Broderick 1985, 1986; CB88) and galaxies in
the {\it Uppsala General Catalogue of Galaxies} (Nilson 1973,
hereafter referred to as UGC).  Our sample consists of all galaxies in
the CB88 sample which meet the following requirements: (1) Hubble type
classification E or S0, (2) recession velocity $< 7000$ \kms , (3)
optical major axis diameter $> 1'$ (this requirement follows from the
UGC), (4) flux density $> 150$ mJy at 1400 MHz, (5) $-5\deg < \delta <
82\deg$, (6) having a `monster' rather than a starburst energy source,
as discerned from the IRAS/Radio flux ratio (CB88) and (7) a radio
size $\geq 10''$ as measured from 1490 MHz VLA A array maps (see Xu
\etal 1999a for a detailed description of the radio properties of the
sample). The last requirement was chosen to ensure we only selected
galaxies with radio jets. The resulting sample consists of 21
galaxies. Distances to the galaxies range between 15 Mpc and 90
Mpc. The sample covers a range of two magnitudes in absolute blue
luminosity and a factor $\sim 800$ in radio-power
(Fig.~\ref{f_lum}). The complete sample of galaxies is listed in
Table~\ref{t_sample}, together with some general properties. All
galaxies are classified as Fanaroff \& Riley (1974) Type I (FR I)
galaxies (Xu \etal 1999a).

\subsection{The Data}
We observed 14 galaxies with the {\tt WFPC2} on board {\tt HST} using
the two broad-band filters {\tt F555W} and {\tt F814W} and 15 galaxies
with a narrow-band filter centered on the \HalphaNII\ emission lines.
No HST data were taken for UGC 7115 and UGC 12064.  For the remaining
galaxies we used archival data with filters matching as closely as
possible our program filters.  The {\tt F555W}/{\tt F547M} and {\tt
F814W}/{\tt F791W} filters correspond closely to the Johnson $V$- and
Cousins $I$-band, respectively.  Thus in total we have broad and
narrow band observations for 19 of our 21 galaxies.  Table \ref{t_obs}
lists the observational setup for all exposures.  The point spread
function (hereafter PSF) of the {\tt WFPC2} images has a typical FWHM
$\sim 0.07''$.  All observations were taken with the telescope guiding
in fine lock, which typically gives an rms tracking-error of
$0.003''$. A detailed description of the {\tt WFPC2} instrument is
given in the WFPC2 Instrument Handbook version 4.0 (Biretta \etal
1996).

\subsection{Image Data Reduction}
All images were calibrated with the standard {\tt WFPC2} pipeline
using the most up-to-date calibration observations \footnote{No
  flatfields are available for the LRF filters. For these observations
  we used the nearest narrow-band flatfield available. This does not
  affect our results in any significant way.}. This pipeline and other
reduction tasks are available in the IRAF STSDAS package. We refer to
the HST Data Handbook version 3.1 (ed. Voit 1998) for a detailed
description of the calibration process.

For each broad-band filter two images were taken back to back to allow
for cosmic-ray removal. For the narrow-band and LRF images, which have
significantly longer exposure times, 3 or 4 images were taken back to
back.  We added the images and removed cosmic-rays with the `crrej'
task using an appropriate noise model for the {\tt WPFC2}. Next we
removed `hot' pixels from the images using the `warmpix' task. We then
interpolated over remaining bad pixels using the `wfixup' task. The
final images are shown in Figure \ref{f_ima}.

\section{Isophotal Analysis}
\label{s_isophotes}

\subsection{Reduction}
We checked the alignment between the $V$ and $I$ band images using
Galactic stars and/or globular clusters. If these objects could not be
used we checked the alignment using the galaxy nucleus. Only a
small sub-pixel shift was performed in one case (NGC 4261).  All other
images were aligned within 0.2 pixels (i.e. $0.01''$) and therefore no
shifts for further alignment were performed.

We used the `ellipse' task to perform isophotal fits to the light
distributions of our targets.  A detailed description of the method is
given by Jedrzejewski (1987).  We identified regions on the CCD which
were not to be considered in the isophotal fitting procedure. These
regions contain foreground stars, companion galaxies, globular
clusters or regions affected by dust. The latter were identified by
eye using the ratio of the $V$ and $I$ images in which local dust
distributions clearly show up due to the differential opacity.
However, this method is not sensitive to global smooth distributions
of dust which might be present. A first fit was performed to the light
distribution allowing center, position angle and ellipticity to vary
freely for each radius. From these fits we determined the center of
the luminosity distribution by taking the average of the center of
ellipses within a certain radial range. A second fit was then
performed but now with the center fixed during the fitting process.
This procedure was performed separately on both the $V$- and $I$-band
image for each galaxy.  Additionally, the isophotal shape parameters
determined for the $I$-band image were then used to determine the
luminosity profile for the $V$-band image using these same parameters.
We compared the resulting position angle (hereafter PA), ellipticity
and Fourier coefficients from these different fitting procedures. Only
insignificant differences were found which are likely due to low-level
small dust features which could not be detected by inspecting the
ratio image of $V$ and $I$.

We derived a model for the galaxy from the isophotal fits, using the
`bmodel' task. We inspected the residuals between these model galaxies
and the real galaxy image, and verified that the isophotal fits were
indeed an accurate representation of the real galaxy. For several
galaxies the dust distributions inhibited a free isophotal fit
to the central regions. In these regions we fixed the isophotal parameters
to the last isophotal fit performed just outside the dust region.
Hence these inner surface brightness profiles are severely affected by
dust obscuration.  For the broad-band images it was not possible to
determine the sky brightness from the {\tt PC} images since the whole
area of the {\tt PC} chip and sometimes the {\tt WF} chips as well
were affected by light from the galaxies. We fitted a de Vaucouleurs
profile to the region where the light contribution from the sky was
negligible and extrapolated this profile to the regions on the {\tt
  WF} chips affected by sky light. Thus the sky counts were defined as
the difference between the observed counts and counts predicted from
the extrapolation. The sky magnitudes have a narrow distribution
around 22.5 mag arcsec$^{-2}$ in $V$ and 21.6 mag/arcsec$^{-2}$ in
$I$.  The error in the sky counts is estimated to be 25\%. We
subtracted the sky counts from the surface brightness profile and
determined $V$ and $I$ magnitudes using the `synphot' task which yield
typical errors in the absolute calibration of $\sim 0.05 \rm{mag}$.
Corrections were made for Galactic foreground extinction using optical
extinction values $A_V$ given by NED\footnote{The NASA/IPAC
  Extragalactic Database (NED) is operated by the Jet Propulsion
  Laboratory, California Institute of Technology, under contract with
  the National Aeronautics and Space Administration.}  which range
between 0.0 and 0.44. The final luminosity profiles agree well with
those inferred from ground-based data (see Appendix \ref{s_galaxies}).
The derived isophotal parameters for the sample are shown in Figure
\ref{f_photoma}.

\subsection{Results}
\label{s_iso_res}

\noindent
{\sl Luminosity Profiles} \\ The luminosity profiles of galaxies with
negligible or no dust affecting their isophotes clearly show a break
radius at which the profile flattens off towards the center. What can
we say about the luminosity profiles affected by dust obscuration? To
get an indication we constructed `un-extincted' luminosity profiles in
Figure \ref{f_photoma}, using several assumptions. We assume that (i)
the dust is distributed in a screen with a fixed fraction $D$ of the
total stellar light along the line of sight in front of it, (ii) the
intrinsic color of the galaxy in the dust affected region is equal to
the one just outside the dusty region (a plausible assumption based on
the color profiles of galaxies without central dust) and (iii) a
Galactic extinction law is valid. We can then use the color excess to
derive the optical depth (see Appendix \ref{s_opt_depth}) and hence
the un-extincted surface brightness. Figure~\ref{f_dustcorrect} shows
the extinction in $V$ magnitudes due to dust obscuration as function
of observed color-excess for different screen positions. This figure
illustrates that the observed color-excess provides a constraint on
the un-extincted surface brightness only when $D$ is known a priori.
In reality, the distribution of the dust will certainly be more
complex than that of a simple screen. To give nevertheless an
indication of the possible corrections to the observed surface
brightness profiles we plotted in Figure \ref{f_photoma} the
un-extincted profiles assuming a front-screen (i.e., $D$=0; dashed
line) and a `mid-screen' (i.e., D=0.5; dotted lines). One of the
mid-screen un-extincted profiles is almost identical to the front
screen approximation (cf. Figure~\ref{f_dustcorrect}). Only color
excess values in the range $[0,0.2]$ can be observed for the
mid-screen assumption (cf. Figure~\ref{f_dustcorrect}. The observed
color excess in the central regions of the sample galaxies lies in
most cases exactly within this range. This indicates that the screen
approximations are at least consistent with the observed
extinction. Our method is formally only correct if we
observe a constant color-excess along each isophote. This is generally
not true. However, simulations show that under the screen assumptions
the method recovers the correct un-extincted isophotal magnitude to
within $0.05^{m}$ for the typical observed color excess variations
along isophotes. The un-extincted profiles show that the majority of
the galaxies which do not show a break likely do have a break in their
profile at some radius inside the dust affected region.  But there are
at least 2 galaxies (NGC 2329 and NGC 4335) for which a power-law all
the way into the central region is probable assuming the central dust
is located near the center of the galaxy. These two galaxies also have
a high central surface brightness which is typical for galaxies with a
steep central cusp. The absolute magnitudes in our sample scatter
around the dividing range of magnitudes between `cuspy' and `shallow'
cores (Faber \etal 1997). Hence it would be interesting to determine
the distribution of central cusp slopes and break radii. This is not
possible with our optical imaging due to the central dusty regions but
requires high resolution near-infrared imaging. From the current data
we can conclude that although obscuration by dust inhibits a direct
determination of central cusp slopes, the data suggest most but not
all galaxies have shallow cores.

\noindent
{\sl Color Profiles} \\ The $V-I$ color profiles
(Fig.~\ref{f_photoma}) provide limited information on the color of the
stellar populations. In the inner regions most of our galaxies are
severely affected by dust while in the outer regions the errors due to
uncertain sky subtraction become large. The central color just outside
the dust distributions in our sample range between 1.25 and 1.45. This
agrees well with $V-I$ colors reported in other studies of giant
ellipticals (Bender and M\"{o}llenhoff 1987; Carollo \etal 1997).
Therefore we can only conclude that the sample galaxies have $V-I$
colors typical for giant ellipticals.

\noindent
{\sl Position Angles} \\ The variation of the position angle of the
stellar isophotes over the measured range of radii is less than
$10\deg$ for $\sim 50\%$ of the galaxies. The other half displays
significantly larger isophotal twists. These could be due to low
ellipticity and/or very shallow luminosity profiles. An additional
possibility is the presence of small low-level dust obscuration which
can not be detected from our $V$-$I$ image and hence were missed
during the masking (e.g., possibly in NGC 5127). Two galaxies, UGC
1841 and NGC 7626, show large twists which are undoubtly real (see
also Sections \ref{s_1841} and \ref{s_7626}). PA twists might be caused
by either triaxiality or gravitational interaction.

\noindent
{\sl Ellipticities} \\ Using galaxy models we tested how the
ellipticity of the inner isophotes is affected by the convolution with
the PSF of {\tt HST}. It turns out that this effect is negligible
outside $r = 0.25''$ for our sample galaxies. In $\sim$ 35\% of the
galaxies the ellipticity varies by 0.1 or more in the inner
$10''$. Another $\sim$ 30\% show variations between 0.05 and 0.1. The
remaining $\sim$ 35\% have variations less than 0.05. We compared this
with ellipticity variations in the inner $10''$ of 41 radio-quiet,
giant elliptical galaxies (i.e., $M_V \lae -19.5$), based on HST
studies published by Lauer \etal (1995) and Carollo \etal (1997). For
this sample of normal galaxies $\sim$ 45\%, $\sim$ 40\% and $\sim$
15\% show variations larger than 0.1, between 0.05 and 0.1 and less
than 0.05, respectively.  Thus radio-loud galaxies tend to have less
variation in their ellipticity profiles in the inner $10''$. However,
this difference might be (partly) caused by the higher frequency of
central dust in the inner few arcseconds in our radio-loud sample
which could conceal variations.

\noindent
{\sl Fourier Coefficients} \\ The fourth order Fourier coefficient,
$c_4$, describes the deviations of the isophotes from perfect
ellipses. Negative values of $c_4$ indicate `boxy' isophotes while
positive values indicate `disky' isophotes.  In our sample departures
from 0 are generally small (i.e.  $< 0.02$) in the inner regions but
sometimes persist over several arcseconds. There are two notable
exceptions: UGC 1841 becomes very 'disky' around $10''$ and NGC 4261
is very boxy outward of $2''$. In our sample there are roughly equal
numbers of galaxies with positive, negative $c_4$ and no
departures. Bender \etal (1989) found that radio-loud galaxies tend to
have `boxy' isophotes at radii larger than several arcseconds.
Interestingly, this is clearly not the case for the central few
arcseconds of our radio-loud galaxies. The sample of Bender \etal
includes all galaxies in our sample which are boxy but does not
include most of the galaxies with elliptical or disky isophotes. We
compared our central $c_4$ values with profiles beyond $\gae 10''$
reported in previous studies (see Section~\ref{s_galaxies}). In all
but one of these the isophotal shapes are either similar or boxier at
$r \gae 10''$ than in the central region. The one exception is NGC
541.

\section{Emission Gas Analysis}
\label{s_emission}

\subsection{Reduction}
\label{s_emi_red}
To derive the flux from the \HalphaNII\ emission lines we subtracted
the contribution of stellar continuum light from the narrow band
(i.e., the `on-band') fluxes. We used the $V$- and $I$-band as
`off-band' images to estimate this continuum. Using these off-bands
poses two potential problems. First, regions with gas emission are
often affected by dust which has an opacity varying with
wavelength. However, by estimating the continuum light distribution in
the narrow-band images from a combination of the broad-band images on
either side of the H$\alpha$ frequency we are able to make a
correction for the effects of dust as explained in more detail
below. Second, the $V$-band filters include several emission lines
(e.g., the far end of the {\tt F555W} filter includes the \HalphaNII\
emission lines). The $I$-band filters do not contain emission
lines. We determined the contribution of emission-line flux to the
stellar flux in the $V$ band filters using our determined \HalphaNII\
fluxes and assuming for each galaxy LINER emission-line ratios as
found for M 87 by Dopita \etal (1997).  The emission flux contributes
at most 2.5\% and is thus negligible. For NGC 4261 we used the {\tt
F675W} $R$-band image from the archive to measure the \HalphaNII\
emission flux since no {\tt HST} narrow-band observation is
available. This passband includes also the [OI]$\lambda\lambda$6300,
6364 and [SII]$\lambda\lambda$6717, 6731 lines. For this galaxy, the
latter lines contribute 50\% of the emission-line flux in {\tt F675W}
(using again the M 87 line ratios). We did not make corrections for
the slightly narrower and broader PSF of the $V$ and $I$ filters,
respectively, compared to the PSF of the narrow-band filter. We
verified that by using the combination of $V$ and $I$ images as
off-band image, these PSF differences alter the emission flux
values by less than 5\%.

First the broad-band images were aligned with the narrow-band image,
using foreground stars and globular clusters or companion galaxies. In
a few cases where these reference points were not available we assumed
that the isophotal center of the narrow- and broad-band images
coincide. For targets with narrow-band images on the {\tt WF2} we
rebinned the broad-band images on the {\tt PC} to the larger {\tt WF}
pixel scale. From isophotal fits we determined the ratio between the
on- and off-band fluxes in regions devoid of emission-line gas and
dust. These ratios could be well fitted as a linear function of
radius. Thus we could extrapolate this linear fit to regions with gas
emission. We determined the observed emission flux $F^{\rm obs}_{\rm
e}$ using the following equation:
\begin{equation}
F^{\rm obs}_{\rm e}=F^{\rm obs}_{NB}- 
\left({{r}_V} \cdot F^{\rm obs}_V \cdot {{\rm 
r}_I} \cdot F^{\rm obs}_I\right)^{1/2}.
\label{e_emi1}
\end{equation}
Here $F^{\rm obs}_j$ is the flux measured through filter $j$ with $NB$
denoting the narrow-band filter. The parameters ${r}_V$ and ${r}_I$
are the ratios of the continuum light in the narrow-band and the $V$-
and $I$-band, respectively, and vary with radius. Next we explain that
by using this combination of $V$ and $I$ as off-band we make a
correction for the differential opacity at $V$, $NB$ and $I$. To
include the obscuration by dust we can write the more general version
of eq.~(\ref{e_emi1}) as:
\begin{equation}
F^{\rm obs}_{\rm e}=F^{\rm obs}_{NB}- 
\left({{r}_V} \cdot F^{\rm obs}_V \cdot {{\rm 
r}_I} \cdot F^{\rm obs}_I\right)^{1/2}
\cdot \left(\frac{{\rm c}(NB) \cdot {\rm c}(NB)}{{\rm c}(V) 
\cdot {\rm c}(I)}\right)^{1/2}.
\label{e_emi2}
\end{equation}
Here the correction factors $c(j)$ indicate the ratio between observed
and intrinsic flux through filter {\sl j}.  If one assumes a `front
screen' of dust $c(j)$ becomes:
\begin{equation}
{\rm c}(j)=e^{-\alpha_j \tau},
\label{e_corr}
\end{equation}
where $\alpha_j$ is the opacity coefficient for filter $j$. If we
define the `total extinction factor' $C$ as:
\begin{equation}
{\rm C}=\left(\frac{{\rm c}(NB) 
\cdot {\rm c}(NB)}{{\rm c}(V) \cdot {\rm c}(I)}\right)^{1/2},
\end{equation}
and we take $\alpha_{\rm V} = 1.0$, $\alpha_{\rm NB}=0.75$,
$\alpha_{\rm I}=0.5$ (consistent with a Galactic extinction law
normalized to $V$), then $C$ equals exactly 1 regardless of optical
depth and eq.~(\ref{e_emi2}) reduces to eq.~(\ref{e_emi1}). Moreover,
$C$ remains within 2\% percent of 1 for reasonable values of $\tau$ if
one assumes the dust screen is placed in the middle of the galaxy
(i.e., 50\% of the stellar light originates in front and behind the
screen) or if one assumes the dust is homogeneously mixed along the
line of sight (see Fig.~\ref{f_extemi}). For the latter two
dust configurations the use of eq.~(\ref{e_emi1}) yields an
under-estimate of the emission flux in areas affected by dust. The
absolute flux calibration for the narrow-band filters for each galaxy
was determined with the `synphot' package which models the throughput
for the {\tt HST} filters and assuming all lines have zero width.

\subsection{Results}
\label{s_emi_res}

The final \HalphaNII\ emission images are shown in Figure \ref{f_ima}.
The derived fluxes are summarized in Table \ref{t_emi}. Derived
luminosities are typical for FR I galaxies (e.g., Zirbel \& Baum
1995). All galaxies clearly show compact \HalphaNII\ emission in their
nucleus. In addition, some galaxies show extended
low-surface-brightness emission features almost always associated with
dust. Individual descriptions of the morphology of the emission gas
can be found in Appendix \ref{s_galaxies}). Four remarks can be made
concerning the final flux values. First, all emission fluxes are {\sl
observed} fluxes, i.e., are not corrected for extinction by dust. We
did not attempt to correct for extinction because we can not reliably
determine the distribution of dust relative to the emission
gas. However, when we assume a front-screen of dust we can determine
the amount of extinction using the observed $V-I$ color-excess per
pixel in each galaxy (see Appendix~\ref{s_opt_depth}. It turns out
that in this configuration the intrinsic flux can be typically at most
40\% larger than the observed flux. Second, given the typical $S/N$ of
the emission images, we are only able to detect emission features with
a surface brightness larger than $\sim 5 \cdot 10^{-15}$
ergs$^{-1}$cm$^{-2}$arcsec$^{-2}$.  Thus extended
low-surface-brightness features below this level will be undetected
but might make up for a significant fraction of the total emission
when integrated over the galaxy. Third, some galaxies reveal a nuclear
point source that is bluer than the surrounding stellar light (see
Section~\ref{s_galaxies}. We did not adjust $r_V$ and $r_I$ for
this. We verified that this results in an over-estimate of the total
absolute flux up to $\sim 20\%$, assuming that the blue point source is
caused by the contribution of a non-stellar continuum flux which
results in the observed color. Fourth, the LRF, {\tt F658}, {\tt F673}
filters have equivalent widths of $\sim 85${\AA}, $28.5${\AA} and
$47.0${\AA}, respectively and therefore a fraction of broad-line
\HalphaNII\ emission, if present, might fall outside the latter two
passbands.

We compared the derived fluxes with values from the literature (see
Table \ref{t_emi}). The fluxes agree reasonably well except for two
galaxies: NGC 741 and NGC 7626. For NGC 741 Macchetto \etal (1996)
report an extended disk of emission.  The average flux-density in this
disk is well below our detection threshold. We could not find a reason
for the discrepancy of NGC 7626.

\section{DUST ANALYSIS}
\label{s_dust}

Dust is detected in 17 out of the 19 radio galaxies. This is
consistent with the frequency of 72\% $\pm$ 16\% found by van Dokkum
\& Franx (1995) for another sample of radio galaxies imaged with
{\tt HST}.

Both galaxies without detected dust, NGC 741 and NGC 2892, are
relatively far away. However, the minimal size of detected dust in our
sample is 200 pc, which would still be resolved with {\tt HST}
($0.6''$ and $0.45''$ for NGC 741 and NGC 2892, respectively). It is
possible that the location and/or optical depth of the dust feature
produces too little reddening to be detected. Both galaxies are round
ellipticals with large shallow cores (similar to M 87 which itself has
little dust). They are also the only two galaxies in the sample with a
very nearby apparent companion (angular separation $\leq 10''$).  Both
companions are featureless round ellipticals. These similarities
suggest there may be similar reasons for not detecting the dust;
either projection effects, or dust destroying processes in the galaxy
itself, or tidal interaction with the companion. In this respect it is
also noteworthy that Macchetto \etal (1996) did report a disk of gas
emission in NGC 741.
 
The dust properties are summarized in Table \ref{t_dust}.  There
appear to be two distinct main classes of apparent (i.e., projected)
dust morphologies: lanes and disks. To what extent is this distinction
intrinsic or just due to different viewing angles? We found several
differences which point in the direction of an intrinsic different
morphology. The classification is robust: observed morphologies fall
clearly in either classification bin.  Only for NGC 5490 is the
classification ambiguous but this is due to the very small size of the
dust feature. The disks have a quite `smooth' appearance. Their
outlines closely resemble ellipses with sometimes spiral structures
inside and small protuberances on the outside. Lanes have a quite
irregular appearance. They show warps or bends and are often
surrounded by secondary filaments and patches of dust.  Dust disks
tend to be closely aligned with the major axis of the stellar
distribution while dust lanes do not display a clear preference of
orientation (Fig.~\ref{f_phi}). In summary: there are various
indications that the morphological difference between dust disks and
lanes is intrinsic.

Physical sizes of the dust features range from 200 pc to 4.5 kpc. Dust
features are typically smaller than $\sim 1 \kpc$. The few features
larger in size all show peculiar morphologies compared to the sub-kpc
morphologies. This suggests that at least some of the larger dust features
are still settling towards stable orbits. Physical sizes do not depend on
the distance of the galaxy which suggests that our capability in
detecting the dust features does not strongly depend on distance.
There is also no indication of a correlation between size and
morphology: both lanes and disks of similar sizes are present.

In general the minor axis of both lanes and disks appears to align
roughly with the radio-jet axis (Fig.~\ref{f_phi}). A notable
exception is NGC 7052 which has a misalignment of $\sim 40^{\circ}$ .
In NGC 3862 (3C 264) the optical jet interacts with the dust (Baum \etal
1998) which implies a large misalignment assuming the dust disk is physically
thin. At the scales of 100 pc or larger, as detected here, the orientation
of the dust lanes is influenced by the gravitational potential and by
the pressure gradients of the hot gas. The relative importance of
these forces, the destruction time scales, and dynamical evolution
(i.e., settling of the dust in stable morphologies) are the factors
determining the observed (mis)alignments (e.g., Quillen \& Bower
1999). An analysis of the dust dynamics falls outside the scope of
this paper and is deferred to a subsequent paper.

The dust masses were determined under the assumption that the dust is
distributed in a screen at a given position with respect to the galaxy
light.  The dust mass $M_{\rm dust}$ is proportional to the optical
extinction $A_V$ (Sadler \& Gerhard 1985; van Dokkum \& Franx 1995):
\begin{equation}
M_{\rm dust}=\Sigma \cdot {\Gamma_V}^{-1} 
\cdot A_V= \Sigma \cdot {\Gamma_V}^{-1} 
\cdot -2.5 \log e^{-\alpha_V \tau}.
\end{equation}
Here $\Sigma$ is the area of the dust feature, $\Gamma_V$ the visual
mass absorption coefficient and $\alpha_V$ the extinction coefficient at
$V$. The optical depth $\tau$ is determined as described in Appendix
\ref{s_opt_depth}. We adopt the Galactic value $\Gamma_V \sim 6 \times
10^6$ mag kpc$^2 M_{\odot}^{-1}$ since differences between the
Galactic and elliptical galaxy extinction curves are generally small
(Goudfrooij \etal 1994b).  Table \ref{t_dust} lists the mass for two
screen positions: in front of and halfway into the galactic light
distribution. The dust mass increases typically by a factor
of 3 by placing the screen halfway instead of in front of the light
distribution. There are two other main uncertainties affecting the
mass estimates. First, the dust geometry is certainly more complex
than a simple screen at one location in the galaxy. Second, there is
no one-to-one connection between dust geometry and color excess (Witt
\etal 1992; Witt \& Gordon 1996 and see also Sec.~\ref{s_iso_res}).
Therefore, the derived dust masses should be treated with caution and
have mainly significance in a relative sense as a measure of the
projected dust mass.  Dust masses are usually smaller than a few times
$10^4 M_\odot$. The more massive features show also the more extended
peculiar morphologies, which are likely still settling down.

\section{SUMMARY AND CONCLUSIONS}
\label{s_summary}
We have presented {\tt HST/WFPC2} broad- and narrow-band observations
of 19 radio-loud early-type nearby galaxies which are part of a
complete sample of 21 nearby Fanaroff \& Riley Type I galaxies. We
have described the data reduction, given isophotal fits and analysed
the central emission-line gas and dust distributions. The general
conclusion is that the centers of these FR I galaxies, apart from the
ubiquitous presence of dust and gas distributions, resemble the
centers of normal bright ellipticals. The galaxies have fairly round
(ellipticity $\leq 0.3$) stellar isophotes. Our radio-loud early-type
galaxies seem to have on average less variation in their ellipticity
profiles in the inner $10''$ compared to normal early-types of similar
luminosity. However, this difference might be caused by the frequent
obscuration by dust of the central regions of our sample galaxies. The
shape of the isophotes is mostly close to perfect ellipsoidal ($|c_4|
< 0.02$). Ground-based studies show that the isophotal shapes either
remain similar or become boxier at radii beyond $10''$. The
dust inhibits a direct measurement of the inner slope of the surface
brightness profile. We applied simple models in which the dust is
represented by a screen to estimate the possible range in slope for
the un-extincted light profile in each galaxy. Most galaxies have
profiles which do flatten off towards the nucleus. Two galaxies have
light profiles which are consistent with a single power law
slope. Thus at this stage both shallow and cuspy cores seem able to
host an active nucleus which produces radio jets.  Dust is detected in
the cores of all but two sample galaxies.  The dust morphology can be
divided in three categories: disks, lanes and irregular
distributions. Both disks and lanes have sizes ranging from a few
hundreds of parsec to a few kiloparsec.  The major axis of the disks
is always aligned within a few degrees with the galaxy major axis
while lanes have misalignments of tens of degrees and often show
twists and warps. Therefore it is suggested that the difference
between disks and lanes is intrinsic and not solely due to different
viewing angles. The apparent major axis of both disks and lanes is
generally roughly perpendicular to the radio jet axis.  Two notable
exceptions are NGC 7052 and NGC 3862. Using the narrow-band
observations, we detected emission-line gas in the nucleus of each
galaxy in our sample. Gas emission is also commonly associated with
the dust distributions.

The dust and gas in the cores of these FR I galaxies provides a
natural fuel supply for the active nucleus.  The question which then
can be raised is why the fraction of normal early-type galaxies which
do show similar dust distributions in their cores do not host an AGN.
Do these galaxies lack a massive BH to accrete the material and
convert the gravitational energy? Or is the fuel, although clearly
present, not funneled effectively to the nucleus due to differences in
the shape of the central gravitational potential well or differences
in the ISM (e.g., pressure)? Thus it is important to perform a
detailed comparitive study of these nearby active cores and normal
quiescent cores. In this context we will obtain HST/STIS observations
for the nuclei of the complete radio-loud sample to study the state
and kinematics of the emission gas and constrain the mass of the BH.


\acknowledgments

Support for this work was provided by NASA through grant number
\#GO-06673.01-95A from the Space Telescope Science Institute, which is
operated by AURA, Inc., under NASA contract NAS5-26555. TdZ gratefully
acknowledges the warm hospitality of STScI. The authors would like to
thank Roeland van der Marel and Marcella Carollo for helpful
discussions and a critical reading of the manuscript, and the referee
for helpful suggestions to improve the manuscript.


\clearpage
\appendix

\section{INDIVIDUAL PROPERTIES OF THE GALAXIES}
\label{s_galaxies}
In this appendix we give a separate description of each galaxy. We
describe its environment, morphological features of the dust
and emission gas apparent from our imaging and mention results from
previous studies. We also compare our isophotal results to previous
measurements. In most cases the results agree well. By this we mean
that (at radii large enough for the isophotal parameters not to be
affected by the point spread function of the observations) differences
in surface brightness are less than $0.1^{m}$, differences in ellipticity
less than 0.02 and differences in PA less than $5\deg$. We also
briefly note the radio morphology on the arcsecond scale and larger.
The radio properties of the sample will be described in more detail by
Xu \etal (1999a) who made observations with the VLA and VLBA in
snapshot mode. A study of the X-ray properties from {\tt ROSAT}
observations will be presented by Xu \etal (1999b).

\subsection{NGC 193}
\label{s_193}
NGC 193 is part of a group of galaxies (CB88) and has an apparent companion,
NGC 204, at $6.7'$. NGC 193 shows a complex central dust distribution
composed of patches and several warped filaments hundreds of parsec in
size.  The dust features are connected to each other. There is a
nuclear emission peak which is marginally resolved.
Low-surface-brightness emission is associated with the dust features.
The center, although apparently obscured by dust, shows an unresolved
peak clearly bluer than its surroundings. NGC 193 has a radio jet with
a weak counter-jet (Xu \etal 1999a).

\subsection{NGC 315}
\label{s_315}
NGC 315 is part of Zwicky cluster 0107.5+3212 (Zwicky \etal 1961)
which is located in the Perseus-Pisces filament. We detect a central
dust disk 820 pc ($2.5''$) in diameter which is close to but not a
perfect ellipse; the north tip of the disk has a small extension. In
addition several mottled patches of dust are detected southwest of the
nucleus out to $5''$ ($\sim 1.5$ kpc). The central part of the dust
disk forms a small bright emission-gas disk which extends into
low-level emission throughout the dust disk. There is also a low-level
emission feature adjacent to the southeast side (i.e., counter-jet
side) of the dust disk which is elongated in the direction of the dust
disk. HI was detected in absorption redshifted by $\sim 400$ \kms\ 
with respect to the systemic velocity (van Gorkom \etal 1989, Knapp
\etal 1990). Ho \etal (1997) found a nuclear H$\alpha$ broad-line
emission component with FWHM 2000 \kms\ in addition to the narrow-line
emission.  Our isophotal analysis is in good agreement with an
analysis by De Juan \etal (1994) from ground-based observations. The
two-sided radio jet of NGC 315 has been well studied (e.g., Venturi
\etal 1993; Bicknell 1994; Mack \etal 1997; Cotton \etal 1999).
  
\subsection{NGC 383 (3C 31)}
\label{s_383}
NGC 383 is the brightest galaxy in the Zwicky cluster 0107.5+3212
(Sakai \etal 1994). NGC 383 is inside a chain of galaxies and forms a
dumbbell pair with Arp 331 (NGC 382) which is at $34''$ (11 kpc). NGC
383 has a nearly face-on central dust disk (diameter 2.3 kpc/$7.4''$)
which shows a spiral structure with a counter-clockwise orientation.
The spiral structure becomes flocculent at the outer edges. Martel
\etal (1999) detected a dust disk in the companion NGC 382 as well.
NGC 383 has a central unresolved bright \HalphaNII\ emission peak with
low-surface-brightness emission extending out to $\sim 1 ''$ (300 pc)
around it. Owen \etal (1990) detected a rotating extended emission
disk from a spectroscopic study.  NGC 383 has a nuclear point source
clearly bluer than its surroundings. Komossa \& B\"{o}hringer (1999)
performed an X-ray study of the NGC 383 group and found indication for
a hard X-ray component which might be associated with the central AGN.
Our isophotal analysis is in very good agreement with previous
ground-based studies by Fraix-Burnet \etal (1991) and De Juan \etal
(1994). The latter found an increase in ellipticity from 0.1 to 0.3
outward of $r \sim 10''$. The radio jet has been extensively studied
in the past (e.g., Lara \etal 1997 and references therein). The radio
jet shows a distorted morphology suggesting gravitational interaction
between NGC 383 and its companion NGC 382 (Blandford \& Icke 1978;
Parma \etal 1991). However, the optical isophotes out to scales larger
than our {\tt HST} imaging show no indication of interaction
(Fraix-Burnet \etal 1991). Evidence for an optical jet was claimed by
Butcher \etal (1980) but not confirmed by later studies (Keel 1988;
Owen \etal 1990; Fraix-Burnet \etal 1991).

\subsection{NGC 541}
\label{s_541}
NGC 541 is located in Abell cluster 194 (Abell \etal 1989). It is
located at $4.5'$ (96 kpc) from the dumbbell pair NGC 545/NGC 547
identified with the radio source 3C 40 . NGC 541 has a central disk
(diameter 640 pc/$1.8''$) which appears to be nearly face-on.  The
disk has an outer rim which is darker, especially on the north side.
There is also an inner ring with radius $\sim 0.22''$ (78 pc) which
shows enhanced obscuration. These darker ring-like structures could be
part of an otherwise hardly visible spiral structure. The $V$-$I$
image shows a hint of a linear dust feature with PA $\sim 267\deg$
sticking out of the disk. This is within $10\deg$ of the VLA radio jet
axis (CB 1988).  At the center of the dust disk there
is a central \HalphaNII\ emission peak. A ring of emission surrounds
the peak just inside the darker rim of the disk. There is a slightly
brighter emission spot at PA $\sim 51\deg$ and on the opposite side in
the ring. The central pixels are slightly bluer than their
surroundings.  Our isophotal analysis agrees well with the study by De
Juan \etal (1994).  They find the abrupt increase in ellipticity
outward of $\sim 10''$ to continue until 0.2 at $20''$ and found a
rise in the fourth order Fourier coefficient $c_4$ to 0.02 at $\sim
25''$. The radio-jets form a small head-tail source (O'Dea \& Owen
1985) which might have triggered the starburst in Minkowski's object,
an irregular dwarf, located in the path of the jet at $45''$ (16 kpc)
NE of NGC 541 (van Breugel \etal 1985).

\subsection{NGC 741}
\label{s_741}
NGC 741 is the brightest galaxy in a group of galaxies (Zabludoff \&
Mulchaey 1998). There is a small apparent elliptical companion at
$\sim 8''$ (2.7 kpc assuming same distance) which is not cataloged in
NED. A second elliptical companion, NGC 742, is located
at $48''$ (16.2 kpc). No dust is detected in NGC 741.  We detect a
central peak of emission with low level emission filaments extending
$\sim 2''$ (680 pc) towards the North, East and South.  Macchetto \etal
(1996) detected an emission-gas disk from the ground with minor and
major diameter of $10.8''$ (3.7 kpc) and $22.6''$ (7.7 kpc)
respectively, and with an average flux density which is significantly below our
detection threshold. This disk fills most of the {\tt PC}.  NGC 741
hosts a head-tail radio source (Birkinshaw \& Davies 1985).

\subsection{UGC 1841 (3C 66B)}
\label{s_1841}
UGC 1841 forms a dumbbell pair with a companion southeast at a
distance of $\sim 24''$ (10 kpc). The pair belongs to a small group of
galaxies close to the Abell cluster 347. UGC 1841 has a central dust
disk (diameter 330 pc/$0.8''$) which appears to be neither a perfect
ellipse nor a circle. There is a protuberance $\sim 0.25''$ ($\sim 100
\pc$) in length with PA $230\deg$. The emission image shows bright
nuclear \HalphaNII\ emission slightly extended towards PA $228\deg$.
Intriguingly, both the slight extension of the \HalphaNII\ emission
and dust protuberance of the disk are in the direction of the counter
jet (Fraix-Burnet 1997). The optical jet was detected by Butcher \etal
(1980) and studied by Keel (1988) and Fraix-Burnet \etal (1989, 1991).
The surface brightness of the optical jet is quite low and therefore
hardly detectable on our images but clearly shows up once a model
galaxy is subtracted. A tentative optical counter-jet is reported by
Fraix-Brunet (1997) from a 15 hour $I$-band exposure. Our isophotal
analysis agrees well with Fraix-Burnet \etal (1991). They found the
same PA twist. We excluded the region affected by the bright star on
the {\tt PC} from the fit. We also confirm the reported offset toward
PA $70\deg$ of the isophotal centers increasing to $3''$ at $r \sim
20''$.  The radio jet displays a distorted double-sided jet structure
(Hardcastle \etal 1996).

\subsection{NGC 2329}
\label{s_2329}
NGC 2329 is part of Abell cluster 569. The galaxy has a small inclined
central dust disk (diameter 740 pc/$2.0''$).  The $V$-$I$ image
shows a few brighter pixels over the disk ending in a lighter spot at
the edge of disk at PA $318\deg$. There also seems to be a small dust
protuberance to the disk in the opposite direction.  The nuclear
emission is slightly resolved and shows a small extension roughly in
the direction of the dust protuberance. The nucleus is clearly bluer
than its surroundings. The galaxy has an exceptional blue $V$-$I$
($\sim 1.20$) color compared to the rest of the sample.  NGC 2329
hosts a wide-angle tailed radio source (Feretti \etal 1985; Xu \etal
1999a).

\subsection{NGC 2892}
\label{s_2892}
NGC 2892 is an isolated galaxy apart from a nearby apparent elliptical
companion at $\sim 10''$ (4.4 kpc assuming same distance) which is not
identified in NED.  No dust is detected in NGC 2892. The emission
image shows a central peak with low-level emission around it out to
$\sim 0.3''$ (130 pc). The $V$-$I$ image shows a blue nuclear point
source. NGC 2892 has a very regular double-sided radio jet (CB88; Xu
\etal 1999a).

\subsection{NGC 3801}
\label{s_3801}
NGC 3801 is inside a group of ten galaxies (Garcia 1993). It has a
very extended and complicated dust distribution. There is a main dust
lane ($\sim 4.4$ kpc/$21''$ ) perpendicular to the major axis of the
galaxy with a warped and stranded structure.  A second filamentary
dust distribution extends roughly perpendicular to the east-side of the
main dust lane while on the west side there are about six thinner
parallel filaments also roughly perpendicular to the main dust lane.
This structure spans $\sim 60''$ (12.6 kpc) and is visible on both the
{\tt PC} and {\tt WF}.  \HalphaNII\ emission is detected $\sim 0.5''$
($\sim 100 \pc$) away from the dust enshrouded nucleus itself.  There
are about a dozen `knots' of emission along the main dust lane.  The
`knots' appear to be connected by `bridges' of low level emission.
Some of the knots are slightly resolved. The brightest `knots'
coincide with slightly brighter small spots in the $V$-band image and
are likely to be star forming regions.  HI is detected in emission
(Heckman \etal 1983) and also in absorption with an estimated mass of
$2.1 \cdot 10^9 \Msun$ (Duprie \etal 1996). Radio maps by Jenkins
(1982) and Xu \etal (1999a) show a warped double-lobed profile roughly
perpendicular to the main dust lane. Our isophotal analysis agrees
with $R$-band photometry by Peletier \etal (1990). There is an
estimated $\sim 0.5''$ uncertainty in the determined center of the
isophotes due to the very large extent of the dust distribution which
affects a large fraction of the {\tt PC} image. Ground-based images
(Heckman \etal 1986) show that the galaxy becomes quite boxy at $r
\sim 30''$.

\subsection{NGC 3862 (3C 264)}
\label{s_3862}
NGC 3862 is part of the cluster Abell 1367 and has a nearby lenticular
companion (IC 2955) at a distance of $54''$ (22 kpc). There is an
apparent faint face-on circular dust disk (Crane \etal 1993) 610 pc
($1.5''$) in diameter which is studied in detail by Baum \etal (1997).
The apparent disk is edge-darkened, especially the southern half. This
might be a projection effect or could indicate that the inner parts of
the dust disk are swept clear by either the jet or some other nuclear
process (Hutchings \etal 1998). The \HalphaNII\ emission is strongly
peaked at the nucleus and diffuse low-level emission is seen
throughout the disk. There is a strong blue nuclear point source. 3C
264 is a head-tail radio source displaying S-shaped wiggles extending
to a long bifurcated tail of diffuse radio emission (e.g., Baum \etal
1988; Parma \etal 1991; Lara \etal 1999). The optical jet,
discovered by Crane \etal (1993), seems to interact with the
surrounding ISM (Baum \etal 1997).

\subsection{NGC 4261 (3C 270)}
\label{s_4261}
NGC 4261 is the main galaxy in a group of 33 galaxies (Nolthenius
1993) in the `Virgo West' cloud. It is a prolate galaxy with stellar
rotation around the apparent major axis (Davies \& Birkenshaw 1986).
NGC 4261 has a small nuclear dust disk (diameter 240 pc/$1.7''$)
discovered by Jaffe \etal (1993). Darker and lighter streaks are
present in the disk suggesting a spiral structure.  The $V$-band image
shows a clear jet-like dust protuberance $\sim 0.65''$ ($\sim 90 \pc$)
in length sticking out of the disk and pointing from the nucleus.
There is a weaker similar dust feature on the opposite side. Both
features were already noted by Ferrarese \etal (1996). They also noted
a small ($\sim 5 $ pc/$0.035''$) offset between the center of the dust
disk and the nucleus on the one hand and center from the isophotes on
the other. By modeling the central gas kinematics they detect a black
hole with mass $\sim 5 \cdot 10^8 \Msun$. M\"{o}llenhoff \& Bender
(1987) detected in addition to the central dust disk a dust feature at
$\sim 15''$ (2.1 kpc) north of the nucleus which falls outside our
image. Jaffe \& McNamara (1994) detected central HI and CO in
absorption. The dust obscured nucleus shows a faint blue point source.
NGC 4261 has a double-sided radio-jet (e.g., Birkenshaw \& Davies
(1985); Jones \& Wehrle 1997). Our isophotal analysis agrees well with
the analysis by Ferrarese \etal (1996) who use the same {\tt HST}
observations and with an analysis from ground-based data (Peletier
\etal 1990). The latter shows a very constant PA out to $\sim 100''$
while the ellipticity decreases to $\sim 0.15$ and the Fourier
coefficient $c_4$ becomes 0.

\subsection{NGC 4335}
\label{s_4335}
NGC 4335 is a field galaxy without nearby companions.  Arc shaped dust
features (with a total diameter of 4 kpc/$13.5''$) surround the
nucleus and appear to be the visible half of a settling dust disk. The
outer parts consist of `arms' of dust while the inner parts are formed
by a small central dust disk (diameter $\sim 240$ pc/$0.8''$) embedded
in a larger dust disk (diameter $\sim 750$ pc/$2.5''$) with similar
axial ratio and PA. There is nuclear \HalphaNII\ emission and
low-level emission is associated with the extended dust features
throughout the galaxy. Knapp \& Rupen (1996) report a tentative
detection of CO in absorption. NGC 4335 hosts a two-sided radio jet
(CB88; Xu \etal 1999a).

\subsection{NGC 4374 (M 84, 3C 272.1)}
\label{s_4374}
NGC 4374 resides in the Virgo cluster and is part of Markarian's
Chain. The two filamentary dust lanes (total size $\sim 1$
kpc/$13.0''$) which flare at the ends have been known for a long time
(e.g., Hansen \etal 1985; van Dokkum \& Franx 1995; Bower \etal 1997).
The PA of the dust lane and the stellar distribution differ by
$50-60\deg$. Quillen \& Bower (1999) recently proposed a model in
which the misalignment and the warps of the dust lanes are caused by
jet-induced pressure gradients. Our emission image agrees well with
results found by Bower \etal (1997) using a different subtraction
technique. NGC 4374 has a central emission-gas disk and lower level
emission which follows the dust distribution. Bower \etal (1998)
derived a black hole mass of $\sim 1.5 \cdot 10^9 \Msun$ from
velocities measured in the central emission-gas disk. Deeper imaging
of the emission gas shows an `S'-shaped twist from PA $\sim 70\deg$ to
$\sim 115\deg$ at $r \sim 7''$ (Hansen \etal 1985; Baum \etal 1988).
Although the nucleus is most likely obscured by dust it is clearly
bluer than its surroundings.  M 84 has a two-sided radio jet (Jones
\etal 1981; Laing and Bridle 1987).

\subsection{NGC 4486 (M 87, 3C 274)}
\label{s_4486}
M 87 is the second brightest galaxy in Virgo. The dust in M 87 is made
up of irregular patches and filaments which can be seen out to $\sim
11''$ (820 pc) from the nucleus. The central emission gas was first
studied with {\tt HST} by Ford \etal (1994) and Harms \etal (1994). An
isophotal fit to our central emission features agrees very well with
their results. Studies of the emission-gas kinematics yield a black
hole with a mass of $\sim 3.2 \cdot 10^9 \Msun$ (Harms \etal 1994;
Macchetto \etal 1997). Irregular spiral arms of emission extend from
the disk and emission gas is associated with all dust features. Dopita
\etal (1997) showed that the off-nucleus emission-gas has a LINER
spectrum and is excited by shocks. Carter \etal (1997) found evidence
for many small dense clouds of gas in the core of M 87 with a velocity
range $\sim 300 $ kms using high resolution ground-based spectroscopy.
The active nucleus in M 87 emits a non-stellar continuum which is
reported to vary in strength over time in the optical (Tsvetanov \etal
1998) and X-rays (Harris \etal 1997). The optical and radio jet have
been studied extensively (e.g., Meisenheimer \etal 1996 for a recent
review; Biretta 1998; Perlman 1999). M 87 has an apparent one-sided
optical and radio jet but there are strong indications for a
counter-jet (Sparks \etal 1992; Stiavelli \etal 1992, 1997). Our
isophotal analysis is consistent with studies by Ferrarese \etal
(1994) and Peletier \etal (1990). The latter find that the ellipticity
increases monotonically to 0.15 between $10''$ and $\sim 100 ''$. This
increase enables a more robust estimate of the PA for this region
which decreases from $\sim 170\deg$ to $145\deg$.

\subsection{NGC 5127}
\label{s_5127}
NGC 5127 is the second brightest galaxy in Zwicky cluster 1319.6+3135.
It has a bended dust lane (size 940 pc/$3.0''$) across the center
which forks in two towards the northeast. There is a compact nuclear
emission peak surrounded by low-surface-brightness emission associated
with the dust lanes. NGC 5127 has a symmetric two-sided VLA radio jet
(Xu \etal 1999a). Our isophotal analysis agrees well with an analysis
by De Juan \etal (1994).  They report the same PA twist and find that
the isophotes become increasingly boxy beyond $r \sim 10''$.

\subsection{NGC 5141}
\label{s_5141}
NGC 5141 is part of a group of six galaxies (Ramella \etal 1989) and
has a companion galaxy, NGC 5142, at $2.3'$ (47 kpc). It has a central
slightly wedge-shaped dust lane with a size of 800 pc ($2.3''$). The
\HalphaNII\ emission peaks at the center and low-level emission is
detected in the dust lane. NGC 5141 has two-sided radio jets with
broad lobes (Xu \etal 1999a). Our isophotal analysis is consistent
with a ground-based study by Gonz\'{a}lez-Serrano \etal (1993). In
contrast with our results, however, they report a sharper decrease in
ellipticity inward of $3''$.  This difference might be caused by their
larger seeing and/or a distortion of the inner isophotes by the
central dust distribution.

\subsection{NGC 5490}
\label{s_5490}
NGC 5490 is the brightest galaxy in Zwicky cluster 1407.6+1750. There
is a very small central dust distribution $0.5''$ (170 pc) in size
with PA $143\deg$. It appears to be either a dust lane or a highly
inclined disk. There is a compact central peak of \HalphaNII\ 
emission. NGC 5490 hosts a double-sided radio jet (CB88; Xu \etal
1999a). Our isophotal analysis agrees with a study by De Juan \etal
(1994).  They found a lower ellipticity in the inner few arcseconds,
but poor seeing is the most likely cause for this. They also note a
$15\deg$ increase of PA outward of $10''$.

\subsection{NGC 7052}
\label{s_7052}
NGC 7052 is a field galaxy with no nearby companion.  There is a large
central dust disk with a diameter of 1 kpc ($4.0''$) (De Juan \etal
1996). The inner part of the dust disk hosts a small emission-gas disk
and low-level emission is detected throughout the dust disk. By
modeling the kinematics of the gas disk van der Marel \& van den Bosch
(1998) detected a black hole with a mass of $\sim 3 \cdot 10^8 \Msun$.
There is a dust protuberance, also seen in the emission image at the
west end of the disk. HI was detected in emission blueshifted by $\sim
500$\kms\ with respect to the systemic velocity of the galaxy
(Huchtmeier \etal 1995). The nucleus which is obscured by dust is
slightly bluer than its surroundings. NGC 7052 hosts a narrow radio
jet with a very weak counter jet (Morganti \etal 1987; CB88; Xu \etal
1999a).  Our isophotal analysis agrees well with previous studies (De
Juan \etal (1994); van den Bosch \& van der Marel 1995; van der Marel
\& van den Bosch 1998). De Juan \etal found the ellipticity to
increase until $\sim 0.55$ at $\sim 30''$ and the isophotes to become
boxy outward of $10''$.

\subsection{NGC 7626}
\label{s_7626}
NGC 7626 has a nearby higher luminosity companion NGC 7619 which is at
$6.8'$ (90 kpc). They are the two brightest ellipticals in the Pegasus
I cluster. NGC 7626 has a counter rotating core (Forbes \etal 1995;
Carollo \etal 1997). There is a small warped dust lane (diameter 230
pc/$1.0''$) across the nucleus. Interestingly the central part of the
dust lane is perpendicular to the radio jet while the warped outer
part aligns with the major axis of the galaxy. There is an unresolved
nuclear emission peak and low-level emission gas is seen along the
dust lane. NGC 7626 has double-sided radio lobes (Xu \etal 1999a). Our
isophotal analysis is in good agreement with previous ground-based
(Peletier \etal 1990; De Juan \etal 1994) and {\tt HST} studies
(Carollo \etal 1997). The PA of the isophotes increases by $20\deg$ in
the central $10''$. Forbes \& Thomson (1992) reported excess light in the
galaxy which might be due to tidal interaction with NGC 7619.

\section{Optical depth}
\label{s_opt_depth}

Following Carollo \etal
(1997), we derive the optical depth under the assumption that the dust
distribution can be represented by a screen. For a dust screen at
position $D$ relative to the distribution of light along the line of
sight ($D=0$ for a screen in front of the galaxy and $D=1$ behind the
galaxy) we can write the ratio between intrinsic and observed fluxes as:  
\begin{equation}
r_j=\frac{F^{\rm obs}_{j}}{F^{\rm int}_{j}}=D+(1-D)e^{-\alpha_j \tau},
\label{e_fluxratio1}
\end{equation}
where $r_j$ is the ratio of $F^{\rm obs}_{j}$ and $F^{\rm int}_{j}$, i.e. of the
observed and intrinsic flux through filter $j$ ($j=V,I$),
respectively.  The extent and morphology of the dust distributions
inhibit a direct estimate of these two ratios from our isophotal
analysis to derive $\tau$. However, the ratio $r^{\rm int}_{VI}$ of the
un-extincted flux in $V$ and in $I$ (i.e., its color) is well fitted as a slowly varying
linear function of radius for the centers of these early-type
galaxies. This fit can be extrapolated into regions affected by dust.
Thus we determine from our imaging the ratio $a_{VI}$ between the un-extincted and observed color which can be written as:
\begin{equation}
a_{VI} \equiv \frac{r^{\rm obs}_{VI}}{r^{\rm int}_{VI}} \equiv \frac{r_{V}}{r_{I}}=\frac{D+(1-D)e^{-\alpha_V \tau}}{D+(1-D)e^{-\alpha_I \tau}}.
\label{e_fluxratio2}
\end{equation}
Taking $\alpha_V=1.0$ and $\alpha_I=0.5$ (consistent with Galactic
extinction) and substituting $t=e^{-0.5 \tau}$, eq.~(\ref{e_fluxratio2})
becomes a second order equation which can be solved analytically for
$\tau$.

\clearpage


\ifsubmode\else
\baselineskip=10pt
\fi


\clearpage


\ifsubmode\else
\baselineskip=14pt
\fi


\newcommand{\figcaplum}{Total luminosities as a function of distance
for the complete sample. The 19 galaxies presented in this paper are
shown as filled dots. Top: total radio luminosity at 1400 MHz. The
solid line indicates the lower cut-off of the radio luminosity set by
our selection. Bottom: absolute photographic magnitude.
\label{f_lum}}

\newcommand{\figcapphotoma}{\footnotesize Isophotal parameters in the
  $I$-band as function of radius along the major axis. From top to
  bottom: surface brightness, $V$-$I$ color, position angle,
  ellipticity and fourth order Fourier coefficient. The galaxy name is
  shown above each column. The top plot also shows the $V$-band
  luminosity profile (i.e. higher $\mu$).  The dashed and dotted lines
  display luminosity profiles corrected for dust obscuration (see
  Section \ref{s_isophotes}). Error bars include the formal error
  given by the isophotal fitting routine and the error in the
  estimated sky counts. The vertical dashed line indicates the maximum
  radius out to which the isophotes are affected by dust obscuration
  and PA and $\epsilon$ are kept fixed.\label{f_photoma}}
\newcommand{\figcapphotomb}{\sl Continued.\label{f_photomb}}
\newcommand{\figcapphotomc}{\sl Continued.\label{f_photomc}}
\newcommand{\figcapphotomd}{\sl Continued.\label{f_photomd}}
\newcommand{\figcapphotome}{\sl Continued.\label{f_photome}}
\newcommand{\figcapphotomf}{\sl Continued.\label{f_photomf}}
\newcommand{\figcapphotomg}{\sl Continued.\label{f_photomg}}
\newcommand{\figcapphotomh}{\sl Continued.\label{f_photomh}}
\newcommand{\figcapphotomi}{\sl Continued.\label{f_photomi}}
\newcommand{\figcapphotomj}{\sl Continued.\label{f_photomj }}

\newcommand{\figcapdustcorrect}{The extinction in $V$ magnitudes
due to obscuration by dust as function of observed color-excess
assuming a dust screen as described in Section~\ref{s_iso_res}. Each
curve gives the relation for a fixed fraction $D$. For a given $D$ the
maximum extinction is $2.5 \log(D^{-1})$. Thus for large optical
depths the magnitude correction increases with decreasing $D$ for a
given color-excess.\label{f_dustcorrect}}

\newcommand{\figcapextemi}{`Total extinction factor' $C$ (see
  Sec.~\ref{s_emi_red}) versus optical depth.  The solid line applies
  for a front screen, the dashed line for a halfway screen position,
  i.e. with 50\% of the light arising in front and behind the screen.
  The dotted line applies if dust and stars are homogeneously
  mixed.\label{f_extemi}}

\newcommand{\figcapphi}{Top: relative apparent angle between dust
  major axis and galaxy major axis as function of the ratio of dust
  minor and major axis. Bottom: relative apparent angle between dust
  major axis dust and radio-jet axis at arcsec scale as function of
  the ratio of dust minor and major axis.\label{f_phi}}

\newcommand{\figcapima}{Left column are $V$-band images and right
  column are \HalphaNII\ emission images. All images have logarithmic
  stretch. North is up and East is to the left.\label{f_ima}}

\newcommand{\figcapimb}{{\sl (cont.)} Left column are $V$-band images
  and right column are \HalphaNII\ emission images. All images have
  logarithmic stretch. North is up and East is to the left.}

\newcommand{\figcapimc}{{\sl (cont.)} Left column are $V$-band images
  and right column are \HalphaNII\ emission images. All images have
  logarithmic stretch. North is up and East is to the left.}

\newcommand{\figcapimd}{{\sl (cont.)} Left column are $V$-band images
  and right column are \HalphaNII\ emission images. All images have
  logarithmic stretch. North is up and East is to the left.}

\newcommand{\figcapime}{{\sl (cont.)} Left column are $V$-band images
  and right column are \HalphaNII\ emission images. All images have
  logarithmic stretch. North is up and East is to the left.}

\newcommand{\figcapimf}{{\sl (cont.)} Left column are $V$-band images
  and right column are \HalphaNII\ emission images. All images have
  logarithmic stretch. North is up and East is to the left.}

\newcommand{\figcapimg}{{\sl (cont.)} Left column are $V$-band images
  and right column are \HalphaNII\ emission images. All images have
  logarithmic stretch. North is up and East is to the left.}

\newcommand{\figcapimh}{{\sl (cont.)} Left column are $V$-band images
  and right column are \HalphaNII\ emission images. All images have
  logarithmic stretch. North is up and East is to the left.}

\newcommand{\figcapimi}{{\sl (cont.)} Left column are $V$-band images
  and right column are \HalphaNII\ emission images. All images have
  logarithmic stretch. North is up and East is to the left.}

\newcommand{\figcapimj}{{\sl (cont.)} Left column are $V$-band images
  and right column are \HalphaNII\ emission images. All images have
  logarithmic stretch. North is up and East is to the left.}


\ifsubmode
\else\printfigtrue\fi

\ifprintfig


\clearpage
\begin{figure}
\centerline{\epsfbox{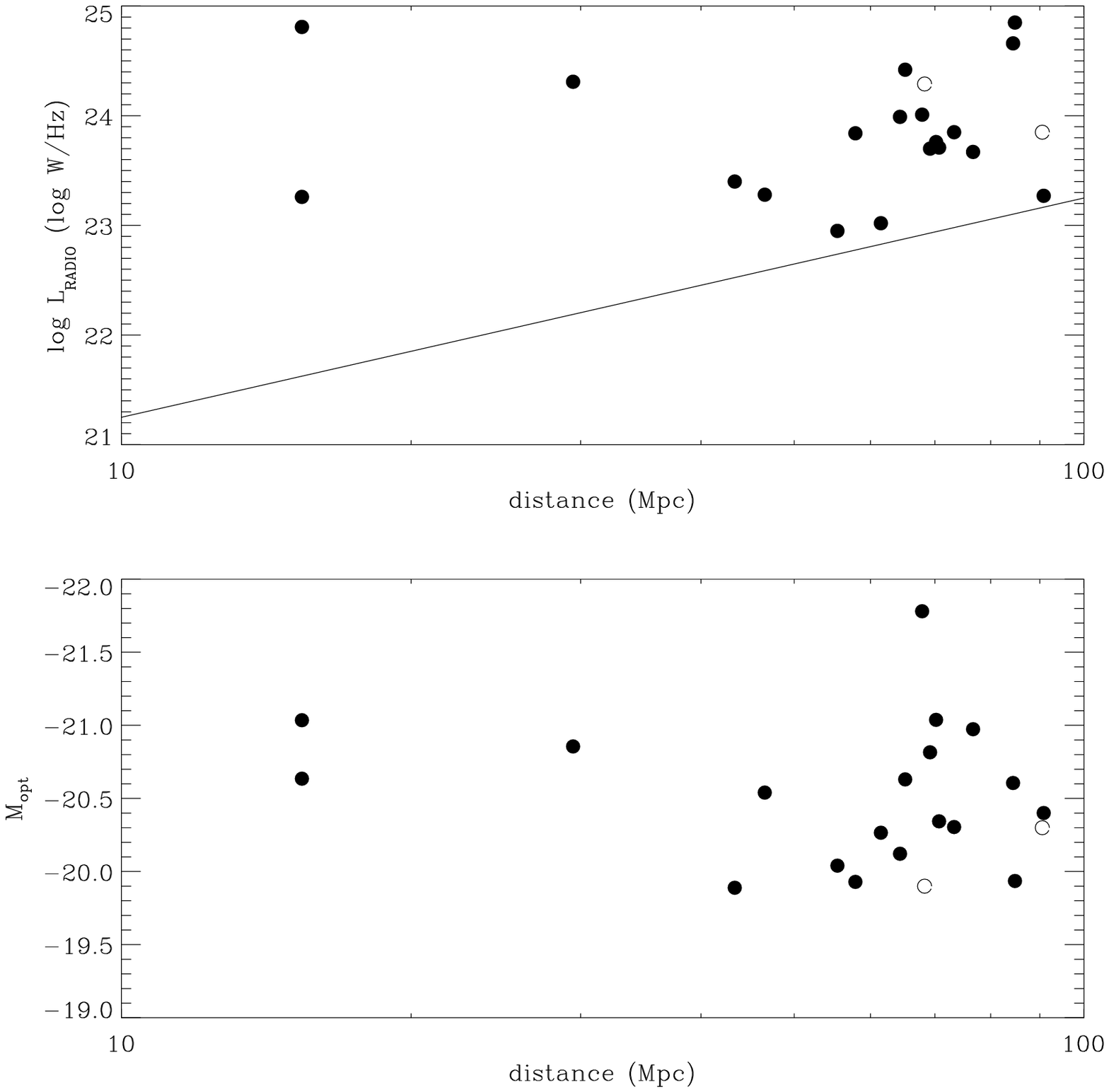}}
\ifsubmode
\else\figcaption{\figcaplum}\fi
\end{figure}

\clearpage
\begin{figure}
\plottwo{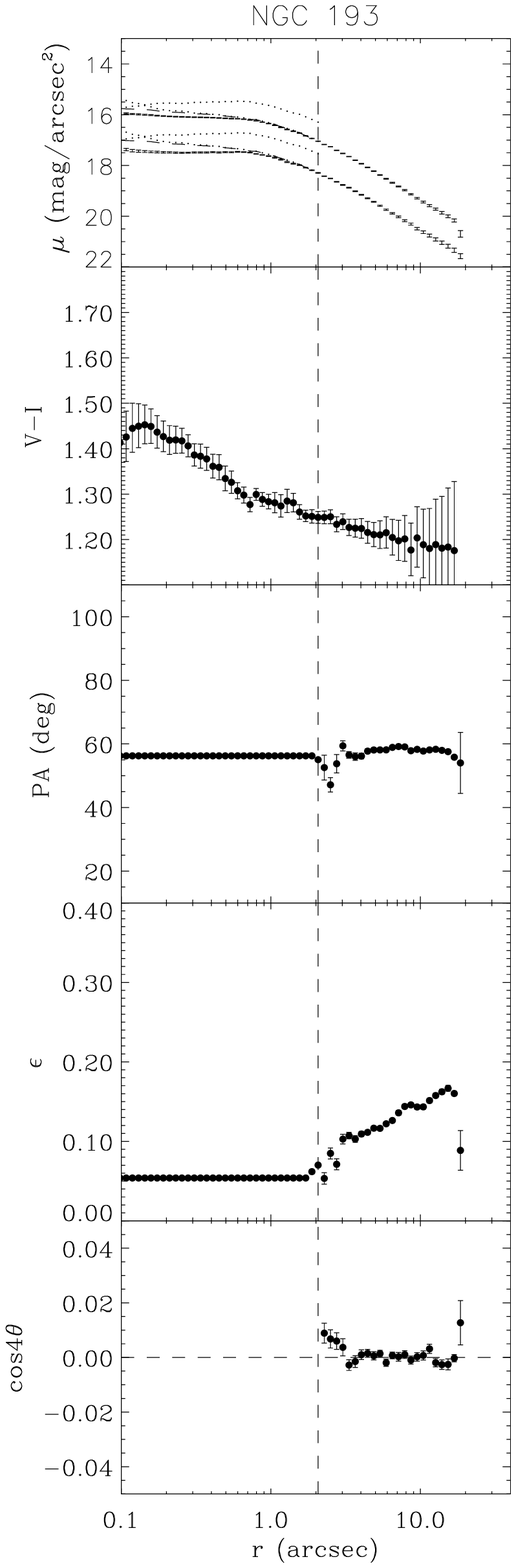}{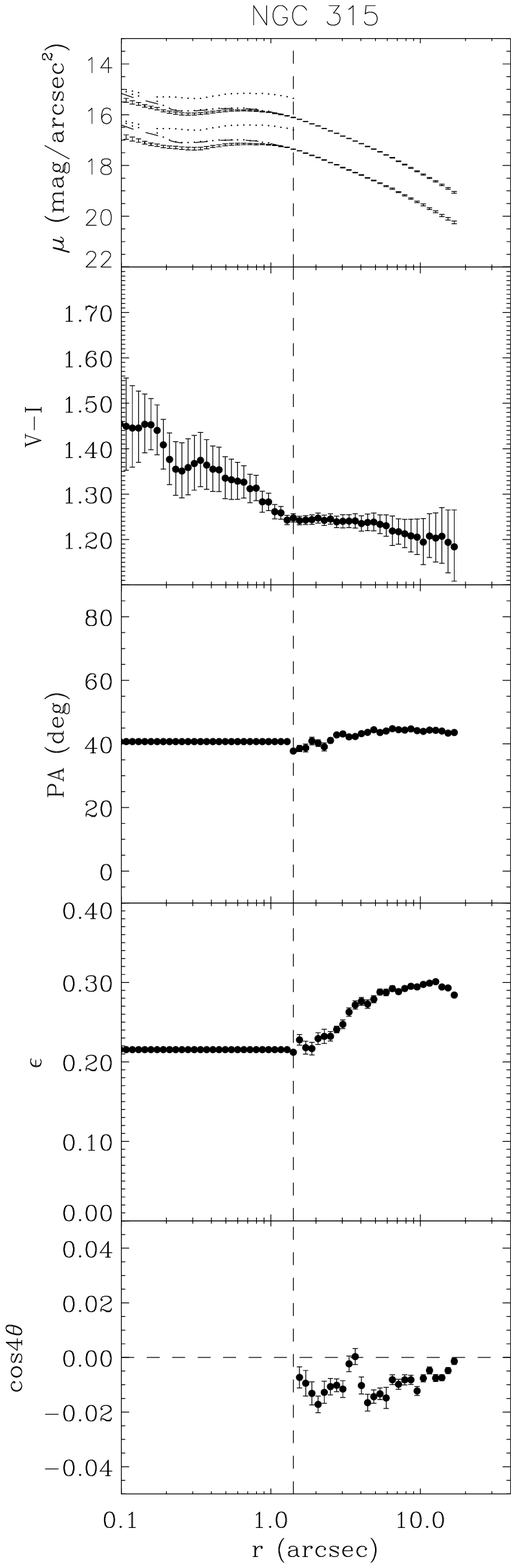}
\ifsubmode
\else\figcaption{\figcapphotoma}\fi
\end{figure}
\addtocounter{figure}{-1}

\clearpage
\begin{figure}
\plottwo{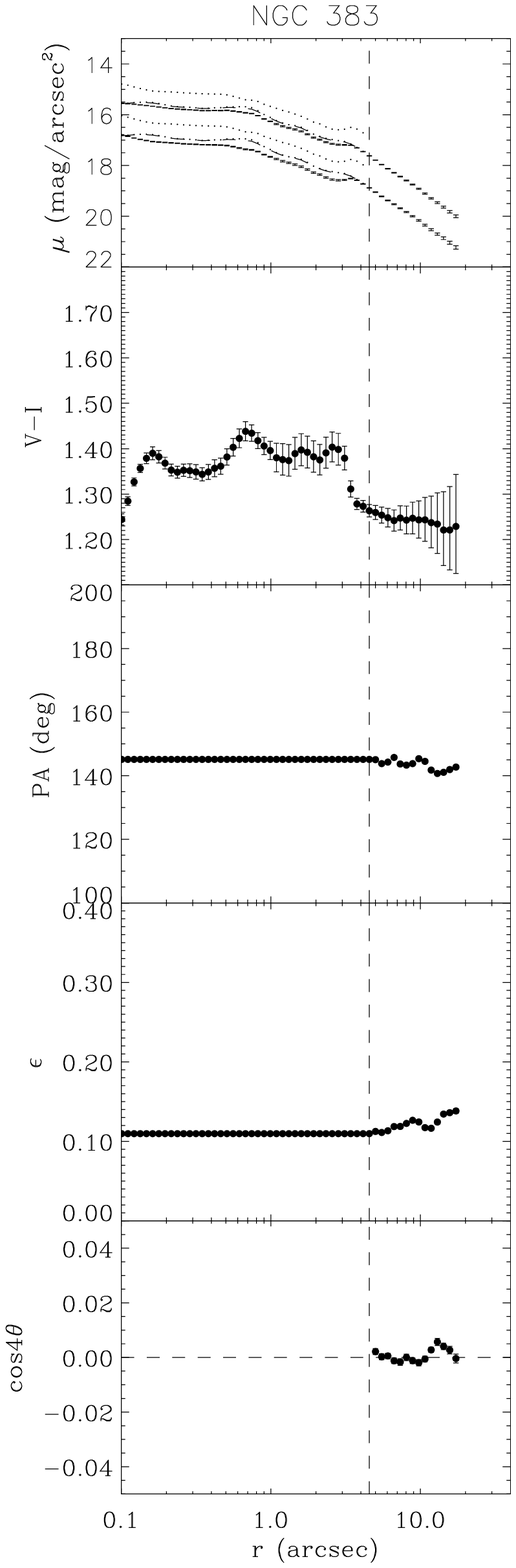}{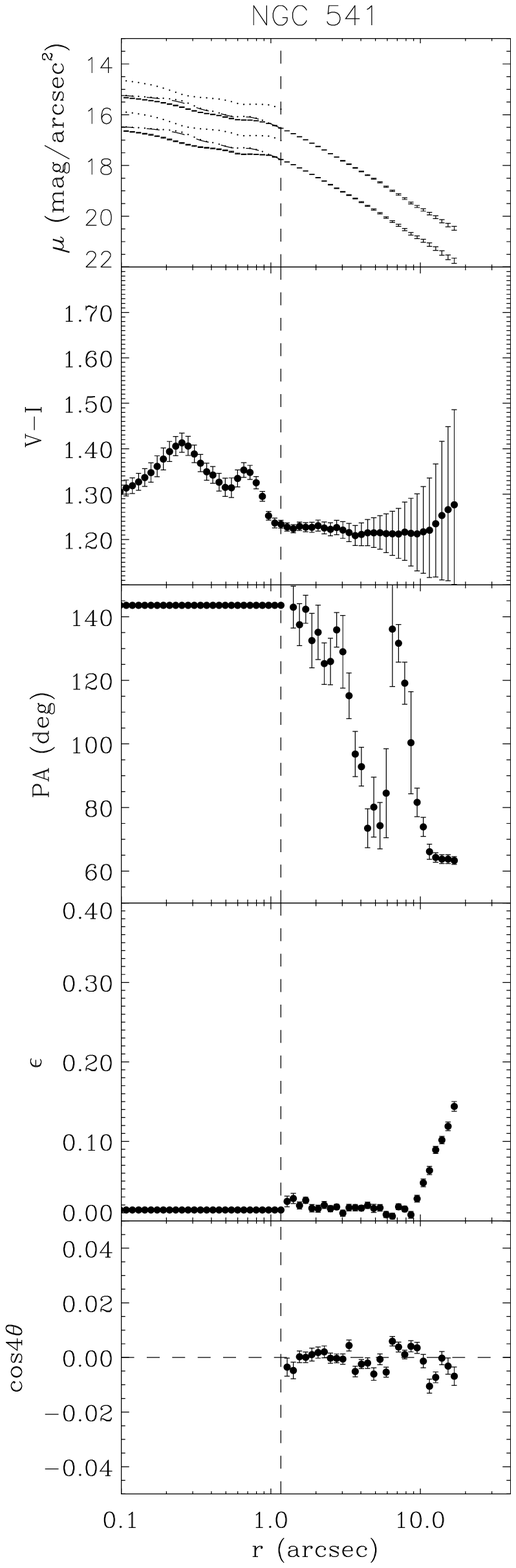}
\ifsubmode
\else\figcaption{\figcapphotomb}\fi
\end{figure}
\addtocounter{figure}{-1}

\clearpage
\begin{figure}
\plottwo{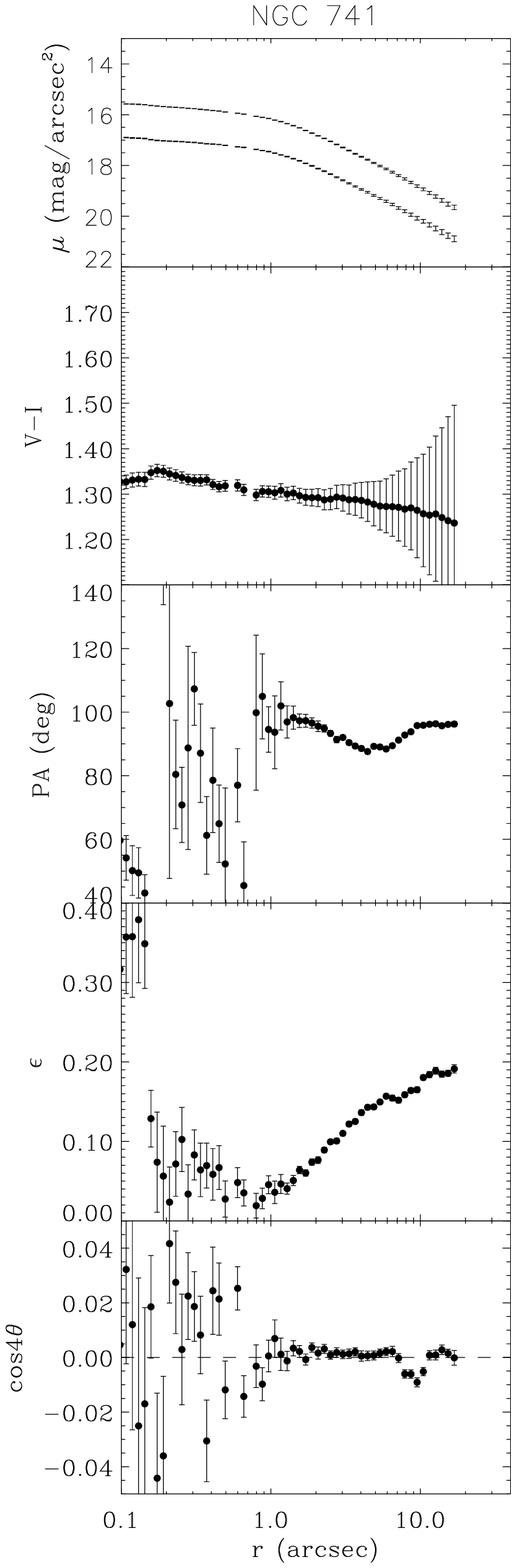}{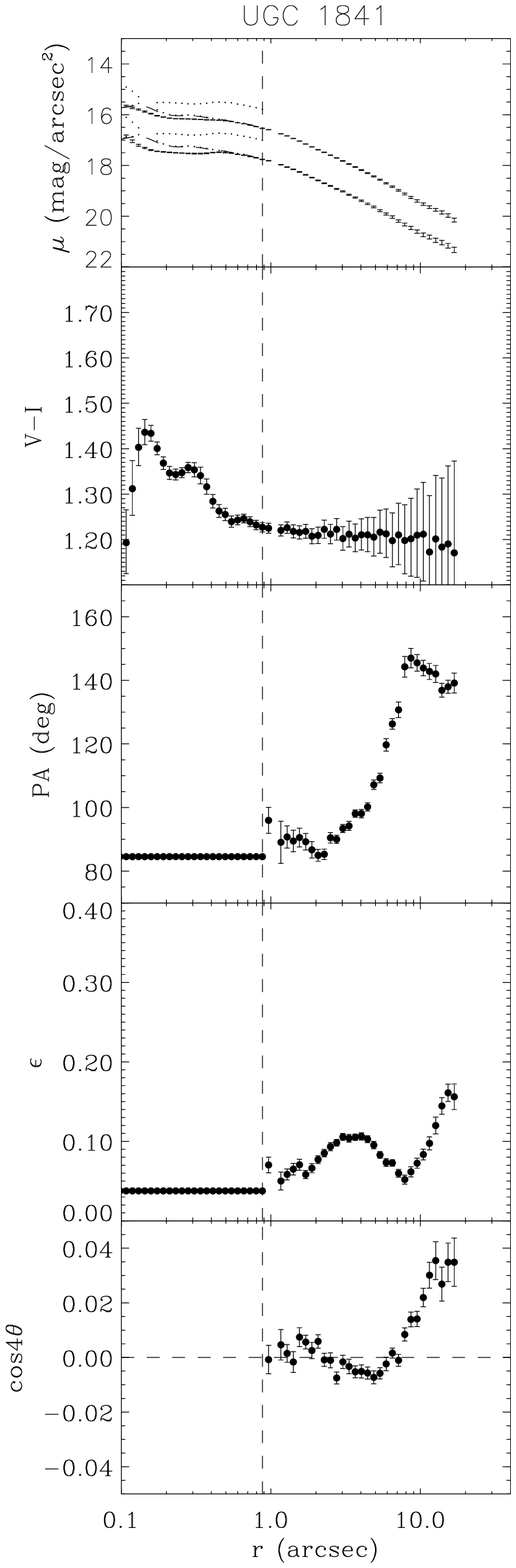}
\ifsubmode
\else\figcaption{\figcapphotomc}\fi
\end{figure}
\addtocounter{figure}{-1}

\clearpage
\begin{figure}
\plottwo{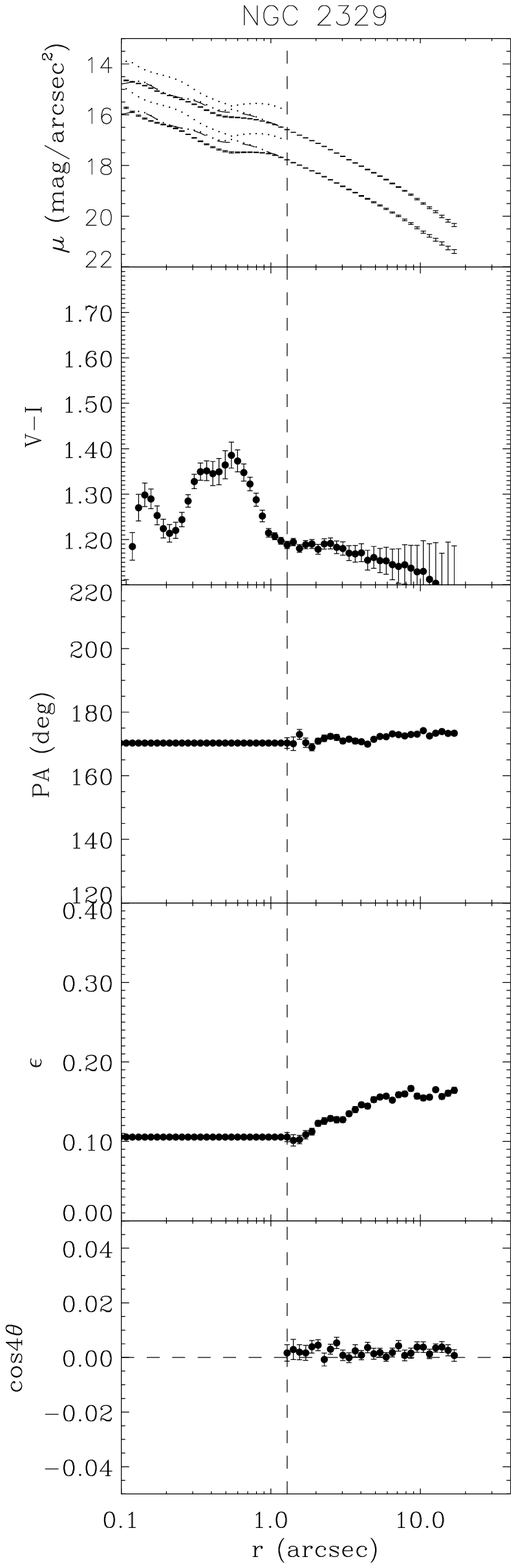}{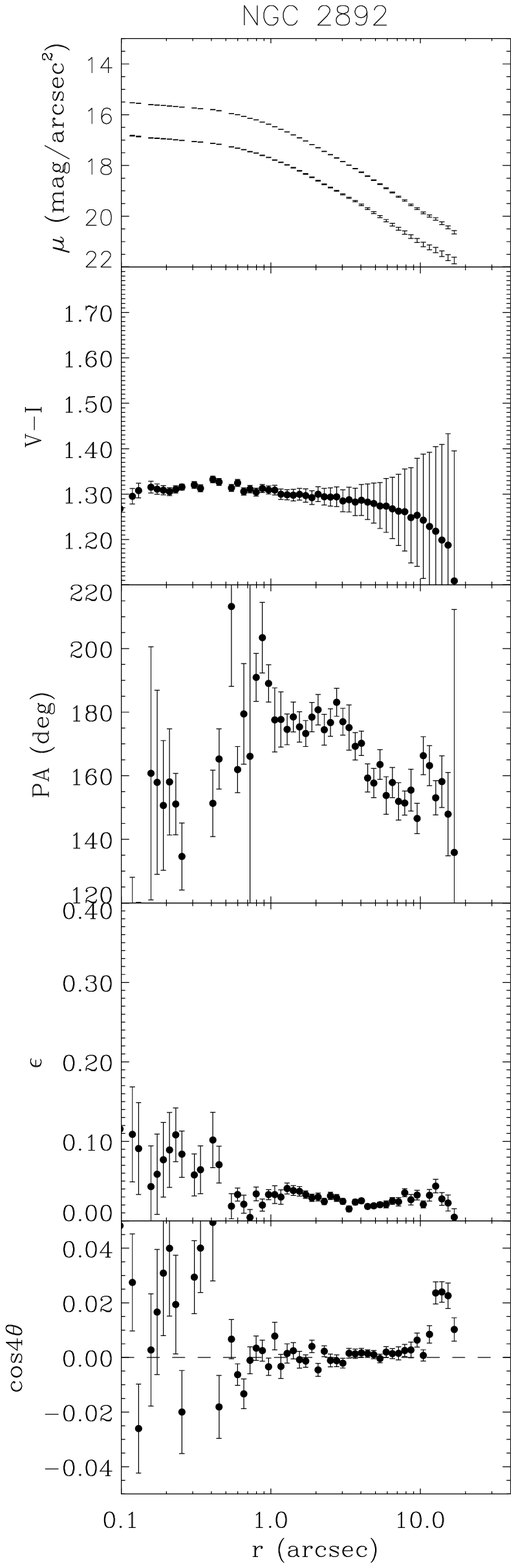}
\ifsubmode
\else\figcaption{\figcapphotomd}\fi
\end{figure}
\addtocounter{figure}{-1}

\clearpage
\begin{figure}
\plottwo{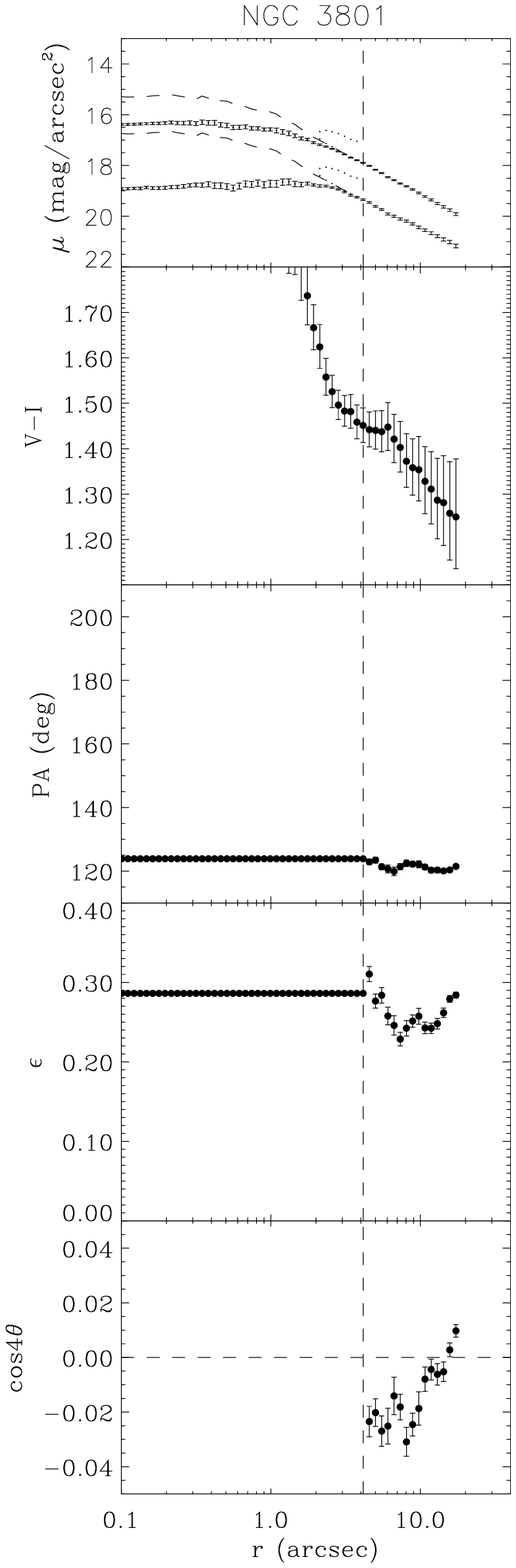}{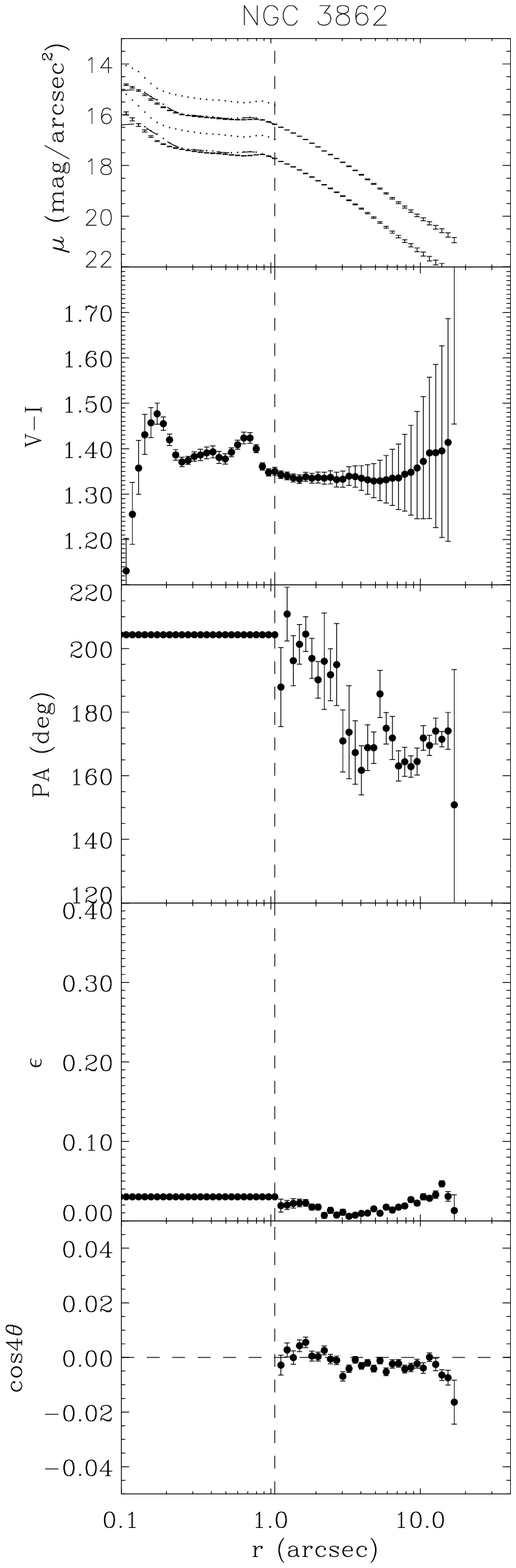}
\ifsubmode
\else\figcaption{\figcapphotome}\fi
\end{figure}
\addtocounter{figure}{-1}

\clearpage
\begin{figure}
\plottwo{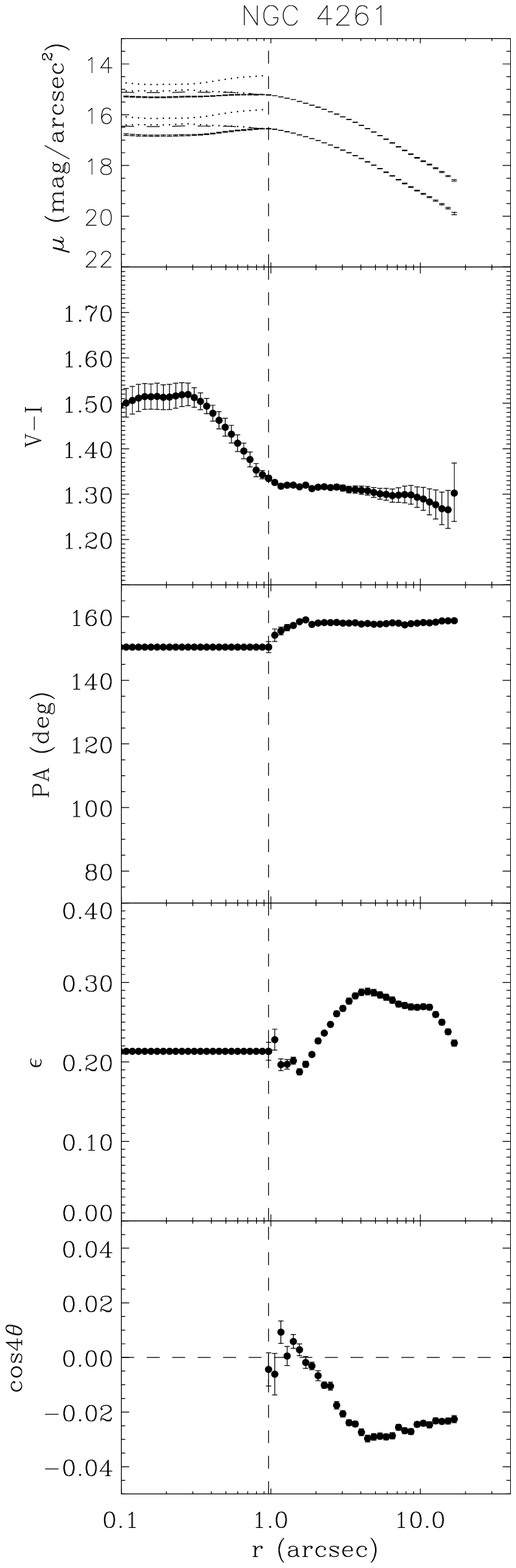}{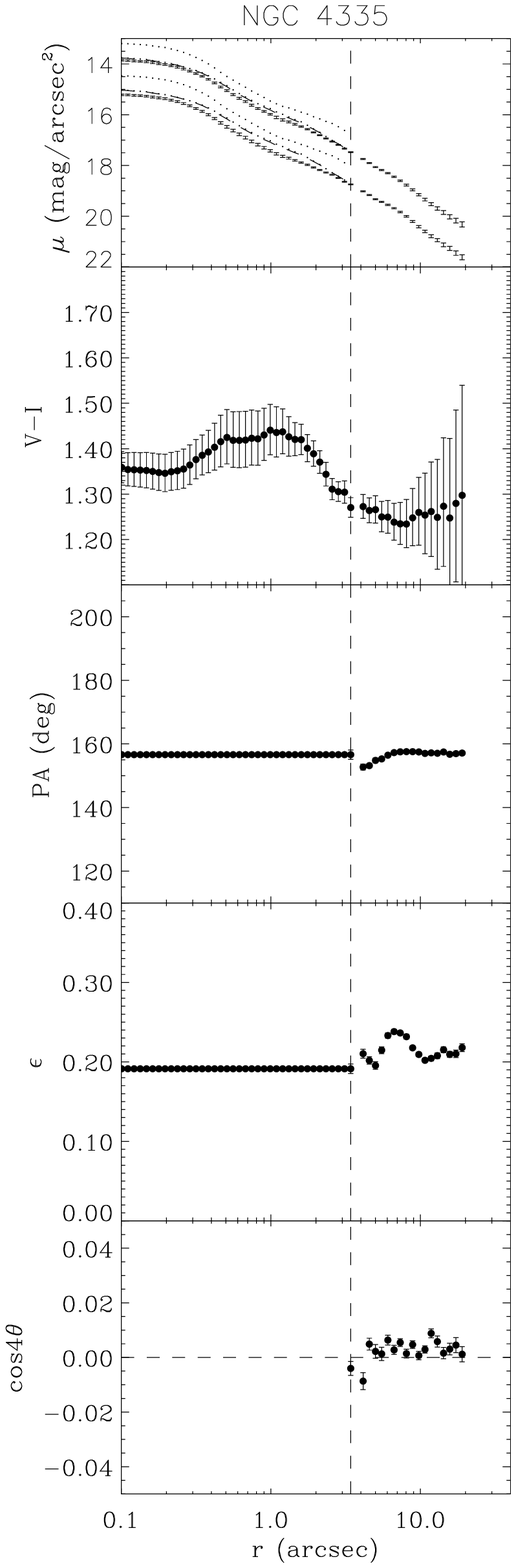}
\ifsubmode
\else\figcaption{\figcapphotomf}\fi
\end{figure}
\addtocounter{figure}{-1}

\clearpage
\begin{figure}
\plottwo{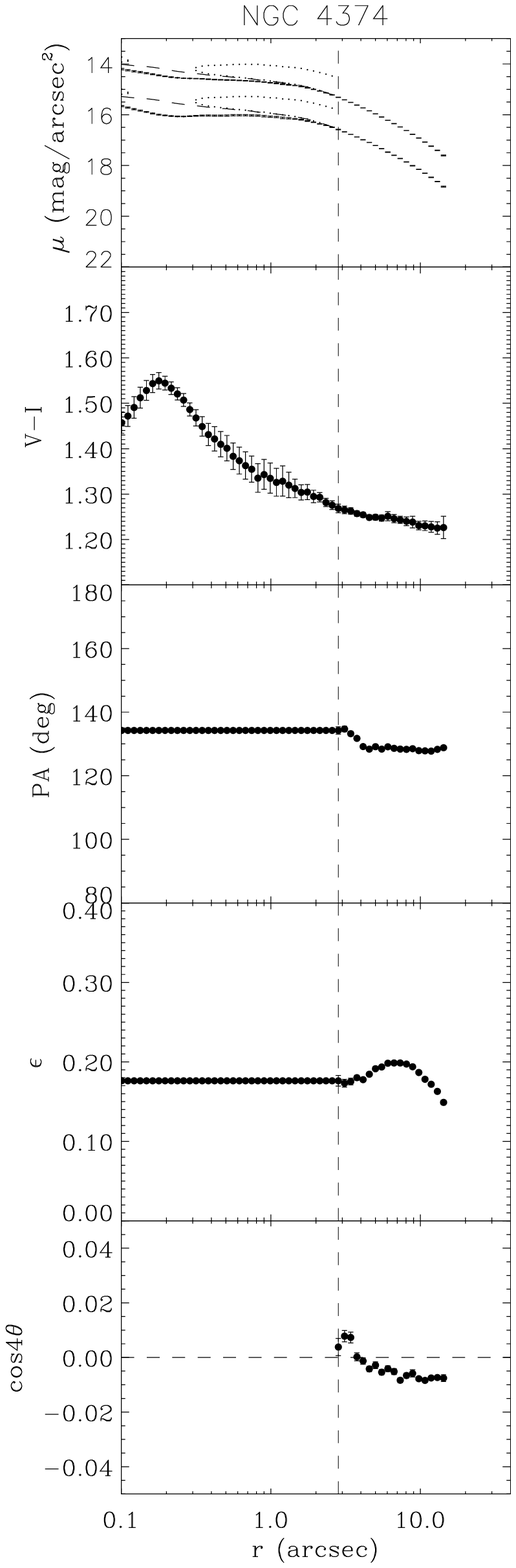}{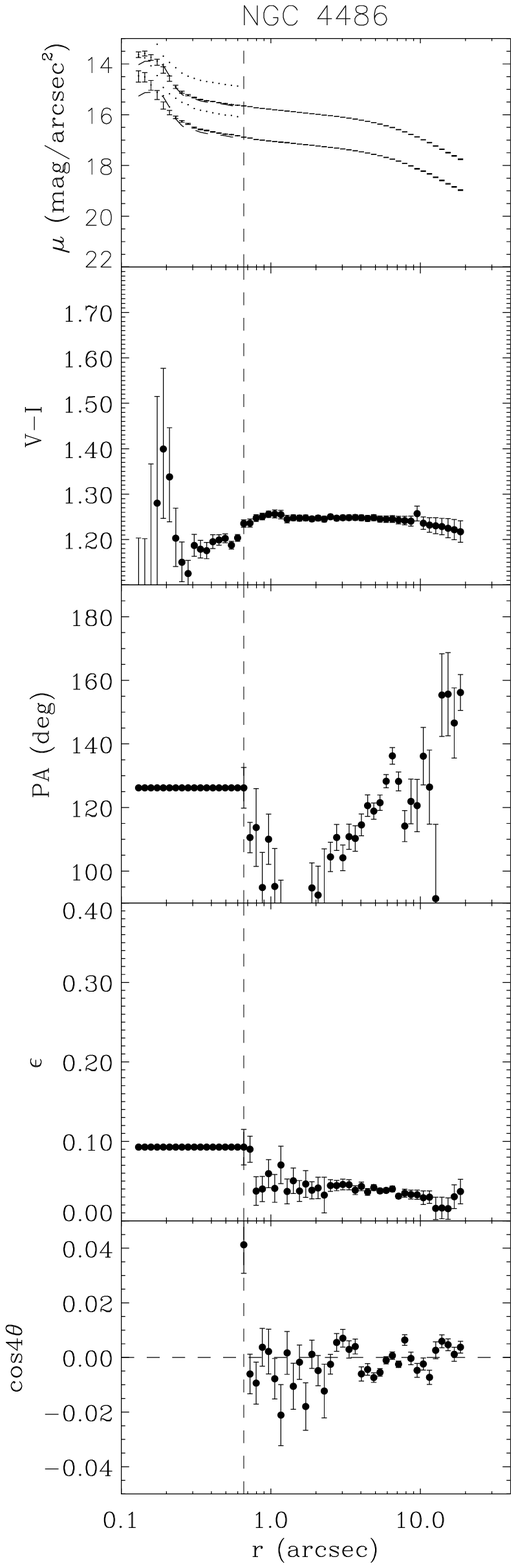}
\ifsubmode
\else\figcaption{\figcapphotomg}\fi
\end{figure}
\addtocounter{figure}{-1}

\clearpage
\begin{figure}
\plottwo{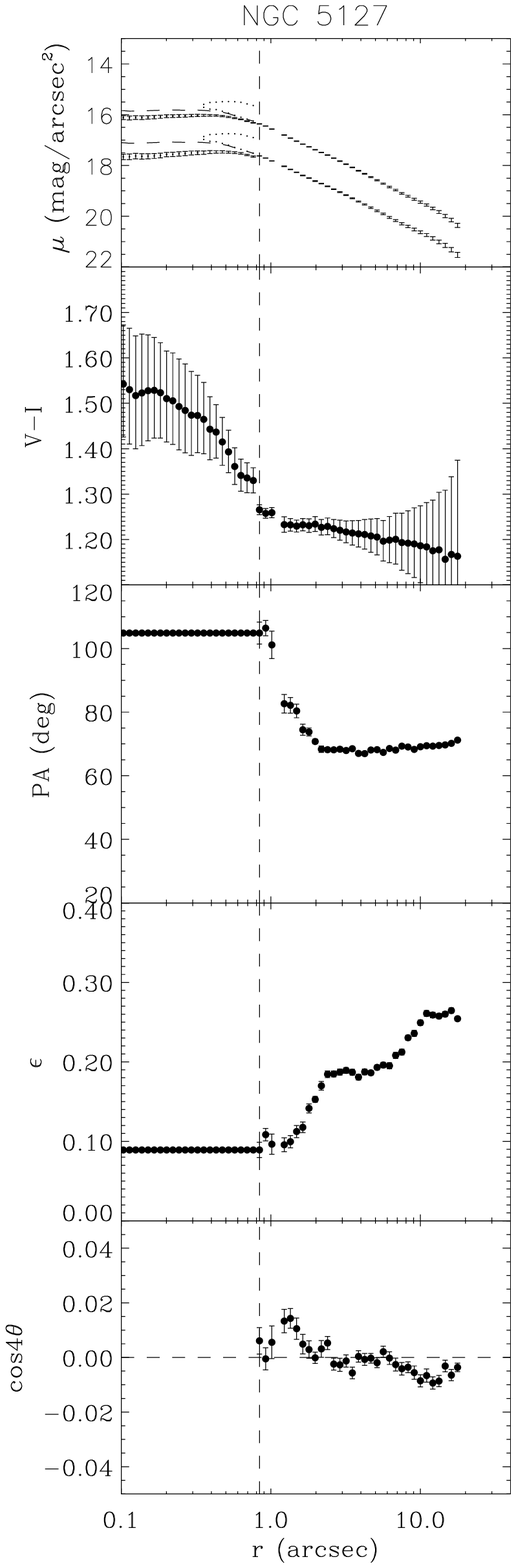}{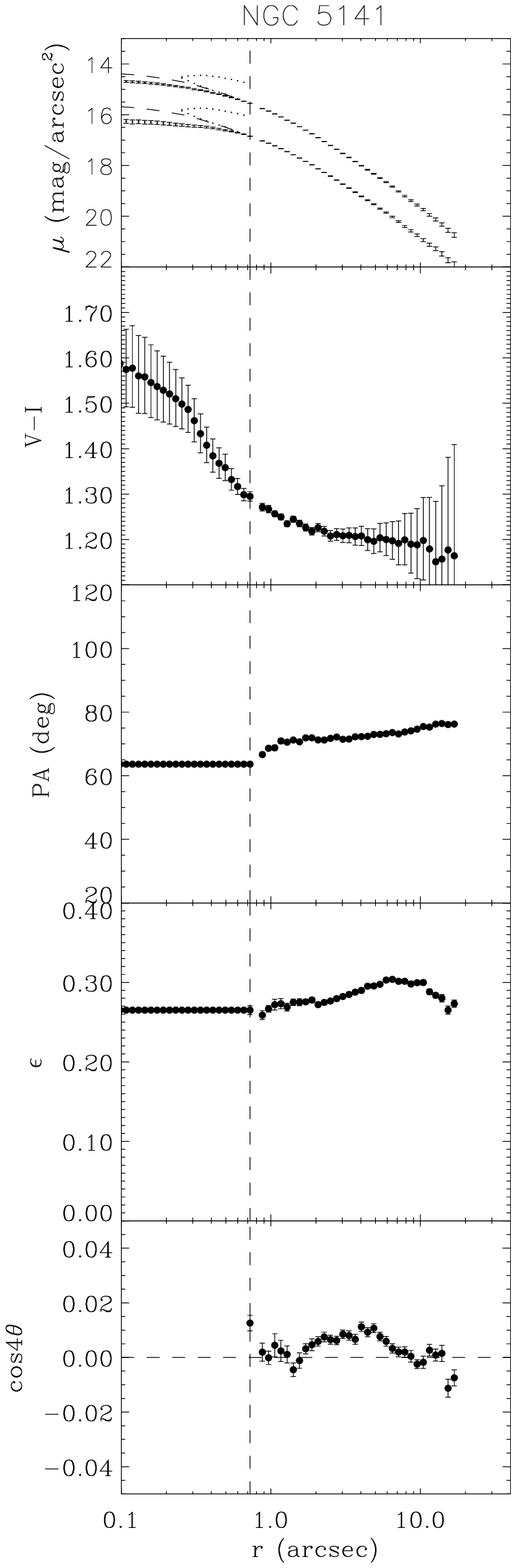}
\ifsubmode
\else\figcaption{\figcapphotomh}\fi
\end{figure}
\addtocounter{figure}{-1}

\clearpage
\begin{figure}
\plottwo{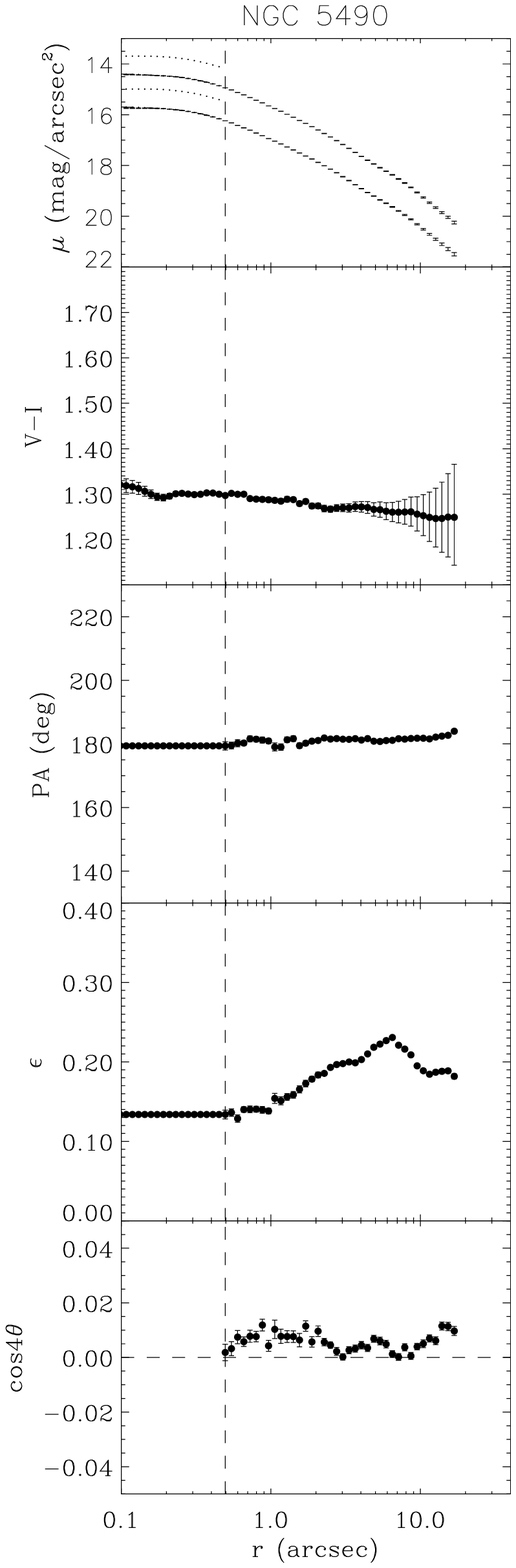}{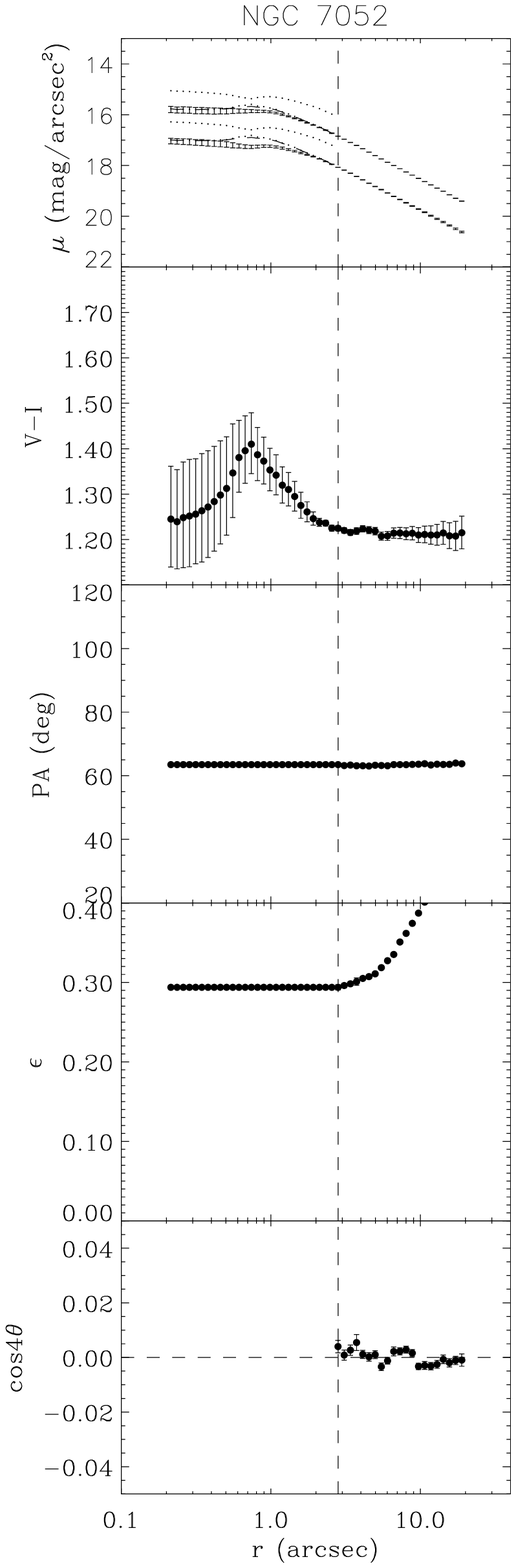}
\ifsubmode
\else\figcaption{\figcapphotomi}\fi
\end{figure}
\addtocounter{figure}{-1}

\clearpage
\begin{figure}
\plottwo{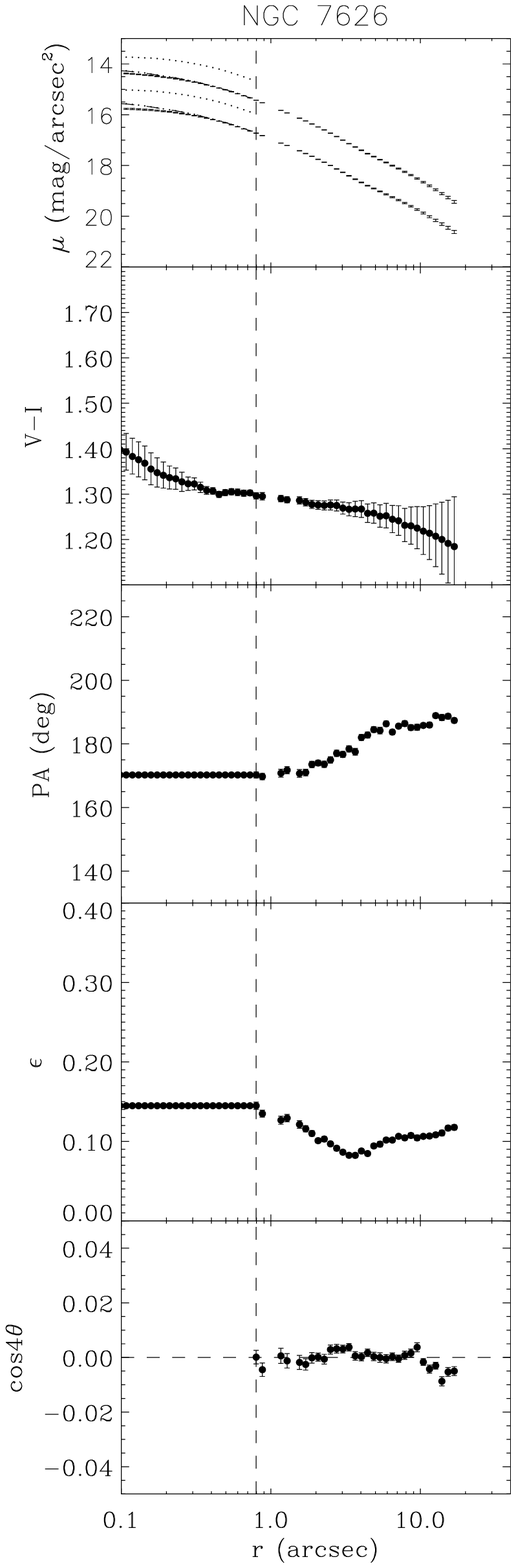}{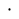}
\ifsubmode
\else\figcaption{\figcapphotomj}\fi
\end{figure}

\clearpage
\begin{figure}
\centerline{\epsfbox{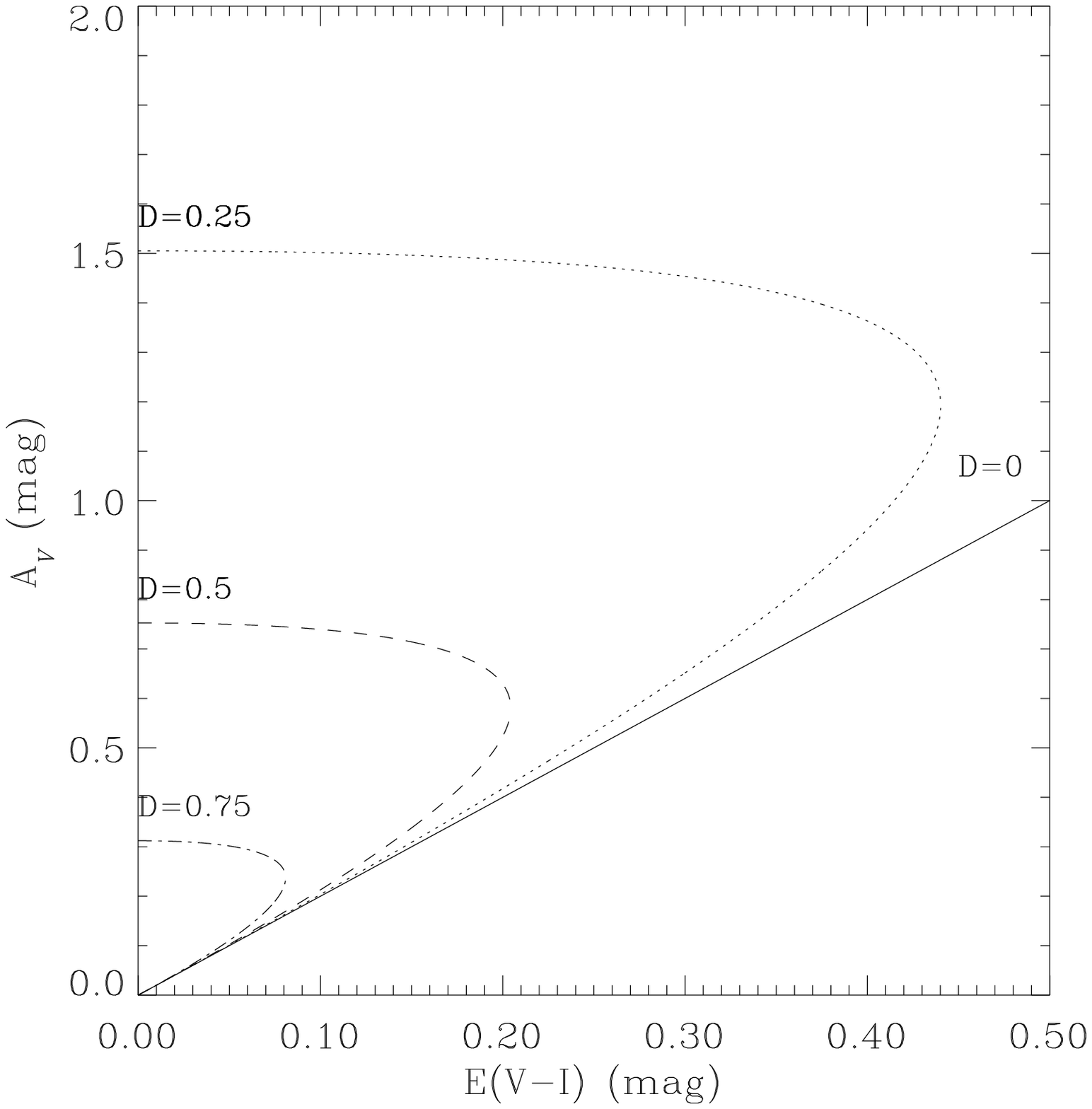}}
\ifsubmode
\else\figcaption{\figcapdustcorrect}\fi
\end{figure}

\clearpage
\begin{figure}
\centerline{\epsfbox{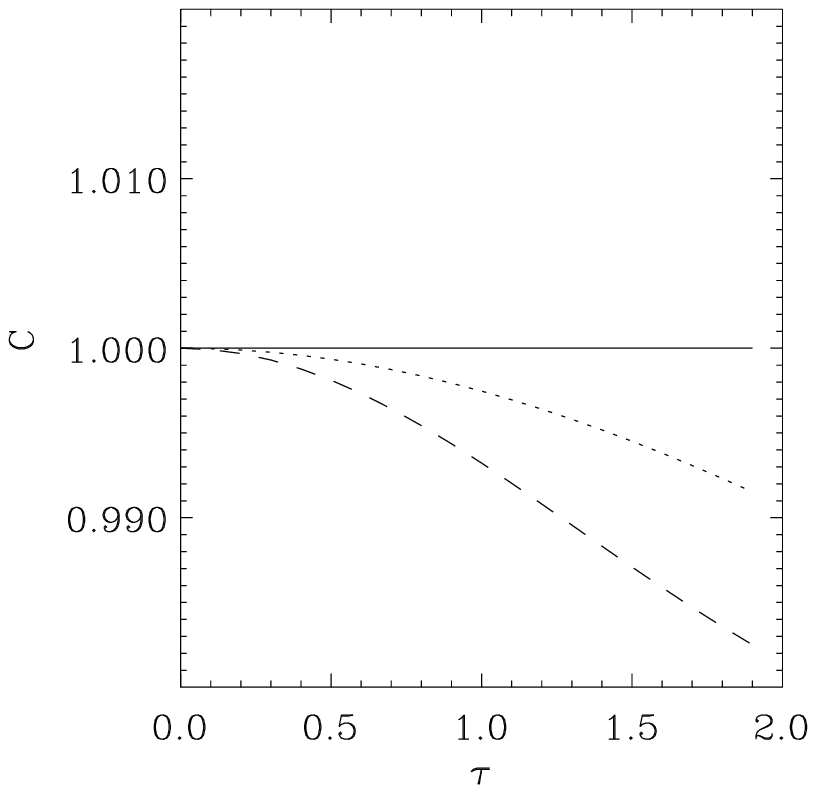}}
\ifsubmode
\else\figcaption{\figcapextemi}\fi
\end{figure}

\clearpage
\begin{figure}
\centerline{\epsfbox{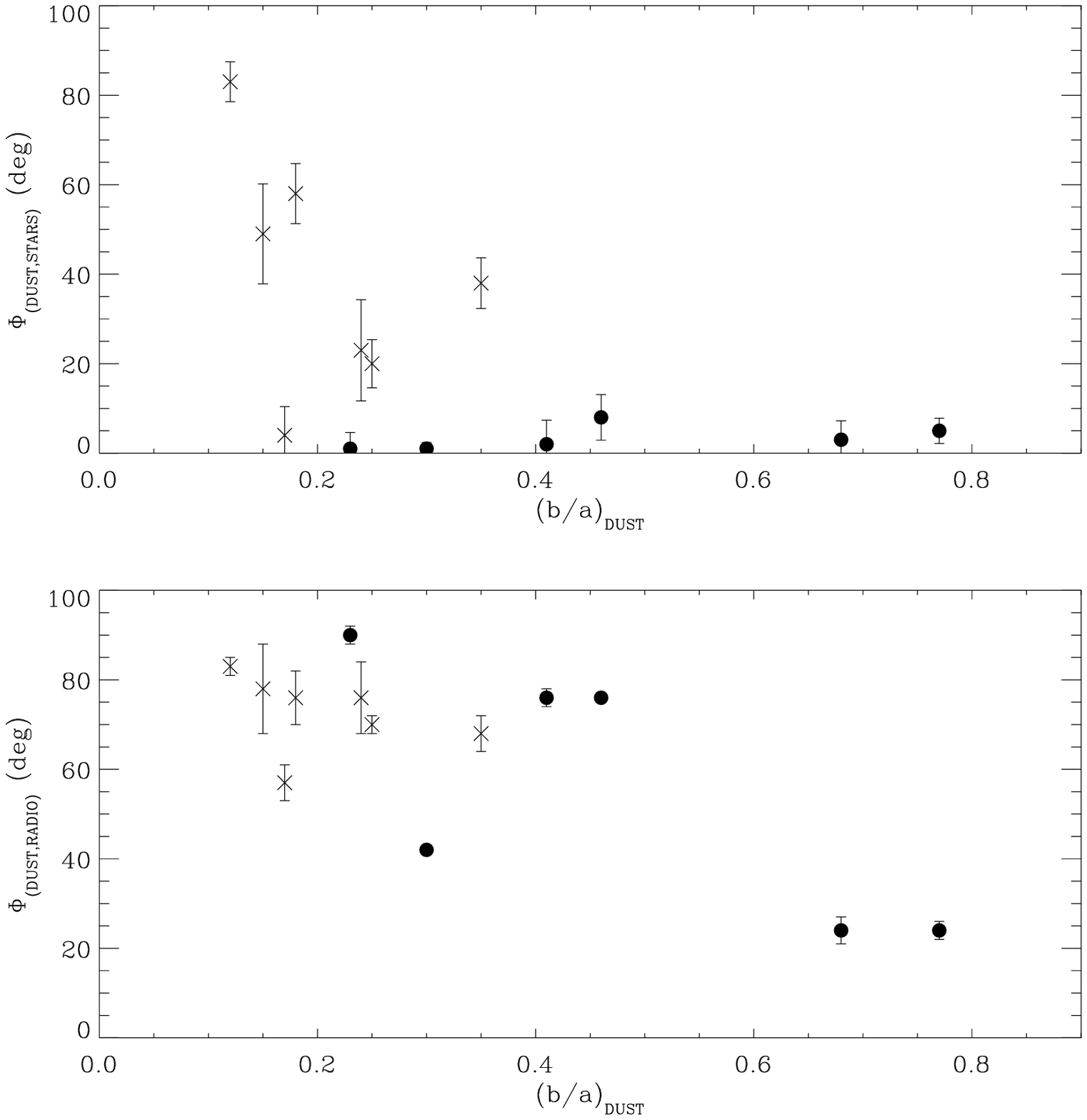}}
\ifsubmode
\else\figcaption{\figcapphi}\fi
\end{figure}

\clearpage
\begin{figure}
\plottwo{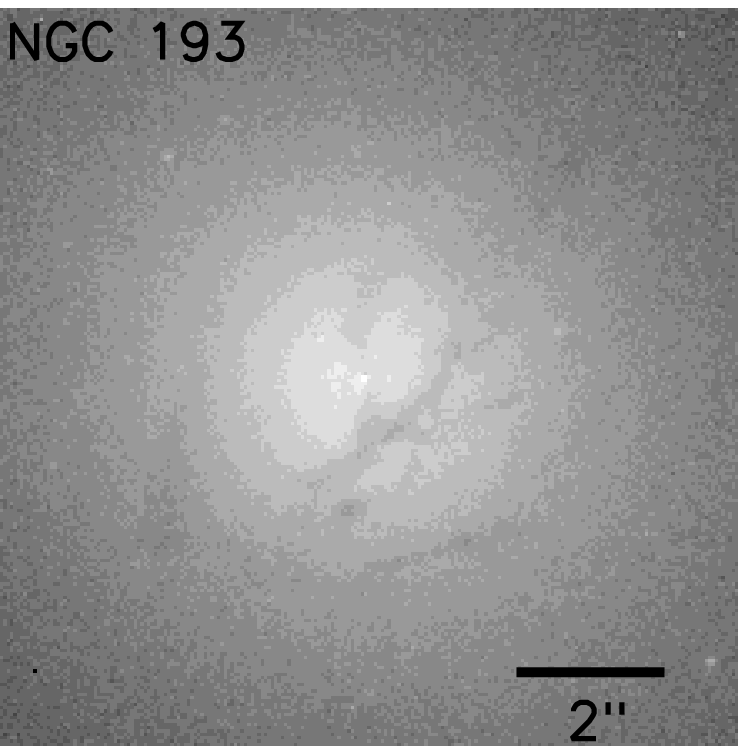}{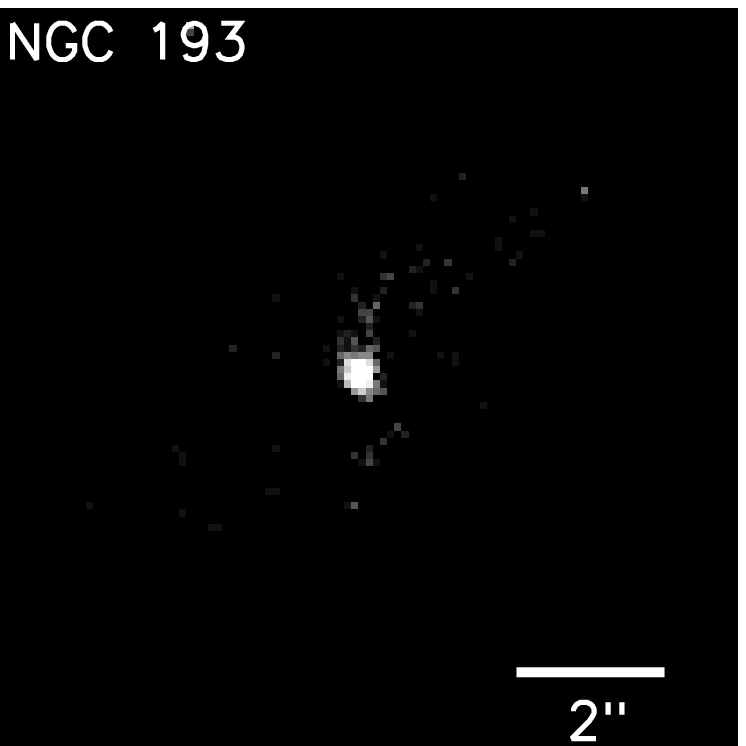}
\vskip0.5truecm
\plottwo{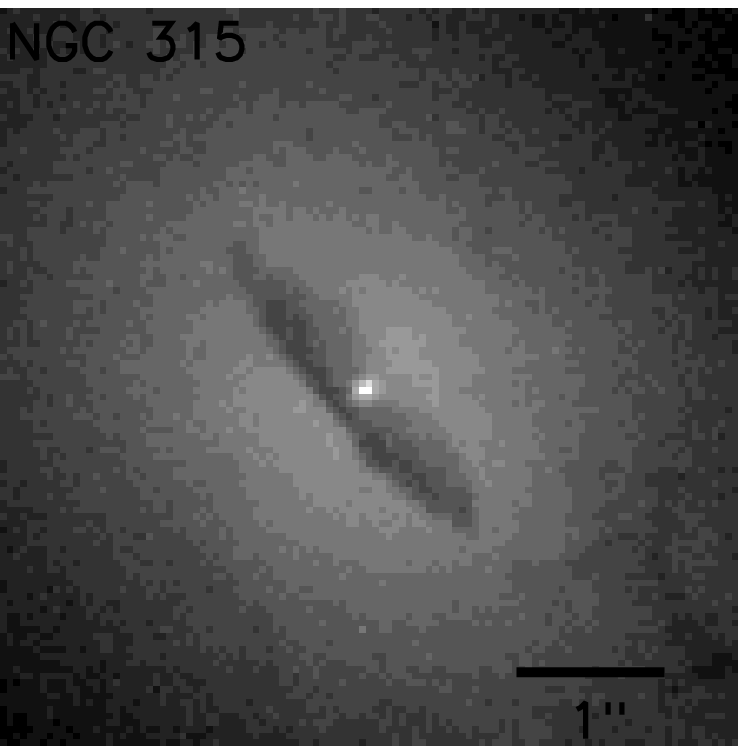}{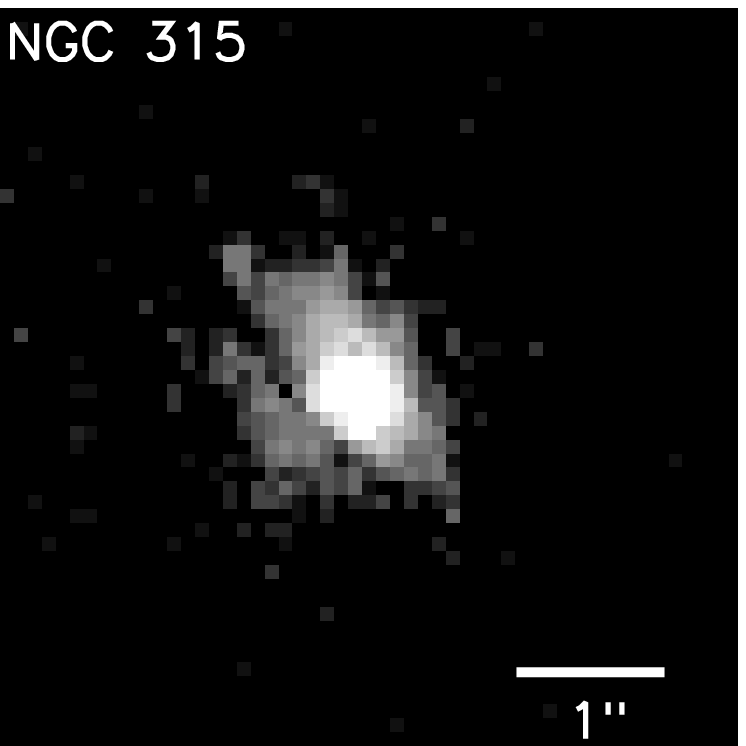}
\ifsubmode
\else\figcaption{\figcapima}\fi
\end{figure}
\addtocounter{figure}{-1}

\clearpage
\begin{figure}
\plottwo{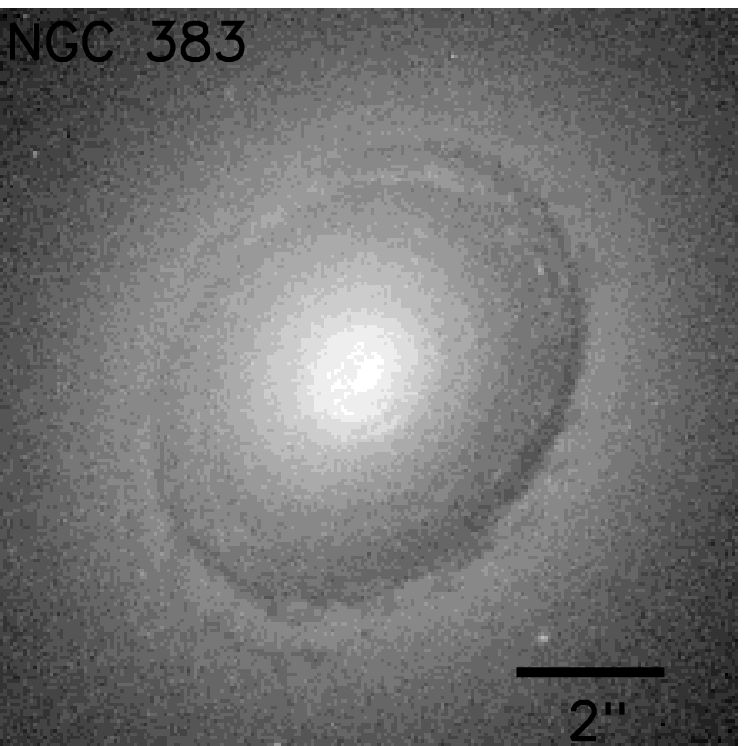}{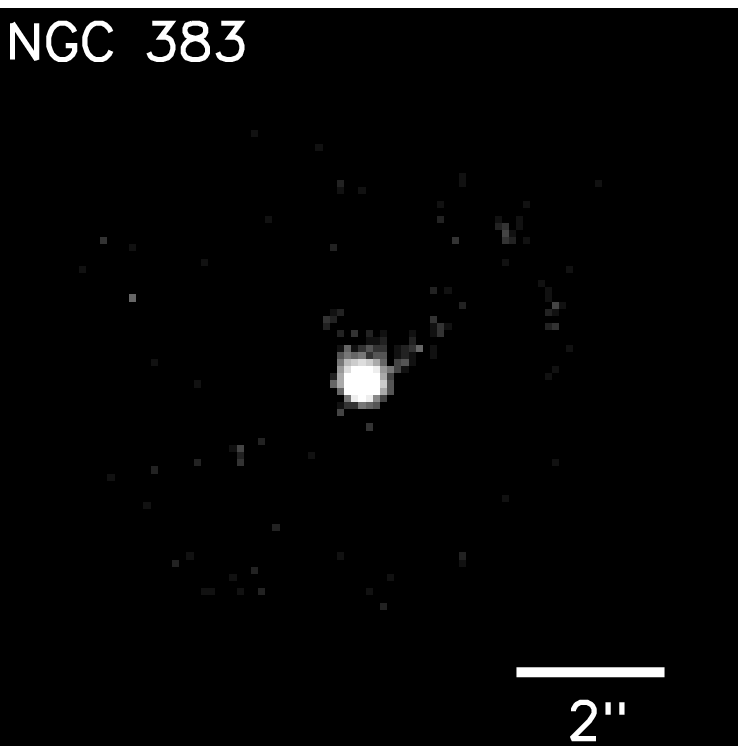}
\vskip0.5truecm
\plottwo{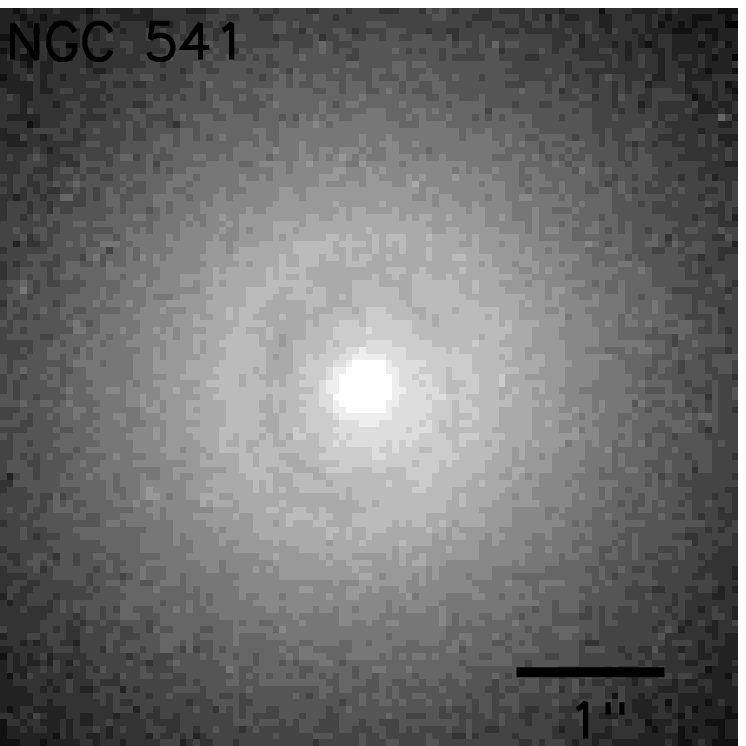}{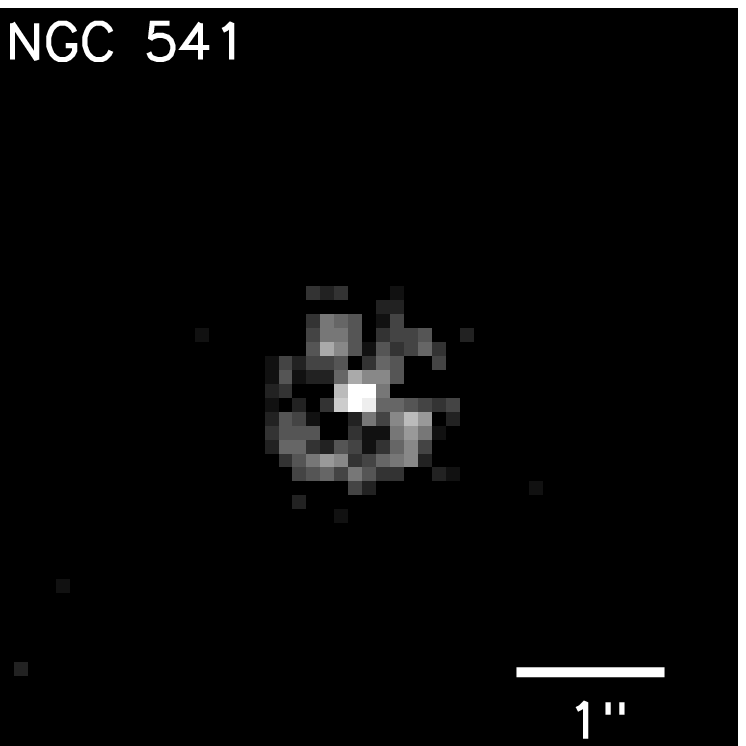}
\ifsubmode
\else\figcaption{\figcapimb}\fi
\end{figure}
\addtocounter{figure}{-1}

\clearpage
\begin{figure}
\plottwo{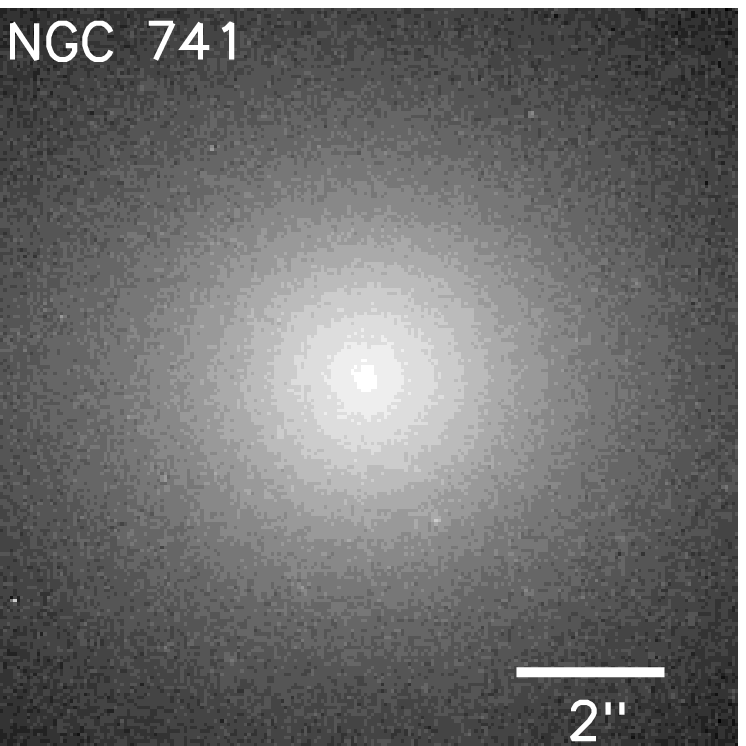}{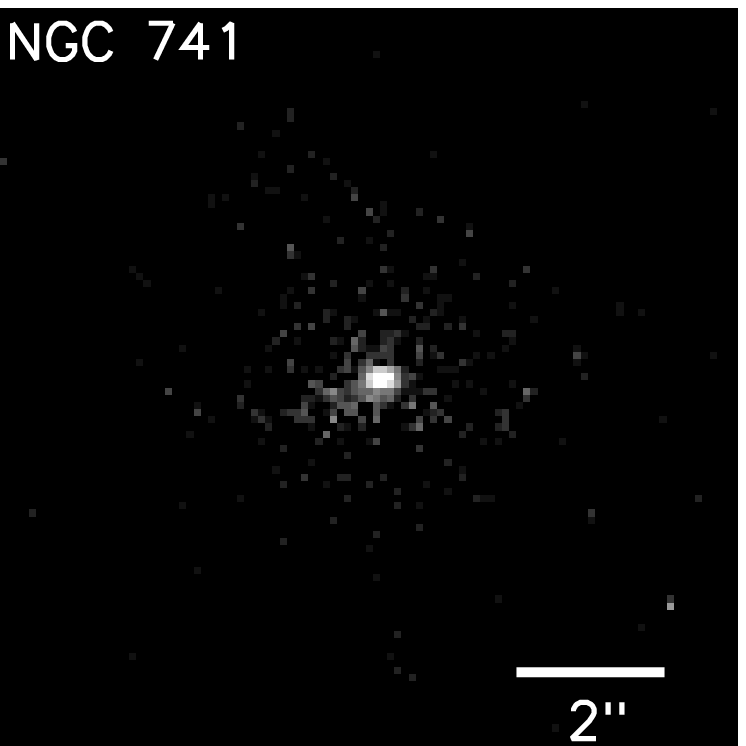}
\vskip0.5truecm
\plottwo{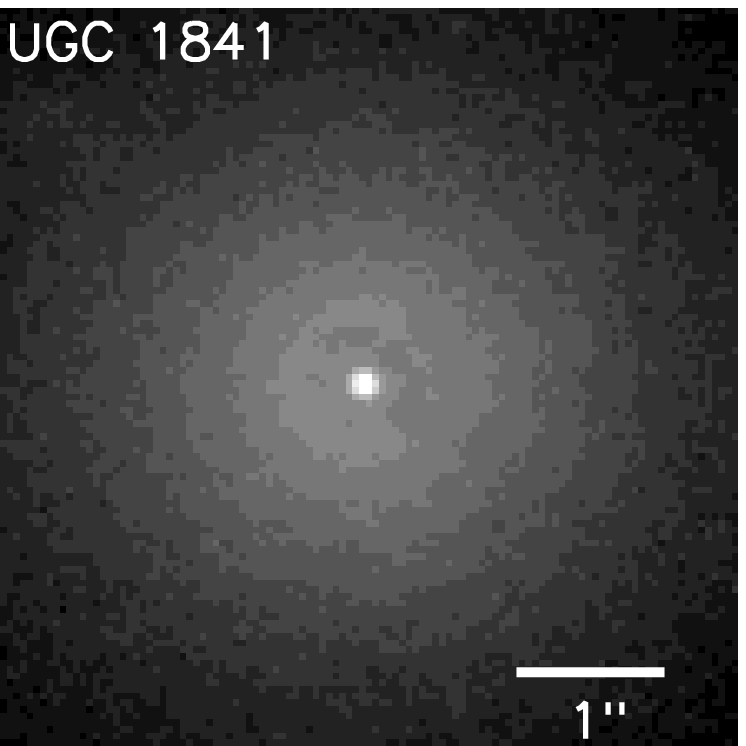}{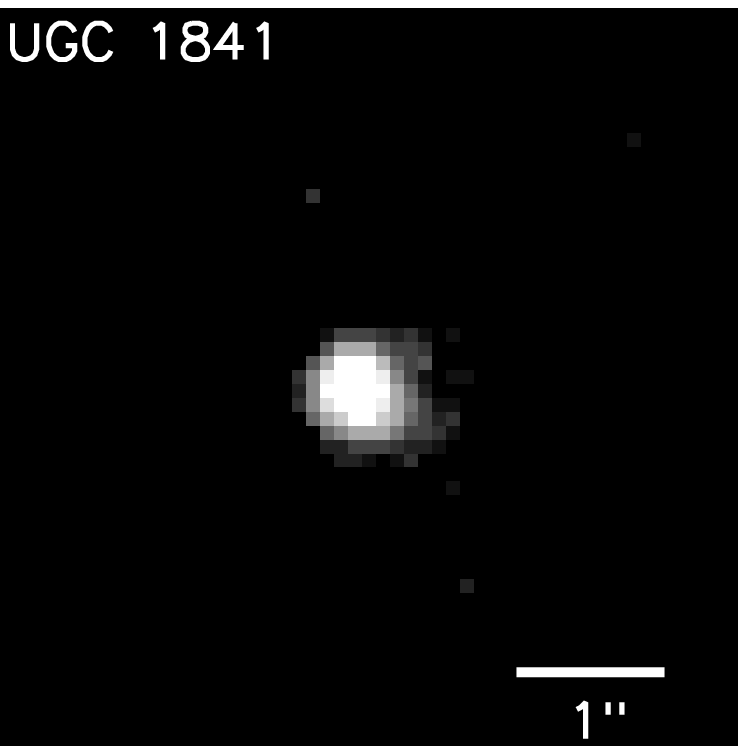}
\ifsubmode
\else\figcaption{\figcapimc}\fi
\end{figure}
\addtocounter{figure}{-1}

\clearpage
\begin{figure}
\plottwo{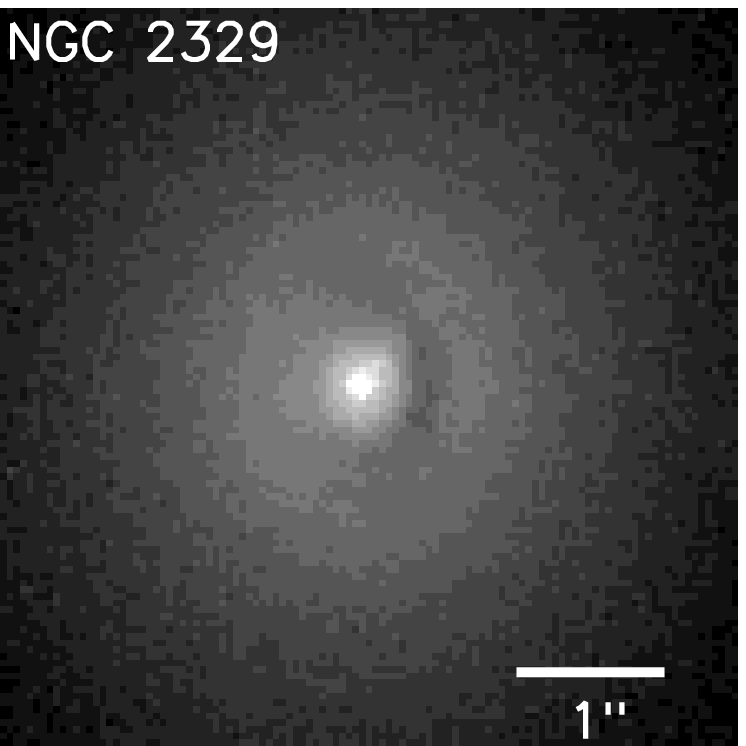}{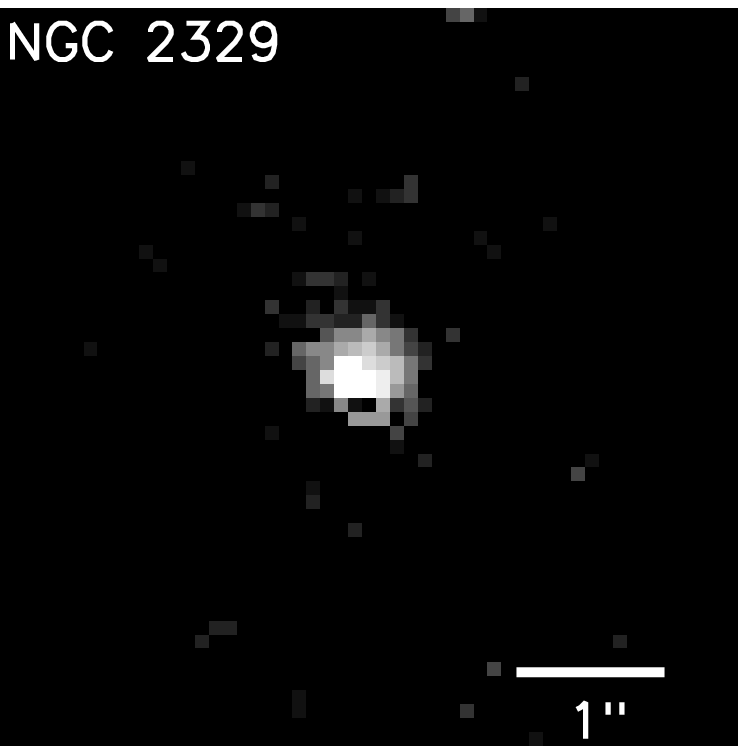}
\vskip0.5truecm
\plottwo{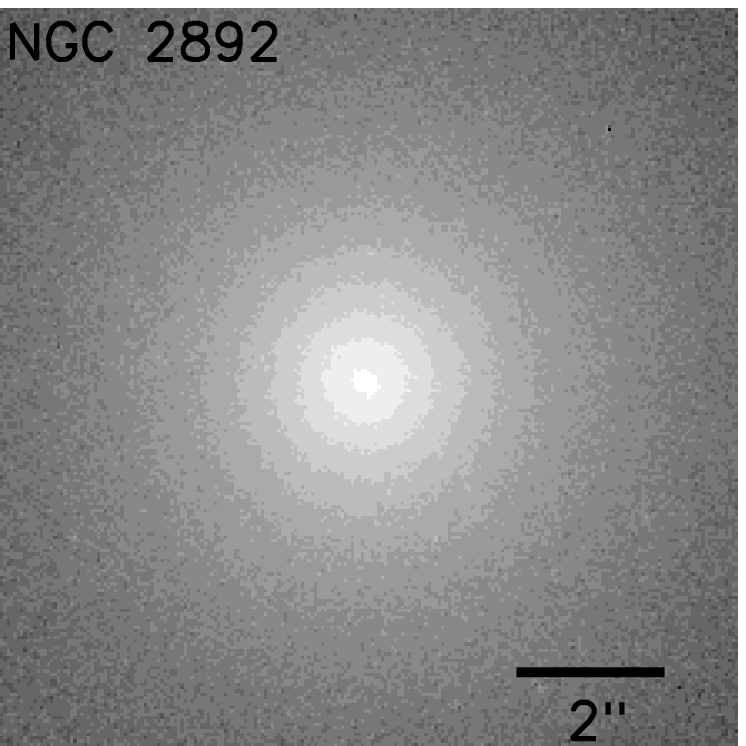}{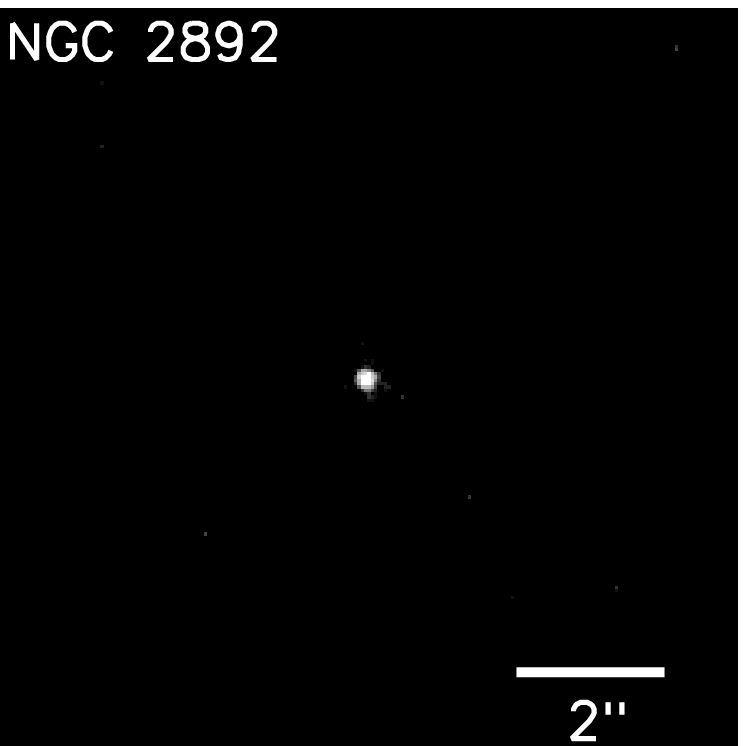}
\ifsubmode
\else\figcaption{\figcapimd}\fi
\end{figure}
\addtocounter{figure}{-1}

\clearpage
\begin{figure}
\plottwo{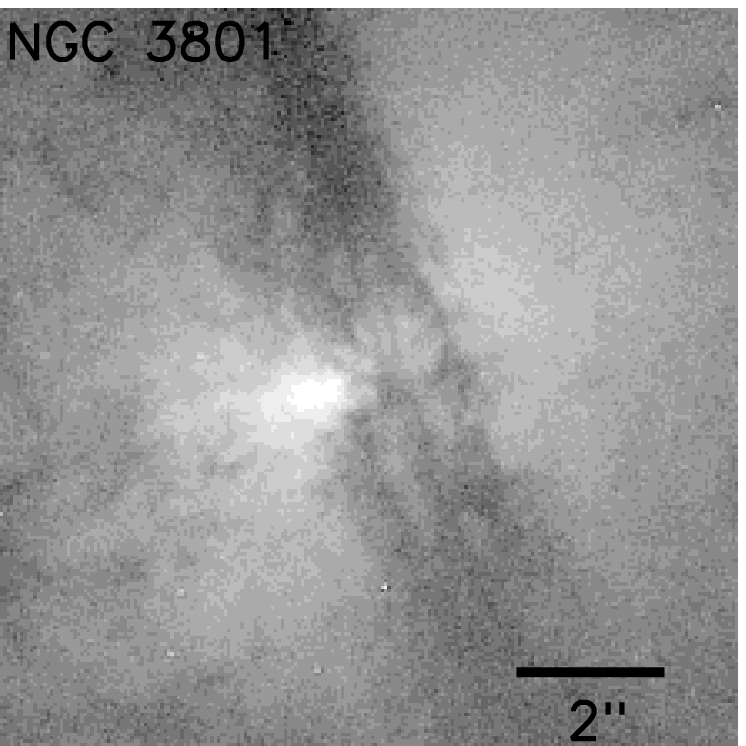}{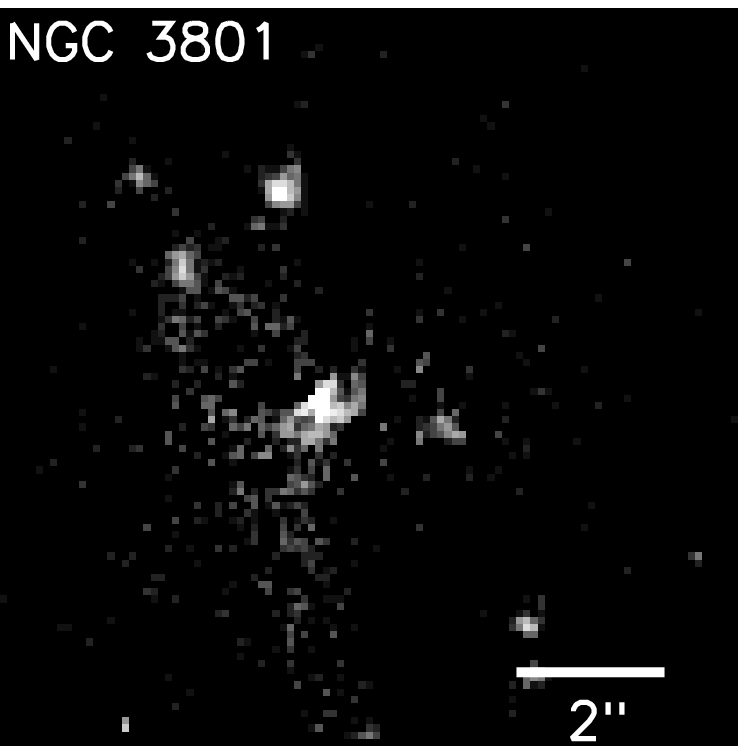}
\vskip0.5truecm
\plottwo{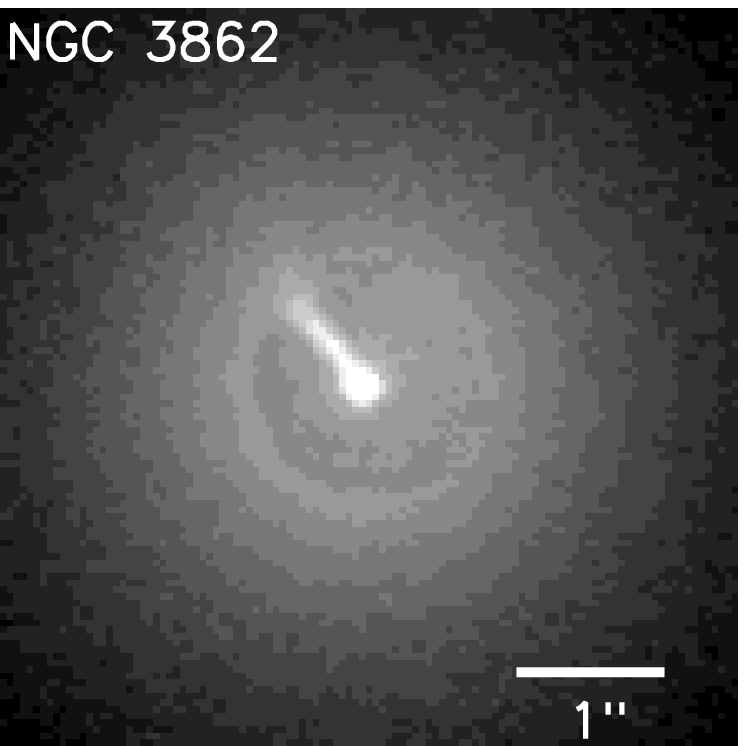}{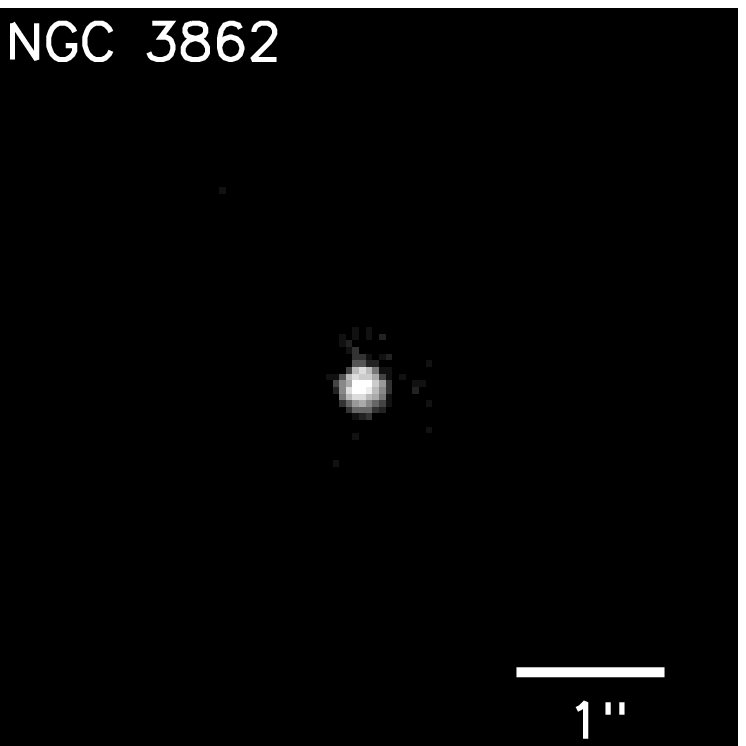}
\ifsubmode
\else\figcaption{\figcapime}\fi
\end{figure}
\addtocounter{figure}{-1}

\clearpage
\begin{figure}
\plottwo{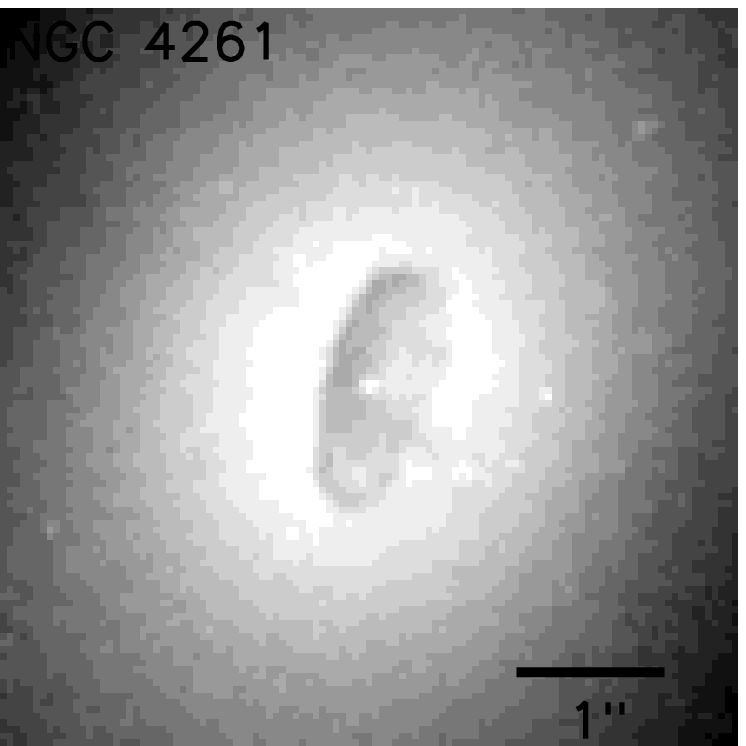}{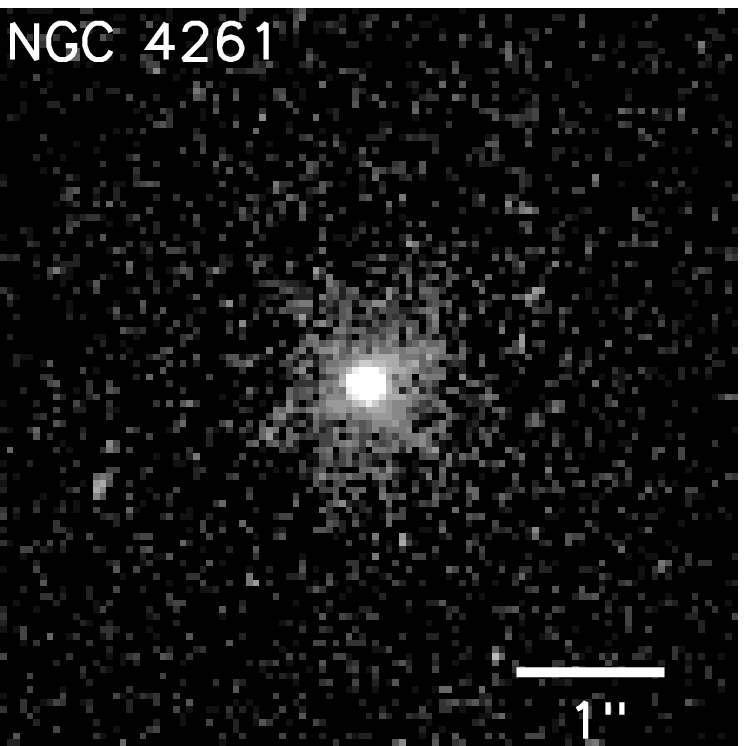}
\vskip0.5truecm
\plottwo{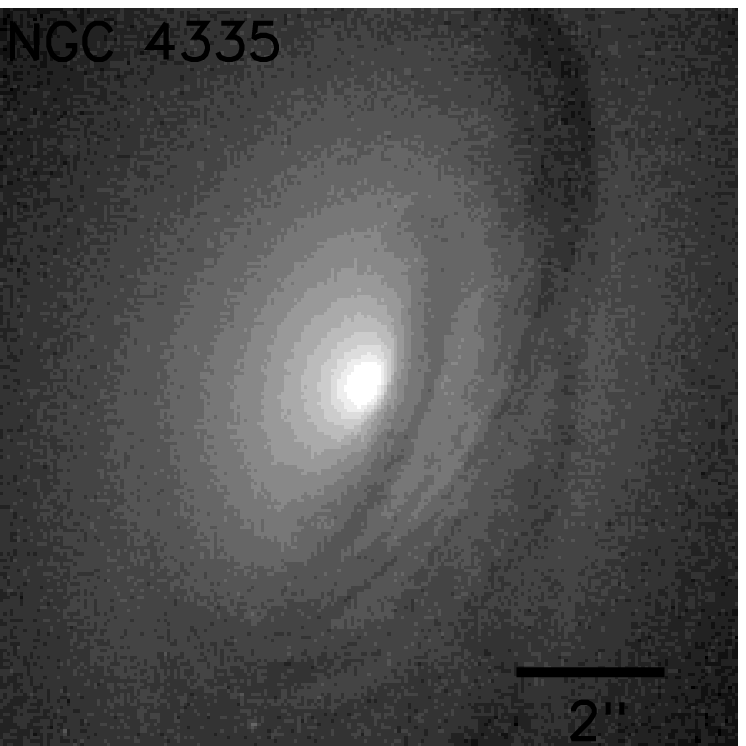}{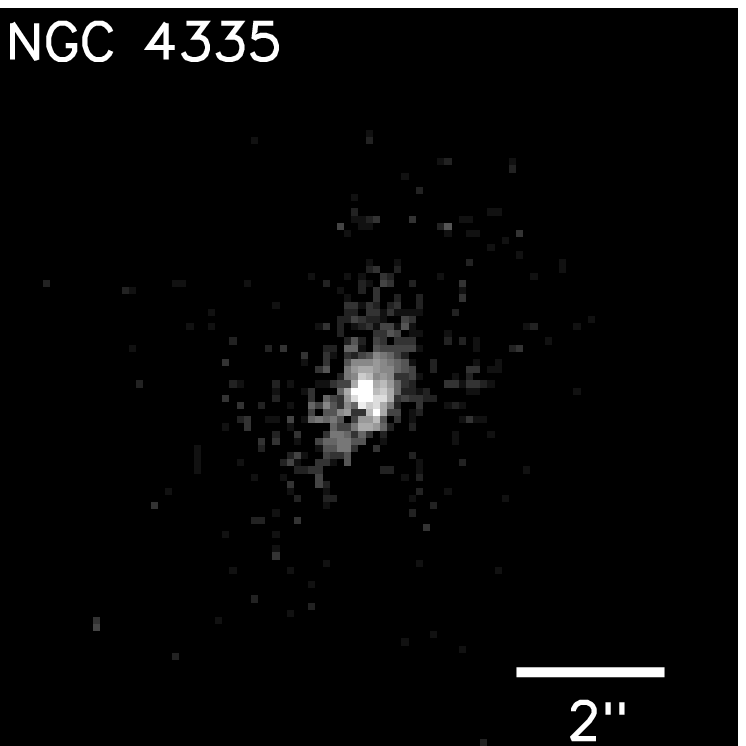}
\ifsubmode
\else\figcaption{\figcapimf}\fi
\end{figure}
\addtocounter{figure}{-1}

\clearpage
\begin{figure}
\plottwo{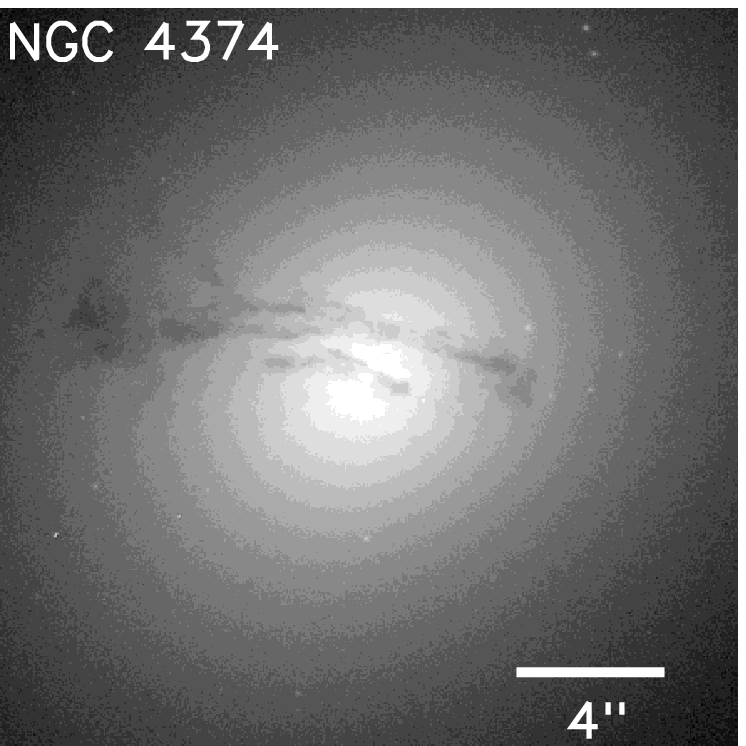}{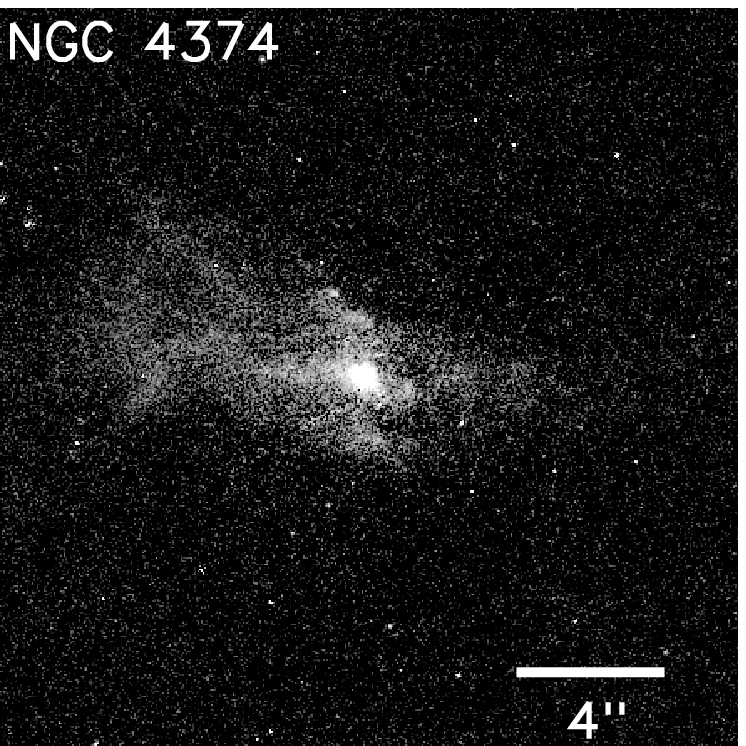}
\vskip0.5truecm
\plottwo{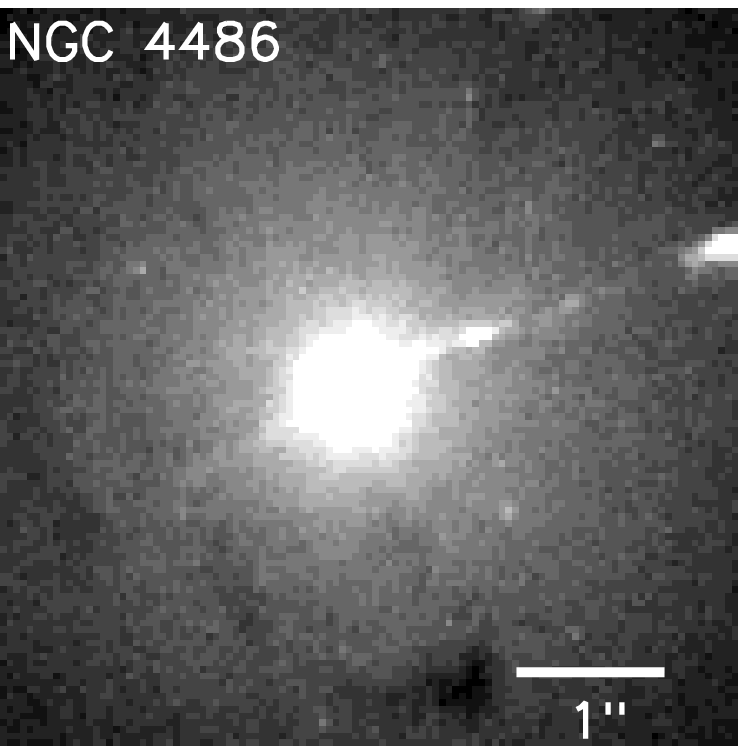}{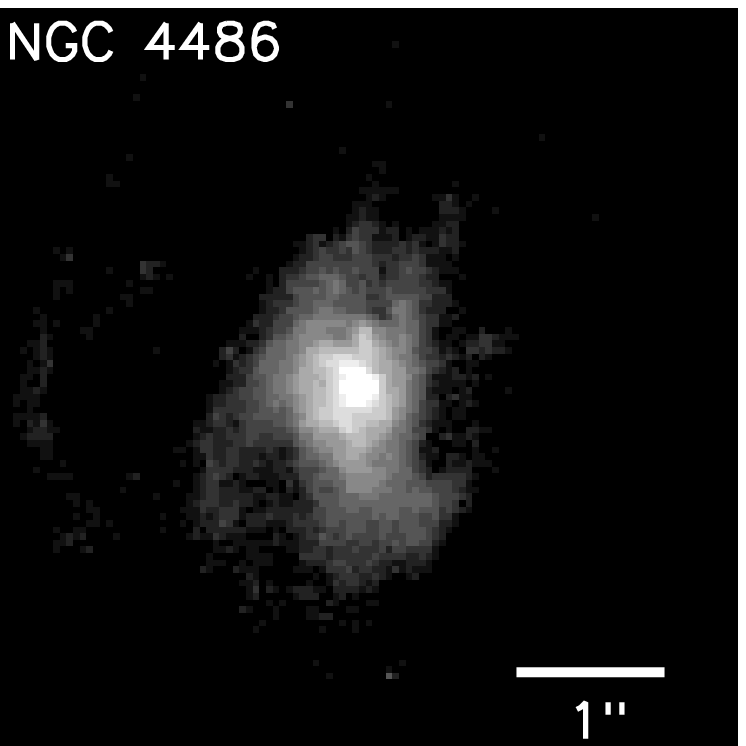}
\ifsubmode
\else\figcaption{\figcapimg}\fi
\end{figure}
\addtocounter{figure}{-1}

\clearpage
\begin{figure}
\plottwo{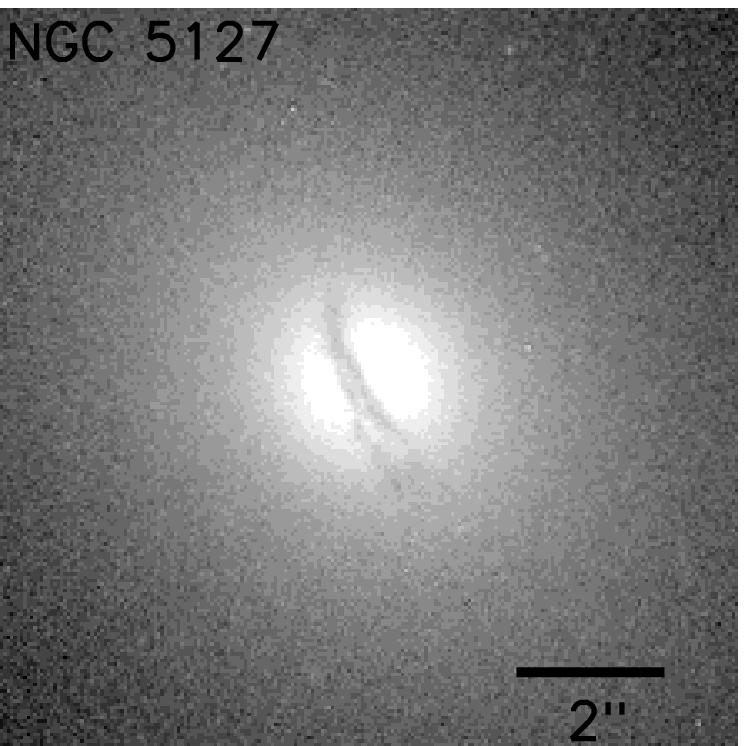}{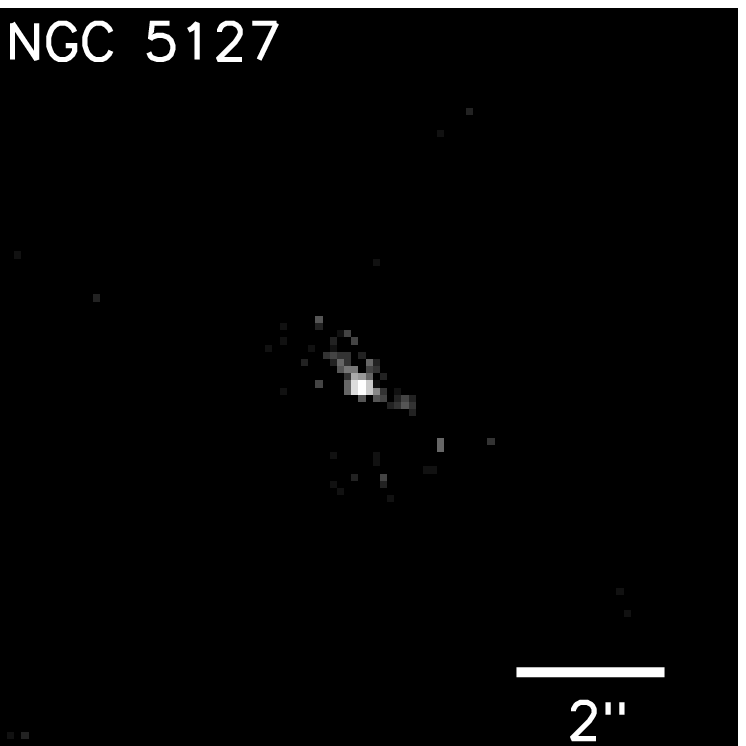}
\vskip0.5truecm
\plottwo{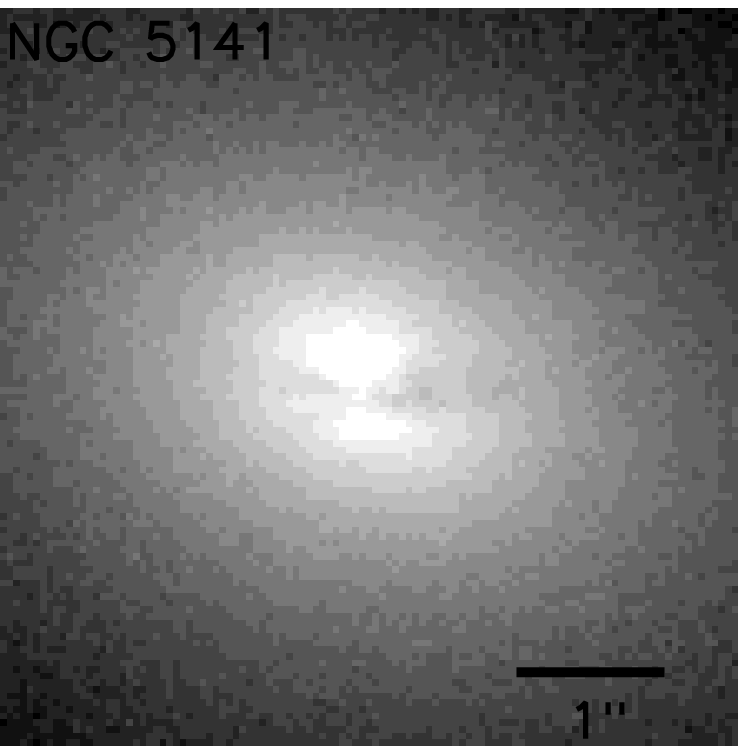}{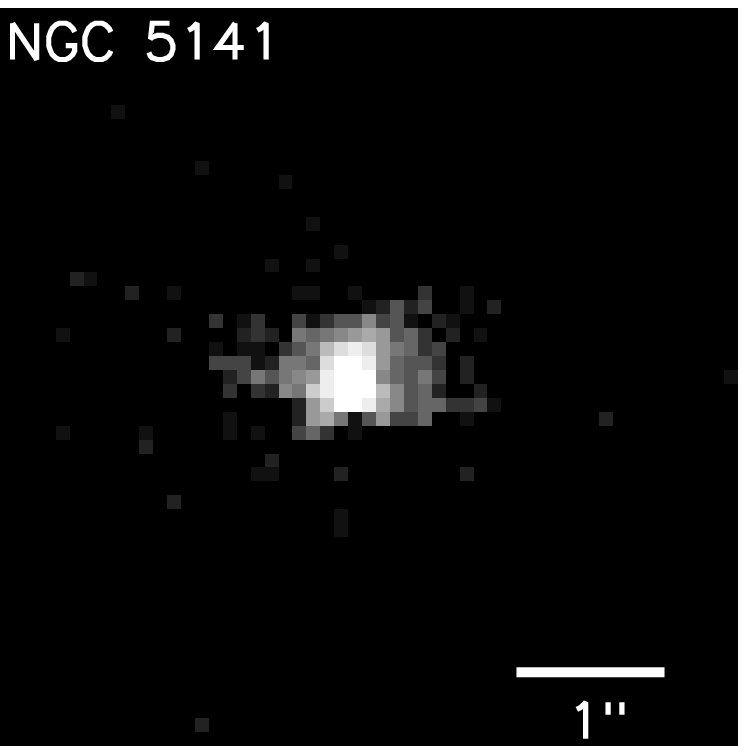}
\ifsubmode
\else\figcaption{\figcapimh}\fi
\end{figure}
\addtocounter{figure}{-1}

\clearpage
\begin{figure}
\plottwo{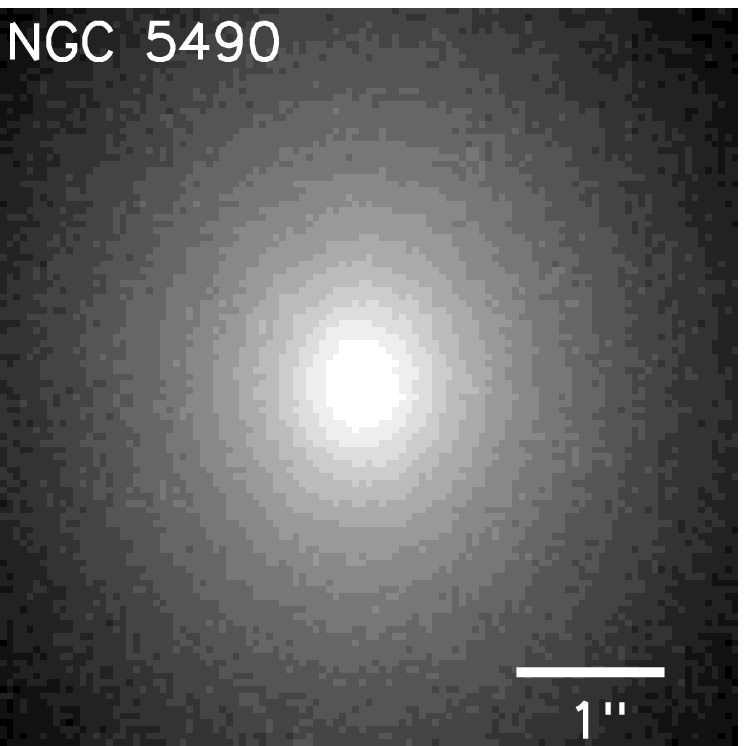}{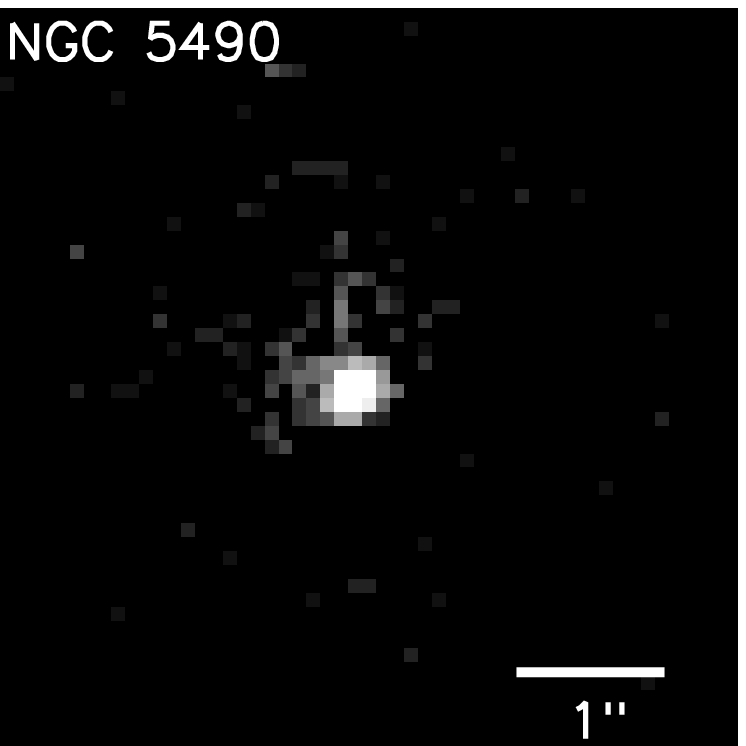}
\vskip0.5truecm
\plottwo{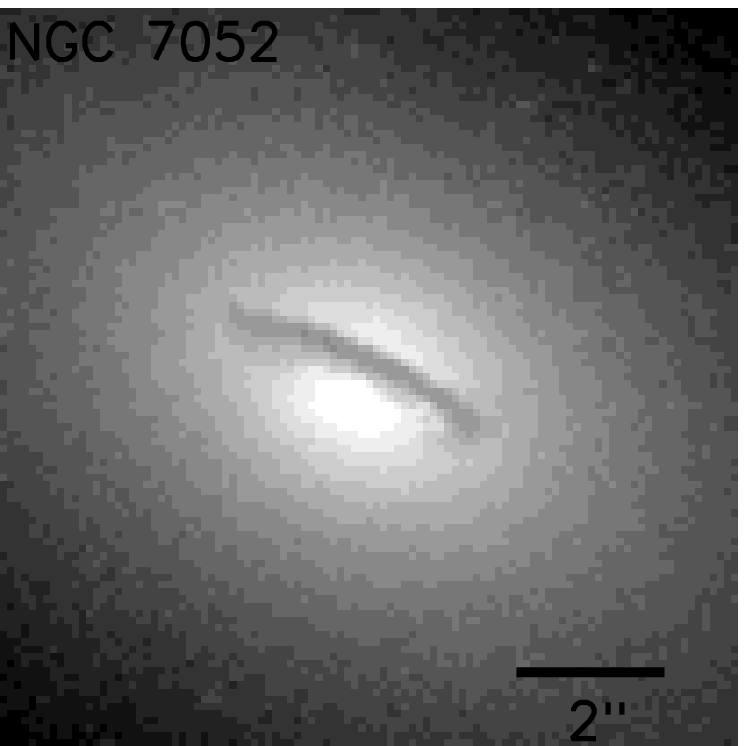}{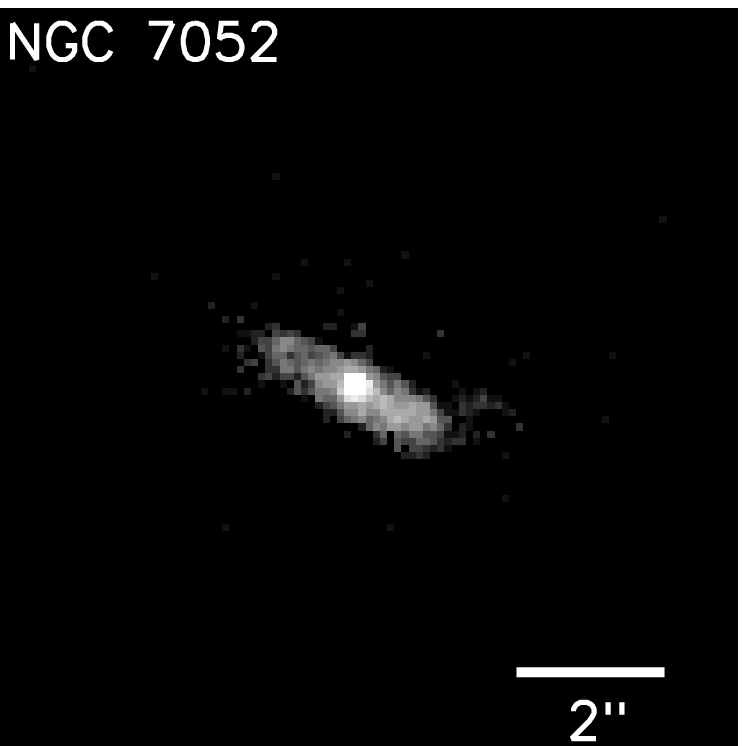}
\ifsubmode
\else\figcaption{\figcapimi}\fi
\end{figure}
\addtocounter{figure}{-1}

\clearpage
\begin{figure}
\plottwo{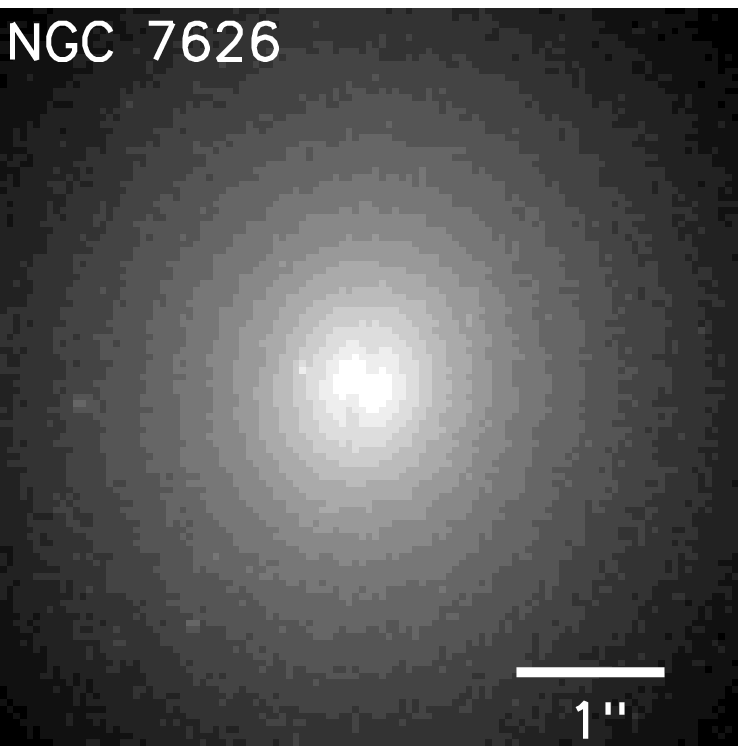}{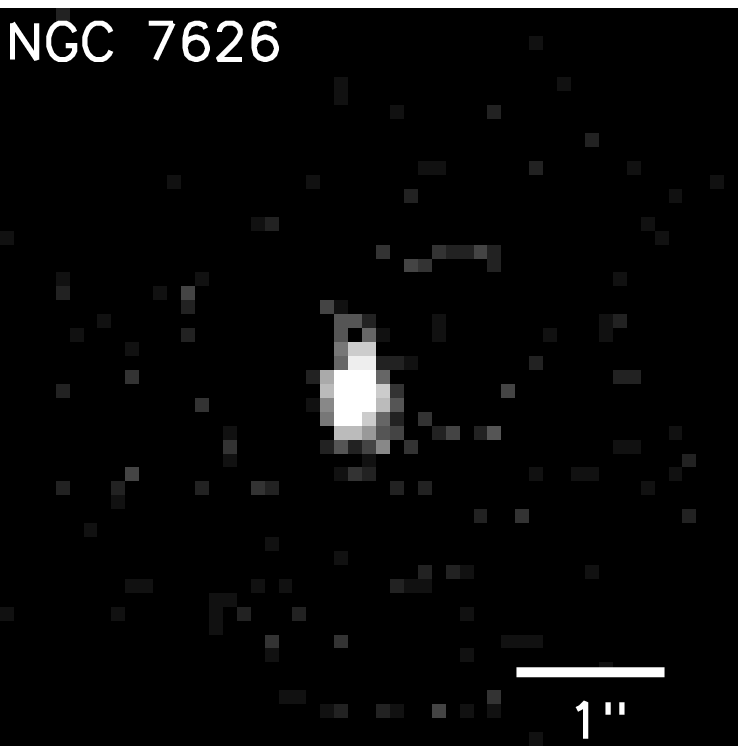}
\vskip0.5truecm
\ifsubmode
\else\figcaption{\figcapimj}\fi
\end{figure}
\addtocounter{figure}{-1}


\fi


\clearpage

\begin{deluxetable}{llcllll}
\tablewidth{0pt}
\tablecaption{Sample\label{t_sample}}
\tablehead{
\colhead{NGC} & \colhead{UGC} & \colhead{Type} & \colhead{D$_{75}$}
& \colhead{M$_{\rm p}$} & \colhead{Log L$_{\rm radio}$} & \colhead{Comment} \\
 & & & \colhead{(Mpc)}
& \colhead{(mag)} & \colhead{(WHz$^{-1}$)} & \\
\colhead{(1)} & \colhead{(2)} & \colhead{(3)} & \colhead{(4)} & \colhead{(5)} & 
\colhead{(6)} & \colhead{(7)} \\}

\startdata
0193 & 00408 & ..  & 57.9 & -19.9  & 23.84 & \nl
0315 & 00597 & E   & 67.9 & -21.8  & 24.01 & \nl
0383 & 00689 & S0  & 65.2 & -20.6  & 24.42 & 3C 31 \nl
0541 & 01004 & E   & 73.3 & -20.3  & 23.85 & \nl
0741 & 01413 & E   & 70.2 & -21.0  & 23.76 & \nl
.... & 01841 & E   & 84.8 & -19.9  & 24.85 & 3C 66B \nl
2329 & 03695 & E-S0& 76.7 & -21.0  & 23.67 & \nl
2892 & 05073 & E   & 90.8 & -20.4  & 23.27 & \nl
3801 & 06635 & S0? & 43.4 & -19.9  & 23.40 & \nl
3862 & 06723 & E   & 84.4 & -20.6  & 24.66 & 3C 264 \nl
.... & 07115 & E   & 90.5 & -20.3  & 23.85 & \nl 
4261 & 07360 & E   & 29.5 & -20.9  & 24.31 & 3C 270 \nl
4335 & 07455 & E   & 61.5 & -20.3  & 23.02 & \nl
4374 & 07494 & S0  & 15.4 & -20.6  & 23.26 & 3C 272.1, M 84 \nl
4486 & 07654 & E   & 15.4 & -21.0  & 24.81 & 3C 274, M 87 \nl
5127 & 08419 & E   & 64.4 & -20.1  & 23.99 & \nl
5141 & 08433 & S0  & 70.7 & -20.3  & 23.71 & \nl
5490 & 09058 & E   & 69.2 & -20.8  & 23.70 & \nl
7052 & 11718 & E   & 55.4 & -20.0  & 22.95 & \nl
.... & 12064 & E-S0& 68.3 & -19.9  & 24.29 & 3C 449 \nl
7626 & 12531 & E   & 46.6 & -20.5  & 23.28 & \nl
\enddata
\tablecomments{\footnotesize Col.~(1): NGC number. Col.~(2): UGC number. Col.~(3):
  Hubble type from CB88. Col.~(4): distances from Faber \etal (1989)
  or if not available directly from recession velocity and $H_0$=75
  kms$^{-1}$Mpc$^{-1}$. Col.~(5): photographic magnitude from CB88.
  Col.~(6): total spectral radio luminosity at 1400 MHz from CB88.
  Col.~(7): 3C and Messier numbers.  }
\end{deluxetable}


\begin{deluxetable}{llrlrlrl}
\tablewidth{0pt}
\tablecaption{Observational Setup\label{t_obs}}
\tablehead{
\colhead{Name}
& \colhead{Filter} & \colhead{T$_{\rm exp}$}
& \colhead{Filter} & \colhead{T$_{\rm exp}$}
& \colhead{Filter} & \colhead{T$_{\rm exp}$} & \colhead{CCD} \\
 & 
& \colhead{(s)}
& & \colhead{(s)}
& & \colhead{(s)} & \\
\colhead{(1)}
& \colhead{(2)} & \colhead{(3)}
& \colhead{(4)} & \colhead{(5)}
& \colhead{(6)} & \colhead{(7)} & \colhead{(8)} \\}

\startdata
NGC  193 & F555W &  460 & F814W &  460 &  LRF680 & 3100 &  WF2 \nl
NGC  315 & F555W &  460 & F814W &  460 &  LRF680 & 3100 &  WF2 \nl
NGC  383 & F555W &  460 & F814W &  460 &  LRF680 & 3100 &  WF2 \nl
NGC  541 & F555W &  460 & F814W &  460 &  LEF680 & 3100 &  WF2 \nl
NGC  741 & F555W &  460 & F814W &  460 &  LRF680 & 3100 &  WF2 \nl
UGC 1841 & F555W &  460 & F814W &  460 &  LRF680 & 3500 &  WF2 \nl
NGC 2329 & F555W &  460 & F814W &  460 &  LRF680 & 3600 &  WF2 \nl
NGC 2892 & F555W &  460 & F814W &  460 &  F673N  & 3800 &  PC  \nl
NGC 3801 & F555W &  460 & F814W &  460 &  LRF680 & 3300 &  WF2 \nl
NGC 3862 & F547M$^{*}$ &  900 & F791W$^{*}$ &  750 &  F673N$^{*}$  & 2500 &  PC \nl
NGC 4261 & F547M$^{*}$ &  800 & F791W$^{*}$ &  800 &  F675W$^{*}$  & 2000 &  PC \nl 
NGC 4335 & F555W &  460 & F814W &  460 &  LRF680 & 3600 &  WF2 \nl
NGC 4374 & F547M$^{*}$ & 1200 & F814W$^{*}$ &  520 &  F658N$^{*}$  & 2600 &  PC \nl 
NGC 4486 & F555W$^{*}$ & 1200 & F814W$^{*}$ & 1200 &  F658N$^{*}$  & 2700 &  PC \nl
NGC 5127 & F555W &  460 & F814W &  460 &  LRF680 & 3300 &  WF2 \nl
NGC 5141 & F555W &  460 & F814W &  460 &  LRF680 & 3400 &  WF2 \nl
NGC 5490 & F555W &  460 & F814W &  460 &  LRF680 & 3300 &  WF2 \nl
NGC 7052 & F547M$^{*}$ &  800 & F814W$^{*}$ & 1400 &  LRF680$^{*}$ & 3700 &  WF2 \nl
NGC 7626 & F555W$^{*}$ & 1000 & F814W$^{*}$ &  460 &  LRF680 & 5100 &  WF2 \nl
\enddata

\tablecomments{
\footnotesize Col.~(1): Target NGC or UGC number.
  Cols.~(2) and (3): name and exposure time for the $V$ band filter.
  Cols.~(4) and (5): name and exposure time for the $I$ band filter.
  Cols.~(6) and (7): name and exposure time for the narrow-band
  filter. Depending on the redshift of the galaxy either the linear
  ramp filter (LRF) {\tt FR680N} or narrow-band filters {\tt F658} and
  {\tt F673} were used in order to have the
  \HalphaNII\ emission lines centered on the passband of the filter.
  NGC 4261 was observed through a broad-band filter since no
  narrow-band filter contained the \HalphaNII\ emission lines. All observations
  marked with a $^{*}$ were not observed in our program but retrieved from the 
{\tt HST} archive. Col.~(8)
  lists the {\tt WFPC2} CCD on which the target was positioned for the
  narrow-band exposure. For the LRF, which is a tunable narrow-band
  filter the positioning depends on the chosen central wavelength. The
  {\tt PC} CCD has a FOV of approximately $36''$ x $36''$ and has a
  pixel size of $0.0455''$. The {\tt WF2} CCD has a FOV of
  approximately $80''$ x $80''$ and has a pixel size of $0.0996''$.
  Targets were positioned on the PC CCD for all broad-band
  observations except one. The {\tt F547M} observation of NGC 7052 was
  positioned on the WF2 CCD.}
\end{deluxetable}

\begin{deluxetable}{lrrrrrrrc}
\scriptsize
\tablewidth{0pt}
\tablecaption{H$\alpha$+[NII]\ emission properties\label{t_emi}}
\tablehead{
\colhead{Name} & \colhead{F(${\rm <1kpc}$)} & \colhead{F($<1''$)} & 
\colhead{F$_{\rm tot}$} & 
\colhead{L(${\rm <1kpc}$)} & \colhead{L($<1''$)} & \colhead{L$_{\rm tot}$} &
\colhead{L$_{\rm lit.}$} & \colhead{Ref.} \\
\colhead{} & \colhead{\tiny ($10^{-14}$erg/s/cm$^{2}$)} & \colhead{\tiny 
($10^{-14}$erg/s/cm$^{2}$)} & \colhead{\tiny 
($10^{-14}$erg/s/cm$^{2}$)} &
 \colhead{\tiny ($10^{39}$ erg/s)} & \colhead{\tiny ($10^{39}$ 
erg/s)} & \colhead{\tiny ($10^{39}$ erg/s)} &
\colhead{\tiny ($10^{39}$ erg/s)} & \colhead{} \\
\colhead{(1)} & \colhead{(2)} & \colhead{(3)} & \colhead{(4)} & \colhead{(5)} & 
\colhead{(6)} & \colhead{(7)} & \colhead{(8)} & \colhead{(9)} \\}

\startdata
NGC 193 &   1.3 &   0.8 &   1.3 &   5.4 &   3.1 &   5.4 \nl
NGC 315 &   4.4 &   3.9 &   4.4 &  24.1 &  21.6 &  24.1 & 26.5 & 1  \nl
NGC 383 &   2.9 &   2.4 &   3.0 &  14.5 &  12.0 &  15.2 \nl
NGC 541 &   0.7 &   0.6 &   0.7 &   4.3 &   3.9 &   4.3 & 15.1 & 2  \nl
NGC 741 &   1.3 &   0.6 &   1.3 &   7.8 &   3.3 &   7.8 \nl
UGC 1841 &   2.3 &   2.2 &   2.3 &  20.0 &  19.2 &  20.0 \nl
NGC 2329 &   0.9 &   0.9 &   0.9 &   6.0 &   6.0 &   6.0 \nl
NGC 2892 &   1.1 &   1.1 &   1.1 &  11.1 &  11.1 &  11.1 \nl
NGC 3801 &   3.4 &   0.3 &   8.9 &   7.6 &   0.8 &  20.1 \nl
NGC 3862 &   6.9 &   6.9 &   6.9 &  58.8 &  58.8 &  58.8 \nl
NGC 4261 &   7.2 &   6.7 &   7.2 &   7.4 &   7.0 &   7.4 &  6.8 & 3  \nl
NGC 4335 &   0.2 &   0.2 &   0.2 &   0.9 &   0.9 &   0.9 \nl
NGC 4374 &  29.8 &   7.5 &  29.8 &   8.5 &   2.1 &   8.5 &  6.1 & 4  \nl
NGC 4486 &  53.6 &  37.6 &  53.6 &  15.2 &  10.7 &  15.2 &  5.7 & 6   \nl
NGC 5127 &   0.6 &   0.3 &   0.6 &   3.2 &   1.3 &   3.2 & $<8.1$ & 5  \nl
NGC 5141 &   1.0 &   0.9 &   1.0 &   6.2 &   5.6 &   6.2 &  5.3 & 5 \nl
NGC 5490 &   0.7 &   0.7 &   0.7 &   4.2 &   4.2 &   4.2 \nl
NGC 7052 &   2.6 &   1.9 &   2.6 &   9.4 &   7.0 &   9.4 &  8.7 & 5  \nl
NGC 7626 &   0.7 &   0.7 &   0.7 &   1.8 &   1.8 &   1.8 & $<0.6$ & 2  \nl
\enddata
\tablecomments{
Col.~(2): \HalphaNII\ flux within a
  circular aperture of 1 kpc. Col.~(3): \HalphaNII\ flux within a
  circular aperture of $1''$. Col.~(4): total \HalphaNII\ flux in
  image. Cols.~(5)-(7): same as Col.~(2)-(4) but converted to
  \HalphaNII\ luminosities. Col.~(8) total \HalphaNII\ luminosities
  from fluxes in the literature. Col.~(9): references used in
  Col.~(8). 1: Ho \etal 1997, 2: Macchetto \etal 1996, 3: Goudfrooij
  \etal 1994a, 4: Bower \etal 1997, 5: Morganti \etal 1992, 6: Ford
  \etal 1994 (measured for $r <1''$).  }
\end{deluxetable}

\begin{deluxetable}{lcrrrrrrrr}
\small
\tablewidth{0pt} 
\tablecaption{Dust Properties\label{t_dust}}
\tablehead{ \colhead{Name} & \colhead{Morph.} & \colhead{Size} &
\colhead{Size} & \colhead{Axis ratio} & \colhead{PA$_{\rm dust}$} &
\colhead{PA$_{\rm gal}$} & \colhead{PA$_{\rm radio}$} &
\colhead{M$_{\rm dust,1}$} & \colhead{M$_{\rm dust,2}$} \\ \colhead{}
& \colhead{} & \colhead{($''$)} & \colhead{(pc)} & \colhead{$b/a$} &
\colhead{($\deg$)} & \colhead{($\deg$)} & \colhead{($\deg$)} &
\colhead{(10$^{4}$ M$_{\odot}$)} & \colhead{(10$^{4}$ M$_{\odot}$)} \\
\colhead{(1)} & \colhead{(2)} & \colhead{(3)} & \colhead{(4)} &
\colhead{(5)} & \colhead{(6)} & \colhead{(7)} & \colhead{(8)} &
\colhead{(9)} & \colhead{(10)} \\ } 

\startdata 
NGC 193  & L & 3.0 & 840 & (0.18)& 0 & 58 & 104 & 4 & 13 \nl
         & I & 5.4 & 1510 & & & & & & \nl 
NGC 315  & D & 2.5 & 820 & 0.23 & 40 & 39 & 130 & 1 & 3 \nl 
NGC 383  & D & 7.4 & 2340 & 0.77 & 138 & 144 & 162 & 13 & 48 \nl 
NGC 541  & D & 1.8 & 640 & 0.91 & - & 143 & 76 & 0.3 & 1 \nl 
NGC 741  & - & - & - & - & - & 92 & - & 0.0 & 0.0 \nl 
UGC 1841 & D & 0.8 & 330 & $\sim$ 0.98 & - & 91 & 50 & 0.4 & 1.3 \nl 
NGC 2329 & D & 2.0 & 740 & 0.68 & 174 & 171 & 150 & 1.4 & 5.3 \nl 
UGC 2892 & - & - & - & - & - & 164 & 52 & 0.0 & 0.0 \nl 
NGC 3801 & L & 21.0 & 4420 & (0.12)& 24 & 121 & 121 & $>$42 & $>$146 \nl
         & I & 60.0 & 12630 & - & 114 & & & & \nl 
NGC 3862 & D & 1.5 & 610 & $\sim$ 0.99 & - & 25 & 30 & 0.5 & 1.5 \nl 
NGC 4261 & D & 1.7 & 240 & 0.46 & 163 & 155 & 87 & 0.2 & 0.6 \nl 
NGC 4335 & D & 2.5 & 750 & 0.41 & 158 & 156 & 82 & 18 & 58.2 \nl
         & I & 13.5 & 4030 & & & & & & \nl 
NGC 4374 & L & 13.1 & 980 & (0.15)& 79 & 128 & 1 & 0.3 & 1.0 \nl 
NGC 4486 & I & 11.0 & 560 & - & - & 128 & 112 & 0.1 & 0.3 \nl 
NGC 5127 & L & 3.0 & 940 & (0.25)& 48 & 68 & 118 & 1.6 & 6.1 \nl
NGC 5141 & L & 2.3 & 550 & (0.25)& 88 & 65 & 12 & 0.8 & 2.6 \nl 
NGC 5490 & L(D?) & 0.5 & 170 & (0.35)& 143 & 1 & 75 & $<$0.1 & $<$0.2 \nl
NGC 7052 & D & 4.0 & 1080 & 0.30 & 65 & 64 & 23 & 1.5 & 4.8 \nl 
NGC 7626 & L & 1.0 & 230 & (0.17)& 167 & 171 & 44 & $<$0.1 & $<$0.2 \nl
\enddata 
\tablecomments{
\footnotesize Col.~(1): NGC / UGC number. Col.~(2): morphology of
  dust feature; D=disk, L=lane, I=irregular.  Col.~(3-4):
  angular and physical length and diameter of lane and disk
  respectively. For NGC 193, NGC 3801 and NGC 4335 we give the size
  of both the central lanes or disk and the large dust filaments.
  Col.~(5): Ratio of the minor and major axis of the disks.  Values in
  parentheses give the ratio of width and length of the dust lanes.
  Col.~(6-8): the position angles of the major axis of the dust,
  of the galaxy major axis just outside the central dust
  distribution and of the radio jet axis at arcsecond scale or
  smaller. Col.~(9): dust mass assuming a front screen of dust.
  Col.~(10): dust mass assuming a screen of dust half way in the
  galaxy.  }
\end{deluxetable}



\end{document}